\documentclass[letterpaper,11pt]{article}

\usepackage{amsmath,amssymb,amstext,mathrsfs}
\usepackage{slashed}
\usepackage{placeins}
\usepackage{float}
\usepackage{array}
\newcolumntype{P}[1]{>{\raggedright\arraybackslash}p{#1}}

\usepackage{graphicx}
\usepackage{tikz}
\usepackage{xcolor}

\usepackage{cite}

\usepackage[colorlinks=true,linkcolor=blue,citecolor=blue,urlcolor=blue]{hyperref}

\renewcommand{\baselinestretch}{1.2}
\allowdisplaybreaks

\newcommand{\cL}{\mathcal{L}}
\newcommand{\rmi}{{\rm i}}
\newcommand{\cV}{\mathcal{V}}
\newcommand{\cZ}{\mathcal{Z}}
\newcommand{\bZ}{\bar Z}
\newcommand{\bcZ}{\bar{\mathcal{Z}}}
\newcommand{\cF}{\mathcal{F}}
\newcommand{\cLW}{\cL_{C^2+\frac16R^2}}

\makeatletter
	\@addtoreset{equation}{section}
	\makeatother

\hypersetup{
  pdftitle={Spectra and Central Charges of Gauged Curvature Squared Supergravities in Five Dimensions},
  pdfauthor={Erol Aksoy, Cagatay Okyay, Mehmet Ozkan, Cemal Berfu Senisik}
}

\begin{document}
		\begin{titlepage}%
			\bigskip
			\begin{center}
				{\LARGE\bfseries Spectra and Central Charges of Gauged\\ Curvature Squared Supergravities in Five Dimensions}
				\\[10mm]
				\textbf{Erol Aksoy, Cagatay Okyay, Mehmet Ozkan and Cemal Berfu Senisik}\\[5mm]
				\vskip 25pt

				{\em  \hskip -.1truecm Department of Physics, İstanbul Technical University,  \\
					Maslak, 34469, İstanbul, T\"urkiye  \vskip 10pt }

				{email: {\tt aksoyer16@itu.edu.tr, okyay15@itu.edu.tr, ozkanmehm@itu.edu.tr, senisik@itu.edu.tr}}

			\end{center}

			\vspace{3ex}

			\begin{center}
				{\bfseries Abstract}
			\end{center}
			\begin{quotation} \noindent
We study the spectra and the central charges of off-shell gauged five-dimensional $\mathcal N=2$ supergravity extended by the Weyl squared, Ricci tensor squared and Ricci scalar squared invariants, and by their Gauss-Bonnet combination. Working exactly in the couplings and keeping all auxiliary fields, we organize the spectra around the maximally supersymmetric AdS$_5$ vacuum into supermultiplets. With the Ricci scalar squared invariant, the spectrum gains a long vector multiplet and is free of ghosts in a finite range of the coupling, while with the Weyl squared or Ricci tensor squared invariant, it gains a long spin-2 multiplet and always contains ghosts or instabilities. The Gauss-Bonnet combination adds no new modes: around its supersymmetric vacuum, the quadratic action is a rescaled copy of that of the two-derivative theory, and this fixes the spectrum of the Ricci tensor squared extension. We also compute the central charges $a$ and $c$ and the flavor central charge exactly in the couplings. The vacuum equation is $a$-maximization, and the kinetic terms of the massless graviton and vector multiplets are proportional to $c$ and to the flavor central charge, so that the critical points, where the massive spin-2 multiplet becomes degenerate with the graviton multiplet, are zeros of $c$. Finally, around supersymmetric AdS$_3\times\Sigma_2$ backgrounds of the Gauss-Bonnet extension, the two-dimensional central charges are reproduced by $c$-extremization, while the spectrum is generically not unitary and contains towers of states whose energies stay bounded at large Kaluza-Klein levels. In part of the parameter space, the tower of spin-2 multiplets has positive norm, so that the spectrum has no finite density of states.
			\end{quotation}

			\vfill
			\flushleft{November 11, 2026}
		\end{titlepage}

		\tableofcontents

\bigskip

\section{Introduction}
\label{sec:intro}
\paragraph{}
Higher-derivative corrections to the two-derivative supergravity actions arise naturally in string theory, where they encode the corrections in the string length to the low-energy effective actions \cite{Zwiebach:1985uq,Gross:1986iv}. In the context of the AdS/CFT correspondence \cite{Maldacena:1997re,Gubser:1998bc,Witten:1998qj}, the four-derivative terms of five-dimensional gauged supergravity capture the leading corrections in the large $N$ expansion of the dual four-dimensional superconformal field theories. In particular, they are responsible for the difference between the central charges $a$ and $c$ of the dual theory, and they modify the transport coefficients of the dual plasma as well as the thermodynamics of black holes \cite{Kats:2007mq,Buchel:2008vz,Castro:2008ne,Cremonini:2008tw,Bobev:2021qxx,Cassani:2022lrk}. Since the dual theories of interest are supersymmetric, it is desirable to work with the supersymmetric completions of the four-derivative terms, and in five dimensions all curvature squared invariants of gauged $\mathcal N=2$ supergravity have now been constructed \cite{Gold:2023ymc,Gold:2025ttt}.

Supersymmetric higher-derivative actions can be constructed by a variety of methods \cite{Ozkan:2024euj}, but broadly speaking, there are two approaches, on-shell and off-shell. In the on-shell approach, the supersymmetry transformation rules depend on the action, and they have to be modified order by order in the couplings of the higher-derivative terms, which makes the construction rather involved. In the off-shell approach, the supersymmetry transformation rules are independent of the action, so that each higher-derivative invariant is separately supersymmetric. The invariants can then be freely combined with each other and with the two-derivative action, and the resulting actions are exact in their couplings. The price to pay is the necessity to introduce auxiliary fields, which may become dynamical in the presence of higher-derivative terms. In five dimensions, the minimal off-shell supergravity multiplet was introduced in superspace in \cite{Howe:1981ev,Howe:1981nz}, and off-shell $\mathcal N=2$ supergravity has since been developed both in superspace \cite{Kuzenko:2007aj,Kuzenko:2007cj,Kuzenko:2007hu,Kuzenko:2008wr,Kuzenko:2014eqa,Butter:2014xxa,Becker:2019fri} and by means of the superconformal tensor calculus \cite{Kugo:2000af,Bergshoeff:2001hc,Fujita:2001kv,Bergshoeff:2002qk,Bergshoeff:2004kh}. The two approaches are related by the conformal superspace of \cite{Butter:2014xxa}, which reduces to the superconformal tensor calculus upon a partial gauge fixing. These formalisms greatly simplify the construction of higher-derivative invariants. In the superconformal tensor calculus, there are two versions of the Weyl multiplet of conformal supergravity, the standard Weyl multiplet and the dilaton Weyl multiplet \cite{Bergshoeff:2001hc}, see also \cite{Hutomo:2022hdi}, and Poincar\'e supergravity is obtained by coupling the Weyl multiplet to compensating matter multiplets and fixing the additional gauge symmetries. In the standard Weyl multiplet, the supersymmetric completion of the Weyl tensor squared was constructed in \cite{Hanaki:2006pj}. In the dilaton Weyl multiplet, the Riemann squared invariant was constructed in \cite{Bergshoeff:2011xn} by means of a map between the Yang-Mills multiplet and the dilaton Weyl multiplet, and the supersymmetric completion of the Weyl tensor squared, whose curvature squared terms become $C_{abcd}C^{abcd}+\tfrac16R^2$ in this multiplet, was constructed in \cite{Ozkan:2013uk}, where the two invariants were combined into the supersymmetric Gauss-Bonnet combination. The Ricci scalar squared invariant was constructed in \cite{Ozkan:2013nwa}. In conformal superspace, the curvature squared invariants of the standard Weyl multiplet were obtained in \cite{Butter:2014xxa}, which also gave the first supersymmetric completion of the Ricci tensor squared. This is a third independent curvature squared invariant, the LOG invariant, so named because it is built from the logarithm of a vector multiplet, whose superspace form was developed further in \cite{Gold:2023dfe} and whose complete component form was given in \cite{Gold:2023ykx}. The internally gauged two-derivative theory, based on the dilaton Weyl multiplet and a linear multiplet compensator, was constructed in \cite{Coomans:2012cf}, see also \cite{Ozkan:2016csy}, and the gauged versions of the three curvature squared invariants were given in \cite{Gold:2023ymc,Gold:2025ttt}.

Higher-derivative terms generically propagate additional modes, some of which are ghosts \cite{Stelle:1977ry}. The Gauss-Bonnet combination is an exception, since it propagates only the massless graviton around maximally symmetric vacua \cite{Zwiebach:1985uq,Boulware:1985wk,Deser:2002jk}. Around AdS vacua, the masses of the additional modes depend on the couplings, and at special values of the couplings, massive modes become degenerate with massless ones. At such critical points, the massive graviton of quadratic gravity is replaced by a logarithmic mode \cite{Lu:2011zk,Deser:2011xc}, the same structure arises in supergravity \cite{Lu:2011mw}, and the inner product on the massless and logarithmic modes is indefinite \cite{Porrati:2011ku}. In off-shell supergravity, the additional modes of the metric are joined by those of the auxiliary fields, which become dynamical, and the complete spectrum is organized into multiplets of the superalgebra of the vacuum. A detailed analysis of this kind was carried out in \cite{Pang:2012xs} for the off-shell gauged six-dimensional $\mathcal N=(1,0)$ supergravity extended by a Riemann squared invariant, around its supersymmetric Minkowski$_4\times S^2$ vacuum, and in five dimensions, the spectrum of the Gauss-Bonnet extension of ungauged supergravity around its Minkowski$_5$ vacuum was obtained in \cite{Ozkan:2013uk}. In six dimensions, the spectrum of Gauss-Bonnet supergravity around its supersymmetric AdS$_3\times S^3$ background was studied in \cite{Novak:2017wqc,Butter:2018wss}. In gauged supergravity, the maximally supersymmetric vacuum is AdS$_5$, and its spectrum determines the operator content of the putative dual superconformal field theory.

In this paper, we study the spectra of off-shell gauged $D=5$, $\mathcal N=2$ supergravity extended by the $C^2+\tfrac16R^2$, LOG and Ricci scalar squared invariants, as well as by their Gauss-Bonnet combination. Our starting point is the general model (\ref{eq:general}), which contains the couplings $\alpha$, $\beta$ and $\zeta$ of the three invariants. We treat these couplings as finite parameters, that is, we neither expand in the couplings nor eliminate the auxiliary fields, and we derive both the bosonic and the fermionic quadratic actions from the complete off-shell Lagrangians. Following \cite{Pang:2012xs}, we determine the masses of all modes and the signs of their kinetic terms, and we organize the spectra into supermultiplets. We give the bosonic Lagrangians of the gauged invariants in terms of the independent fields, including the terms that arise from the gauged Bianchi identity of the three-form field strength. As we will show, these terms are needed for the supermultiplet structure of the spectrum around backgrounds with fluxes. The complete Lagrangians up to quadratic order in fermions are lengthy, and we make them available in \cite{repo}, together with their quadratic actions around the supersymmetric AdS$_5$ vacuum and the resulting spectra. In deriving the fermionic terms, we have also found that some of the fermionic terms of the composite expressions given in \cite{Ozkan:2013uk} are not correct, and we give the corrected expressions in Appendix~\ref{app:composite}.

We first consider the maximally supersymmetric AdS$_5$ vacuum. As we will show, it depends on the couplings only through the combination $\tau=(\beta+6\zeta)g^2$, where $g$ is the gauge coupling, and it exists for $\tau\leq2$. The $C^2+\tfrac16R^2$ invariant does not contribute to the vacuum, since the Weyl tensor and the other covariant quantities of the standard Weyl multiplet, from which it is built, vanish on every maximally supersymmetric AdS$_5$ vacuum. It contributes, however, to the quadratic action, and the four models have rather different spectra. In addition to the massless graviton and vector multiplets of the two-derivative theory, which contain $12+12$ states, the Ricci scalar squared extension propagates a long vector multiplet with $8+8$ states, and its spectrum is free of ghosts and tachyons for $0<\zeta g^2<\tfrac14$. The $C^2+\tfrac16R^2$ extension propagates a long spin-2 multiplet with $32+32$ states, and there is no range of its coupling in which the spectrum is free of both ghosts and instabilities. At $\alpha g^2=-2$, the long multiplet becomes degenerate with the massless graviton multiplet, and logarithmic partners appear, as in critical gravity, but the components of the long multiplet that do not combine with the graviton multiplet retain negative norm. The Gauss-Bonnet extension, on the other hand, shifts the vacuum but propagates no additional modes. Around its supersymmetric vacuum, its complete quadratic action, including the fermions, is a rescaled copy of that of the two-derivative theory, so that it is free of ghosts for all $\lambda g^2<1$, where $\lambda$ is its coupling. Since the LOG invariant is a linear combination of the Gauss-Bonnet and the $C^2+\tfrac16R^2$ invariants, this result determines the spectrum of the LOG extension in terms of that of the $C^2+\tfrac16R^2$ extension with a related coupling, which we also confirm by a direct computation in both sectors.

Our second main result concerns the central charges of the putative dual $\mathcal N=1$ superconformal field theory, which give these results a simple interpretation. We compute the central charges $a$ and $c$, as well as the two-point coefficient $C_F$ of the flavor current that is dual to the massless vector multiplet, exactly in the couplings. The kinetic terms of the massless graviton multiplet are proportional to $c$ and those of the massless vector multiplet to $C_F$, so that the points at which these kinetic terms vanish, which include the critical points at which the long spin-2 multiplet becomes degenerate with the graviton multiplet, are zeros of $c$ or of $C_F$. Moreover, the Lagrangian evaluated on the supersymmetric configurations, with the scalar of the linear multiplet left free, is the trial $a$-function of the dual theory, obtained from the 't~Hooft anomalies that follow from the Chern-Simons terms, and the vacuum equation is $a$-maximization \cite{Intriligator:2003jj,Tachikawa:2005tq}, as was first shown in \cite{Hanaki:2006pj} in the presence of the Weyl tensor squared invariant and as follows from the general analysis of \cite{Hristov:2025ygn}. At first order in the couplings, our central charges agree with those of \cite{Gold:2023ymc}. For the Gauss-Bonnet extension, $a$ and $c$ coincide with those of Einstein-Gauss-Bonnet gravity, and the bounds of \cite{Hofman:2008ar} on $a/c$ hold in a range of the coupling that lies strictly inside the range in which the spectrum is free of ghosts. Our spectra also illustrate the fact, emphasized in \cite{Pang:2012xs}, that a supersymmetric vacuum need not be stable, since the positivity of energy relies on a positive-definite norm, which is violated by the ghosts that arise from the higher-derivative terms. In fact, the unitarity of a representation of the superalgebra and the signs of the norms of its states are independent conditions. For $-2<\alpha g^2<0$, for instance, the long spin-2 multiplet of the $C^2+\tfrac16R^2$ extension is a unitary representation of $SU(2,2|1)$, but all of its states have negative norm.

The simplicity of the Gauss-Bonnet extension around the AdS$_5$ vacuum is partly due to the vacuum itself, on which all vector and tensor fields vanish together with the covariant quantities of the standard Weyl multiplet, so that most of the couplings of the invariant do not contribute to the quadratic action. In order to study the Gauss-Bonnet extension in a less symmetric setting, we consider its supersymmetric AdS$_3\times\Sigma_2$ backgrounds, where $\Sigma_2$ is a sphere, a torus or a compact quotient of the hyperbolic plane, supported by the fluxes of the two vector fields through $\Sigma_2$ and by a three-form flux along AdS$_3$. These backgrounds preserve four real supersymmetries, and they are candidate near-horizon geometries of supersymmetric black strings \cite{Maldacena:2000mw,Benini:2013cda}. We expand the fluctuations in harmonics on $\Sigma_2$ and determine the spectrum at every Kaluza-Klein level from the quadratic action evaluated on modes of fixed scaling dimension. The spectrum turns out to be very different from that of the two-derivative theory. At a generic level, it consists of ten long multiplets of $su(1,1|1)$, compared to six in the two-derivative theory, and one of them is a spin-2 multiplet that originates from the gravitational Chern-Simons term, as in topologically massive gravity \cite{Deser:1981wh,Deser:1982vy}. The spectrum is generically not unitary, since it contains negative-norm modes for every nonzero value of the coupling away from a few exceptional loci, and at large levels, towers of states accumulate at a finite energy; for a range of the parameters that we determine in closed form and that includes the two-derivative limit, the tower of spin-2 multiplets has positive norm, so that, if all modes are retained, the spectrum on a compact $\Sigma_2$ has no finite density of states. Thus the same invariant that is free of ghosts around AdS$_5$ is not unitary around its flux backgrounds, and the difference is due to the Chern-Simons couplings to the fluxes, which do not contribute on AdS$_5$. The central charges of the dual two-dimensional theory, which we obtain in closed form for all values of the fluxes and of the coupling, are reproduced by $c$-extremization \cite{Benini:2013cda}. We have also computed the fermionic spectrum directly at a number of points and found that it completes the bosonic multiplets. This tests the two-fermion terms of the Gauss-Bonnet invariant, most of which do not contribute around the AdS$_5$ vacuum.

This paper is organized as follows. In Section~\ref{sec:theory}, we present the off-shell gauged $D=5$, $\mathcal N=2$ supergravity and the three curvature squared invariants, give their bosonic Lagrangians in terms of the independent fields, and introduce the general model and the Gauss-Bonnet combination. In Section~\ref{sec:AdS5}, we study the spectra around the supersymmetric AdS$_5$ vacuum. In Section~\ref{sec:vacuum}, we determine the vacuum, set up the linearized analysis and recall the multiplets of $SU(2,2|1)$, and we summarize the results of the section in Figure~\ref{fig:phases} and Table~\ref{tab:summary}. We start with the Ricci scalar squared extension in Section~\ref{sec:R2}, since its additional modes form a long vector multiplet, the smallest of the multiplets we encounter, and we then turn to the $C^2+\tfrac16R^2$ extension in Section~\ref{sec:W}, which propagates a long spin-2 multiplet. In both cases, we study the bosonic and the fermionic sectors separately and then arrange the modes into supermultiplets, as in \cite{Pang:2012xs}. We study the Gauss-Bonnet extension in Section~\ref{sec:GB}, before the LOG extension in Section~\ref{sec:LOG}, although it is a combination of the $C^2+\tfrac16R^2$ and LOG invariants, because the spectrum of the LOG extension follows from those of the $C^2+\tfrac16R^2$ and Gauss-Bonnet extensions. For the same reason, the bosonic and the fermionic sectors of these two extensions are not treated separately. In Section~\ref{sec:central}, we compute the central charges of the general model and relate them to the special points of the spectra. In Section~\ref{sec:GBprod}, we study the Gauss-Bonnet extension around supersymmetric AdS$_3\times\Sigma_2$ backgrounds. We determine the backgrounds in Section~\ref{sec:GBprodbg} and describe our method in Section~\ref{sec:GBprodfluc}. We give the spectrum at generic levels in Section~\ref{sec:GBprodgeneric} and at special levels, together with the central charges of the dual two-dimensional theory, in Section~\ref{sec:GBprodspecial}. We discuss the norms in Section~\ref{sec:GBprodnorms} and compare the fermionic spectrum with the bosonic one in Section~\ref{sec:GBprodferm}. We present our conclusions in Section~\ref{sec:conclusions}. Our conventions are given in Appendix~\ref{app:conventions}, the corrected composite expressions in Appendix~\ref{app:composite}, and the longer polynomials of Section~\ref{sec:GBprod}, including the one that determines the $(1,0)$ multiplets, in Appendix~\ref{app:quintic}.

\section{\texorpdfstring{Off-Shell Gauged $D=5$, $\mathcal N=2$ Supergravity and Curvature Squared Invariants}{Off-Shell Gauged D=5, N=2 Supergravity and Curvature Squared Invariants}} \label{sec:theory}
\paragraph{}

In this section, we present the off-shell supergravity models whose spectra are studied in this paper. These are the off-shell gauged $D=5$, $\mathcal N=2$ supergravity \cite{Coomans:2012cf,Ozkan:2016csy,Gold:2025ttt} and its extensions by three curvature squared invariants: the invariant that arises from the supersymmetric completion of the Weyl tensor squared \cite{Hanaki:2006pj,Ozkan:2013uk,Butter:2014xxa}, the LOG invariant \cite{Butter:2014xxa,Gold:2023dfe,Gold:2023ykx,Gold:2025ttt} and the Ricci scalar squared invariant \cite{Ozkan:2013nwa,Ozkan:2016csy,Butter:2014xxa,Gold:2023dfe}. We present the bosonic parts of the Lagrangians in terms of the independent fields, together with the supersymmetry transformation rules of the fermions. The complete Lagrangians up to quadratic order in fermions, which retain their full nonlinear bosonic dependence, are lengthy, and we make them available in \cite{repo}. Our conventions are summarized in Appendix~\ref{app:conventions}.

\subsection{Two-Derivative Gauged Supergravity}
\label{sec:twoderivative}

The off-shell $D=5$, $\mathcal N=2$ gauged supergravity multiplet consists of the fields \cite{Coomans:2012cf,Ozkan:2016csy,Gold:2025ttt}
\begin{equation}
\left(e_\mu{}^a,\ \psi_\mu^i,\ C_\mu,\ B_{\mu\nu},\ E_{\mu\nu\rho},\ L,\ N,\ V_\mu,\ V^{\prime}_\mu{}^{ij},\ \varphi^i\right),
\label{eq:fields}
\end{equation}
where $B_{\mu\nu}$ and $E_{\mu\nu\rho}$ are antisymmetric tensor fields, $L$ and $N$ are real scalars, $V^{\prime}_\mu{}^{ij}$ is symmetric in its $SU(2)$ indices and satisfies $\delta_{ij}V^{\prime}_\mu{}^{ij}=0$, and the spinors are symplectic Majorana. The above fields have $(10,32,4,6,4,1,1,4,10,8)$ degrees of freedom. Of these, $L$, $\varphi^i$, $E_{\mu\nu\rho}$ and $N$ are the fields of the linear multiplet compensator, with its $SU(2)$ triplet $L^{ij}$ gauge fixed to the singlet $L$ \cite{Coomans:2012cf}, see Appendix~\ref{app:composite}, and the remaining fields belong to the dilaton Weyl multiplet. The vector fields $V_\mu$ and $V^{\prime}_\mu{}^{ij}$ combine into the $SU(2)$ gauge field $V_\mu{}^{ij}=V^{\prime}_\mu{}^{ij}+\tfrac12\delta^{ij}V_\mu$, of which $V_\mu$ gauges the $U(1)$ R-symmetry. In the bosonic Lagrangians, it is convenient to combine the two independent components of $V^{\prime}_\mu{}^{ij}$ into the complex vector
\begin{equation}
Z_\mu=V^{\prime}_\mu{}^{12}-\rmi V^{\prime}_\mu{}^{11}\,,
\label{eq:Z}
\end{equation}
which carries charge under $V_\mu$.

The bosonic part of the Lagrangian is given by
\begin{align}
e^{-1}\cL_{LR}={}&L\Big(R-\tfrac12G^2+4H^2\Big)+L^{-1}\partial_aL\partial^aL+4LZ_a\bZ^a-2L^{-1}N^2-2L^{-1}E_aE^a\nonumber\\*
&-2\sqrt2E^aV_a-4gE^aC_a-2gLN-4gN+2g^2L^2-\tfrac12g^2L^3\,,\label{eq:LLR}
\end{align}
where $g$ is the gauge coupling constant. In (\ref{eq:LLR}) and in the higher-derivative invariants below, the $g$-independent terms are given first, followed by the terms of increasing order in $g$. The field strengths in (\ref{eq:LLR}) are defined as
\begin{equation}
G_{\mu\nu}=2\partial_{[\mu}C_{\nu]}\,,\qquad
H_{ab}=-\tfrac{1}{12}\epsilon_{abcde}H^{cde}\,,\qquad
E^\mu=-\tfrac{1}{12}e^{-1}\epsilon^{\mu\nu\rho\sigma\lambda}\partial_\nu E_{\rho\sigma\lambda}\,,
\label{eq:strengths}
\end{equation}
where the three-form field strength is given by
\begin{equation}
H_{\mu\nu\rho}=3\partial_{[\mu}B_{\nu\rho]}+\tfrac32C_{[\mu}G_{\nu\rho]}+\tfrac12gE_{\mu\nu\rho}\,.
\label{eq:H3}
\end{equation}
Here and in the following, we use the notation
\begin{equation}
G^2=G_{ab}G^{ab}\,,\quad G^4=G_{ab}G^{bc}G_{cd}G^{da}\,,\quad H^2=H_{ab}H^{ab}\,,\quad H^4=H_{ab}H^{bc}H_{cd}H^{da}\,.
\label{eq:notation}
\end{equation}
Notice that the gauging deforms the three-form field strength by the three-form of the linear multiplet. As a consequence, the Bianchi identity for the dual field strength reads
\begin{equation}
\nabla^bH_{ab}=-\tfrac{1}{16}\epsilon_{abcde}G^{bc}G^{de}+\tfrac12gE_a\,,
\label{eq:Bianchi}
\end{equation}
which we use in the higher-derivative invariants below. The bosonic gauge transformations of the form fields are
\begin{equation}
\delta C_\mu=\partial_\mu\Lambda\,,\qquad
\delta B_{\mu\nu}=2\partial_{[\mu}\Lambda_{\nu]}-\tfrac12\Lambda G_{\mu\nu}-\tfrac12g\Lambda_{\mu\nu}\,,\qquad
\delta E_{\mu\nu\rho}=3\partial_{[\mu}\Lambda_{\nu\rho]}\,,
\label{eq:formgauge}
\end{equation}
and $Z_\mu$ transforms as $\delta Z_\mu=\rmi\theta Z_\mu$ under $\delta V_\mu=\partial_\mu\theta$, so that its covariant derivative is $D_\mu Z_\nu=\nabla_\mu Z_\nu-\rmi V_\mu Z_\nu$. The supersymmetry transformation rules of the fermions, up to terms bilinear in fermions, are
\begin{align}
\delta\psi_\mu^i={}&\big(\partial_\mu+\tfrac14\omega_\mu{}^{ab}\gamma_{ab}\big)\epsilon^i-V_\mu{}^{ij}\epsilon_j-\tfrac12\rmi H_{ab}\gamma_\mu{}^{ab}\epsilon^i-\tfrac12\rmi G_{\mu a}\gamma^a\epsilon^i+\tfrac{1}{2\sqrt2}\rmi gL\delta^{ij}\gamma_\mu\epsilon_j\,,\nonumber\\*
\delta\varphi^i={}&\tfrac12\Big(N+\tfrac32gL^2\Big)\epsilon^i-\tfrac12\rmi\gamma^aE_a\epsilon^i-\tfrac{1}{2\sqrt2}\rmi\gamma^a\partial_aL\,\delta^{ij}\epsilon_j\nonumber\\*
&-\tfrac{1}{\sqrt2}\rmi L\gamma^aV^{\prime}_a{}^{(i}{}_k\delta^{j)k}\epsilon_j+\tfrac{1}{4\sqrt2}L\gamma^{ab}\big(G_{ab}+4H_{ab}\big)\delta^{ij}\epsilon_j\,.\label{eq:susy}
\end{align}
Note that the term $\tfrac34gL^2\epsilon^i$ in $\delta\varphi^i$, which arises from the gauging, plays a crucial role in the determination of the supersymmetric vacua in Section~\ref{sec:AdS5}.

The fields $N$, $V_\mu$ and $Z_\mu$ appear algebraically in (\ref{eq:LLR}). Their field equations give
\begin{equation}
N=-\tfrac12gL(L+2)\,,\qquad E^a=0\,,\qquad Z_\mu=0\,.
\end{equation}
The second equation implies that $E_{\mu\nu\rho}$ is pure gauge, and it can be absorbed into $B_{\mu\nu}$ by means of (\ref{eq:formgauge}). Upon eliminating these fields, the model describes $\mathcal N=2$ gauged supergravity coupled to a single vector multiplet whose scalar is $L$ \cite{Coomans:2012cf}. In the presence of the curvature squared invariants, these fields generically acquire kinetic terms, and we keep them off-shell throughout the paper.

\subsection{\texorpdfstring{$C^2+\tfrac16R^2$ Invariant}{C2 + R2/6 Invariant}}
\label{sec:weyl}

The supersymmetric completion of the Weyl tensor squared was first constructed in the standard Weyl multiplet background \cite{Hanaki:2006pj}. In the dilaton Weyl multiplet background, the auxiliary fields of the standard Weyl multiplet become composite and pick up terms that depend on the Ricci scalar \cite{Bergshoeff:2001hc,Ozkan:2013uk}. As a result, the curvature squared terms of the invariant change from the Weyl tensor squared to the combination $C_{abcd}C^{abcd}+\tfrac16R^2$, where $C_{abcd}$ is the Weyl tensor, and we refer to it as the $C^2+\tfrac16R^2$ invariant. Its gauged version was given in \cite{Gold:2023ymc,Gold:2025ttt}. In terms of the fields (\ref{eq:fields}), its bosonic part is given, up to a total derivative, by
\begin{align}
e^{-1}\cLW={}&\tfrac18R_{abcd}R^{abcd}-\tfrac16R_{ab}R^{ab}+\tfrac{1}{24}R^2-\tfrac16\cV_{ab}\cV^{ab}-\tfrac23\cZ_{ab}\bcZ^{ab}\nonumber\\*
&+\tfrac12R_{abcd}H^{ab}G^{cd}+\tfrac13R_{abcd}H^{ab}H^{cd}-\tfrac13R_{ab}G^{ac}G^b{}_c\nonumber\\*
&+\tfrac23R_{ab}H^{ac}G^b{}_c-\tfrac83R_{ab}H^{ac}H^b{}_c+R\Big(-\tfrac16G_{ab}H^{ab}+\tfrac23H^2\Big)\nonumber\\*
&-\tfrac{1}{24}(G^2)^2+2(H^2)^2+\tfrac14G^4+4H^4-\tfrac43G^2H^2+\tfrac16G^2\,G_{ab}H^{ab}\nonumber\\*
&+\tfrac83H^2\,G_{ab}H^{ab}-\tfrac13(G_{ab}H^{ab})^2-G_{ab}G^{bc}G_{cd}H^{da}\nonumber\\*
&+\tfrac83G_{ab}G^{bc}H_{cd}H^{da}+\tfrac23G_{ab}H^{bc}G_{cd}H^{da}-\tfrac{20}{3}G_{ab}H^{bc}H_{cd}H^{da}\nonumber\\*
&+\tfrac13\nabla_bG^{ab}\nabla^cG_{ac}-\tfrac43\nabla_aH_{bc}\nabla^aH^{bc}-\tfrac23\epsilon^{abcde}H_{ab}H_{cd}\nabla^fG_{ef}\nonumber\\*
&+\epsilon^{abcde}H_b{}^f\nabla_aH_{cf}G_{de}+\tfrac13\epsilon^{abcde}H_{ab}\nabla^fG_{cf}G_{de}\nonumber\\*
&+\tfrac{1}{48}\epsilon^{abcde}\nabla^fG_{af}G_{bc}G_{de}+\tfrac{1}{16}\epsilon^{abcde}C_aR_{bc}{}^{fg}R_{defg}-\tfrac{1}{24}\epsilon^{abcde}C_a\cV_{bc}\cV_{de}\nonumber\\*
&-\tfrac16\epsilon^{abcde}C_a\cZ_{bc}\bcZ_{de}+gN\Big(-\tfrac13R+\tfrac23G^2-\tfrac23G_{ab}H^{ab}-4H^2\Big)\nonumber\\*
&-\tfrac{1}{3\sqrt2}gL\,\cV_{ab}\big(G^{ab}-2H^{ab}\big)-\tfrac23gE_a\nabla_bG^{ab}-\tfrac{5}{24}g\epsilon^{abcde}E_aG_{bc}G_{de}\nonumber\\*
&+\tfrac13g\epsilon^{abcde}E_aH_{bc}H_{de}+g^2L^2\Big(-\tfrac{1}{12}R+\tfrac16G^2-\tfrac16G_{ab}H^{ab}-H^2\Big)\nonumber\\*
&+\tfrac43g^2N^2+g^2E_aE^a+\tfrac23g^3NL^2+\tfrac{1}{12}g^4L^4\,,\label{eq:LC2}
\end{align}
where
\begin{equation}
\cV_{ab}=2\partial_{[a}V_{b]}+2\rmi\big(\bZ_aZ_b-Z_a\bZ_b\big)\,,\qquad \cZ_{ab}=2D_{[a}Z_{b]}\,.
\label{eq:VZ}
\end{equation}
These are the $U(1)$ and the remaining components of the $SU(2)$ field strength $V_{ab}{}^{ij}$, such that $V_{ab}{}^{ij}V^{ab}{}_{ij}=\tfrac12\cV_{ab}\cV^{ab}+2\cZ_{ab}\bcZ^{ab}$. In (\ref{eq:LC2}), we have used the Bianchi identity of $G_{\mu\nu}$ to write the terms quadratic in its derivatives in terms of $\nabla^bG_{ab}$, and the same form is used for the LOG invariant below. A few comments are in order. First, the curvature squared terms combine into $\tfrac18\big(C_{abcd}C^{abcd}+\tfrac16R^2\big)$, in agreement with \cite{Ozkan:2013uk,Ozkan:2013nwa}. Second, the gauging generates a scalar potential, curvature couplings of $N$ and $L$, and the terms that contain $E_a$, which arise from the gauged Bianchi identity (\ref{eq:Bianchi}).\footnote{These terms, and the corresponding terms in (\ref{eq:Llog}), arise from the $g$-dependent term in (\ref{eq:Bianchi}), and they are absent in the bosonic Lagrangians given in \cite{Gold:2023ymc,Gold:2025ttt}. The expressions (\ref{eq:LC2}) and (\ref{eq:Llog}) are the corrected ones. Since we have used the Bianchi identities to simplify (\ref{eq:LC2}) and (\ref{eq:Llog}), the comparison with \cite{Gold:2023ymc,Gold:2025ttt} has to be made after the same rewriting.\label{fn:Eterms}} Third, the term $\tfrac{1}{16}\epsilon^{abcde}C_aR_{bc}{}^{fg}R_{defg}$ is the mixed gauge-gravitational Chern-Simons term. Finally, the $SU(2)$ field strength squared provides kinetic terms for $V_\mu$ and $Z_\mu$, which are auxiliary in the two-derivative theory.

\subsection{LOG Invariant}
\label{sec:log}

The LOG invariant was constructed in conformal superspace in \cite{Butter:2014xxa,Gold:2023dfe}, and in components in \cite{Gold:2023ykx}. Its bosonic part in the gauged dilaton Weyl background was given in \cite{Gold:2023ymc,Gold:2025ttt}. In terms of the fields (\ref{eq:fields}), it is given, up to a total derivative, by
\begin{align}
e^{-1}\cL_{\log}={}&-\tfrac16R_{ab}R^{ab}+\tfrac{1}{24}R^2+\tfrac{1}{12}\cV_{ab}\cV^{ab}+\tfrac13\cZ_{ab}\bcZ^{ab}-\tfrac23R_{abcd}H^{ab}H^{cd}\nonumber\\*
&+\tfrac16R_{ab}G^{ac}G^b{}_c-\tfrac43R_{ab}G^{ac}H^b{}_c+\tfrac43R_{ab}H^{ac}H^b{}_c+R\Big(\tfrac13G_{ab}H^{ab}-\tfrac13H^2\Big)\nonumber\\*
&+\tfrac{1}{48}(G^2)^2-2(H^2)^2-\tfrac18G^4+\tfrac23G^2H^2-\tfrac13G^2\,G_{ab}H^{ab}\nonumber\\*
&-\tfrac43H^2\,G_{ab}H^{ab}+\tfrac23(G_{ab}H^{ab})^2+G_{ab}G^{bc}G_{cd}H^{da}-\tfrac43G_{ab}G^{bc}H_{cd}H^{da}\nonumber\\*
&-\tfrac43G_{ab}H^{bc}G_{cd}H^{da}+\tfrac{16}{3}G_{ab}H^{bc}H_{cd}H^{da}-\tfrac16\nabla_bG^{ab}\nabla^cG_{ac}\nonumber\\*
&+\tfrac23\nabla_aH_{bc}\nabla^aH^{bc}+\tfrac{1}{48}\epsilon^{acdef}\nabla^bG_{ab}G_{cd}G_{ef}+\tfrac13\epsilon^{acdef}\nabla^bG_{ab}H_{cd}H_{ef}\nonumber\\*
&-\tfrac16\epsilon^{acdef}\nabla^bG_{ab}H_{cd}G_{ef}-\tfrac{1}{24}\epsilon^{abcde}C_a\cV_{bc}\cV_{de}-\tfrac16\epsilon^{abcde}C_a\cZ_{bc}\bcZ_{de}\nonumber\\*
&+gN\Big(\tfrac16R-\tfrac13G^2-\tfrac23G_{ab}H^{ab}+2H^2\Big)+\tfrac{1}{6\sqrt2}gL\,\cV_{ab}\big(G^{ab}+4H^{ab}\big)\nonumber\\*
&+\tfrac13gE_a\nabla_bG^{ab}+\tfrac{1}{24}g\epsilon^{abcde}E_aG_{bc}G_{de}-\tfrac23g\epsilon^{abcde}E_aH_{bc}H_{de}\nonumber\\*
&+g^2L^2\Big(-\tfrac{1}{12}R+\tfrac{1}{24}G^2-\tfrac16G_{ab}H^{ab}-H^2-Z_a\bZ^a\Big)\nonumber\\*
&-\tfrac14g^2\partial_aL\partial^aL-\tfrac16g^2N^2+\tfrac23g^3NL^2+\tfrac{5}{24}g^4L^4\,.\label{eq:Llog}
\end{align}
Here, we have used the Bianchi identity (\ref{eq:Bianchi}) to write the terms quadratic in the derivatives of $H_{ab}$ in the same form as in (\ref{eq:LC2}). In this process, the divergence $\nabla^bH_{ab}$ is replaced by its value in (\ref{eq:Bianchi}), and as a result, the term proportional to $g^2E_aE^a$ cancels. As in the $C^2+\tfrac16R^2$ invariant, the terms that contain $E_a$ arise from the gauged Bianchi identity (\ref{eq:Bianchi}). Unlike the $C^2+\tfrac16R^2$ invariant, the LOG invariant contains no Riemann tensor squared term, and its $SU(2)$ field strength squared term has the opposite sign.

\subsection{Ricci Scalar Squared Invariant}
\label{sec:r2}

The supersymmetric completion of the Ricci scalar squared was constructed in \cite{Ozkan:2013nwa} by means of a composite vector multiplet built out of a linear multiplet, and in the present formulation the linear multiplet can be taken to be the compensator \cite{Ozkan:2016csy,Gold:2025ttt}. In terms of the fields (\ref{eq:fields}), its bosonic part is given, up to a total derivative, by
\begin{align}
e^{-1}\cL_{R^2}={}&\Bigl(-\tfrac12R+\tfrac14G^2-2H^2-L^{-2}N^2-L^{-2}E_aE^a+L^{-1}\Box L-\tfrac12L^{-2}\partial_aL\partial^aL-2Z_a\bZ^a\Bigr)^2\nonumber\\
&+4L^{-2}\Big|\sqrt2E^aZ_a+\rmi D^a\big(LZ_a\big)\Big|^2-\tfrac14\cF_{ab}\cF^{ab}+L^{-1}N\cF_{ab}\big(G^{ab}-4H^{ab}\big)\nonumber\\
&-L^{-2}N^2G_{ab}\big(G^{ab}-4H^{ab}\big)-2\partial_a\big(L^{-1}N\big)\partial^a\big(L^{-1}N\big)\nonumber\\
&-\tfrac18\epsilon^{abcde}\cF_{ab}\cF_{cd}C_e-4gL^{-2}N^3+g^2L^2\Big(-\tfrac14R+\tfrac18G^2-H^2-Z_a\bZ^a\Big)\nonumber\\
&-\tfrac34g^2\partial_aL\partial^aL-\tfrac52g^2N^2-\tfrac12g^2E_aE^a+\tfrac{1}{16}g^4L^4\,,\label{eq:LR2}
\end{align}
where $\Box=\nabla^a\nabla_a$ and
\begin{equation}
\cF_{ab}=4\partial_{[a}\big(L^{-1}E_{b]}\big)+2\sqrt2\,\partial_{[a}V_{b]}\,.
\label{eq:calF}
\end{equation}
The first term gives rise to the Ricci scalar squared, $e^{-1}\cL_{R^2}\supset\tfrac14R^2$. Unlike the $C^2+\tfrac16R^2$ and the LOG invariants, the Ricci scalar squared invariant contains no $SU(2)$ field strength squared term. Instead, it provides a kinetic term for the scalar $N$ and for the combination of $V_\mu$ and $E^\mu$ that appears in $\cF_{ab}$.

\subsection{The General Action and the Gauss-Bonnet Combination}
\label{sec:general}

The most general off-shell model we consider in this paper is given by
\begin{equation}
S=\frac{1}{2\kappa^2}\int d^5x\,\Big(\cL_{LR}+\alpha\,\cLW+\beta\,\cL_{\log}+\zeta\,\cL_{R^2}\Big)\,,
\label{eq:general}
\end{equation}
where the couplings $\alpha$, $\beta$ and $\zeta$ have dimension of length squared. We treat these couplings as finite parameters, i.e.\ we neither expand in powers of the couplings nor eliminate the auxiliary fields, and all background and fluctuation equations are derived from the complete off-shell action (\ref{eq:general}). Note that the three invariants are independent already at the level of their curvature squared terms.

The Gauss-Bonnet combination corresponds to the choice $\beta=2\alpha$ and $\zeta=0$. We denote the corresponding coupling by $\lambda$, so that the model is given by $\cL_{LR}+\lambda\,\cL_{\rm GB}$ with $\cL_{\rm GB}\equiv\cLW+2\cL_{\log}$, where the combination applies to the bosonic and the fermionic parts of the two invariants. Using (\ref{eq:LC2}) and (\ref{eq:Llog}), the bosonic part of the gauged Gauss-Bonnet invariant is given, up to a total derivative, by
\begin{align}
e^{-1}\cL_{\rm GB}={}&\tfrac18\Big(R_{abcd}R^{abcd}-4R_{ab}R^{ab}+R^2\Big)+\tfrac12R_{abcd}H^{ab}G^{cd}-R_{abcd}H^{ab}H^{cd}\nonumber\\
&-2R_{ab}H^{ac}G^b{}_c+\tfrac12R\,G_{ab}H^{ab}-2(H^2)^2+4H^4-\tfrac12G^2\,G_{ab}H^{ab}\nonumber\\
&+(G_{ab}H^{ab})^2+G_{ab}G^{bc}G_{cd}H^{da}-2G_{ab}H^{bc}G_{cd}H^{da}\nonumber\\
&+4G_{ab}H^{bc}H_{cd}H^{da}+\epsilon^{abcde}H_b{}^f\nabla_aH_{cf}G_{de}+\tfrac{1}{16}\epsilon^{abcde}\nabla^fG_{af}G_{bc}G_{de}\nonumber\\
&+\tfrac{1}{16}\epsilon^{abcde}C_aR_{bc}{}^{fg}R_{defg}-\tfrac18\epsilon^{abcde}C_a\cV_{bc}\cV_{de}-\tfrac12\epsilon^{abcde}C_a\cZ_{bc}\bcZ_{de}\nonumber\\
&+\sqrt2gL\,\cV_{ab}H^{ab}-2gN\,G_{ab}H^{ab}-\tfrac18g\epsilon^{abcde}E_aG_{bc}G_{de}\nonumber\\
&-g\epsilon^{abcde}E_aH_{bc}H_{de}+g^2L^2\Big(-\tfrac14R+\tfrac14G^2-\tfrac12G_{ab}H^{ab}-3H^2-2Z_a\bZ^a\Big)\nonumber\\
&-\tfrac12g^2\partial_aL\partial^aL+g^2N^2+g^2E_aE^a+2g^3NL^2+\tfrac12g^4L^4\,.\label{eq:GB}
\end{align}
The curvature squared terms form the Gauss-Bonnet combination, and, as can be seen from (\ref{eq:GB}), the $SU(2)$ field strength squared terms, as well as the terms quadratic in the derivatives of $G_{ab}$ and $H_{ab}$, cancel. Therefore, unlike the $C^2+\tfrac16R^2$ and the LOG invariants, the Gauss-Bonnet invariant contains no Maxwell-type kinetic terms for $V_\mu$ and $Z_\mu$, and these fields appear with derivatives only through the Chern-Simons couplings and the term $\sqrt2gL\,\cV_{ab}H^{ab}$. Although the Gauss-Bonnet invariant is a combination of the $C^2+\tfrac16R^2$ and LOG invariants, it plays a central role in this paper. Around the supersymmetric AdS$_5$ vacuum, it propagates only the modes of the two-derivative theory, and this is also what makes the spectrum of the LOG extension tractable, see Sections~\ref{sec:GB} and \ref{sec:LOG}. In Section~\ref{sec:GBprod}, we study it around supersymmetric AdS$_3\times\Sigma_2$ backgrounds.

Finally, let us comment on the fermionic sector. The complete Lagrangians of (\ref{eq:general}) up to quadratic order in fermions are given in \cite{repo}. The fermionic terms of the two-derivative Lagrangian and of the Ricci scalar squared invariant are determined by the composite expressions for a vector multiplet in terms of the linear multiplet. Since some of the fermionic terms in the expressions given in \cite{Ozkan:2013uk} are not correct, we present the corrected expressions in Appendix~\ref{app:composite}.

\section{Spectra around the Supersymmetric AdS\texorpdfstring{$_5$}{5} Vacuum} \label{sec:AdS5}
\paragraph{}
In this section, we study the spectra of the four models around their common maximally supersymmetric AdS$_5$ vacuum. Section~\ref{sec:vacuum} collects what is common to all of them, namely the vacuum, which depends on the couplings only through a single combination, the linearized analysis around it and the supermultiplets of the AdS$_5$ superalgebra, and it ends with a summary of the results of this section in Figure~\ref{fig:phases} and Table~\ref{tab:summary}. Following \cite{Pang:2012xs}, we then study the models one by one in Sections~\ref{sec:R2}--\ref{sec:LOG}, in the order R$^2$, $C^2+\tfrac16R^2$, Gauss-Bonnet and LOG, separating the bosonic and the fermionic sectors and arranging the resulting spectra into supermultiplets. The Gauss-Bonnet extension is studied before the LOG extension for the reason explained at the beginning of Section~\ref{sec:GB}, and for these two extensions the two sectors are not treated separately. In Section~\ref{sec:central}, we relate the special values of the couplings found in these models to the central charges of the putative dual theory.

\subsection{Preliminaries}
\label{sec:vacuum}

\subsubsection{Supersymmetric vacuum}
\label{sec:vacuumsolution}

We consider maximally symmetric configurations in which the scalars take constant values and all vector and tensor fields vanish up to gauge transformations,
\begin{align}
&\bar R_{\mu\nu\rho\sigma}=-\ell^{-2}\big(\bar g_{\mu\rho}\bar g_{\nu\sigma}-\bar g_{\mu\sigma}\bar g_{\nu\rho}\big)\,,\qquad
\bar L=L_0\,,\qquad \bar N=N_0\,,\nonumber\\*
&\bar C_\mu=\bar B_{\mu\nu}=\bar E_{\mu\nu\rho}=\bar V_\mu=\bar Z_\mu=0\,,
\label{eq:background}
\end{align}
where we use ``bar'' to denote quantities evaluated in the background, $\ell$ is the radius of AdS$_5$ and $L_0>0$. On such a configuration, the supersymmetry transformation rules (\ref{eq:susy}) reduce to
\begin{equation}
\delta\varphi^i=\tfrac12\Big(N_0+\tfrac32gL_0^2\Big)\epsilon^i\,,\qquad
\delta\psi_\mu^i=\bar D_\mu\epsilon^i\equiv\bar\nabla_\mu\epsilon^i+\tfrac{1}{2\sqrt2}\rmi gL_0\,\delta^{ij}\gamma_\mu\epsilon_j\,.
\label{eq:Killing}
\end{equation}
The first equation implies that a supersymmetric background requires $N_0=-\tfrac32gL_0^2$, and the integrability condition of the Killing spinor equation $\bar D_\mu\epsilon^i=0$ fixes the radius of AdS$_5$. Thus, we obtain
\begin{equation}
N_0=-\tfrac32gL_0^2\,,\qquad \ell^{-2}=\tfrac12g^2L_0^2\,,
\label{eq:susyconditions}
\end{equation}
in which case the Killing spinor equation admits eight independent solutions. Therefore, the background preserves all eight supersymmetries, and its isometry supergroup is $SU(2,2|1)$.

The conditions (\ref{eq:susyconditions}) follow from supersymmetry alone and hold for all values of the couplings, whereas the value of $L_0$ is determined by the field equations. Substituting (\ref{eq:background}) and (\ref{eq:susyconditions}) into the field equations of (\ref{eq:general}), we find that the field equations of $N$ and $L$ and the trace of the Einstein equation are all proportional to the same expression, and the vacuum condition reads
\begin{equation}
\tau L_0^2-8L_0+8=0\,,\qquad \tau\equiv\big(\beta+6\zeta\big)g^2\,,
\label{eq:vacuum}
\end{equation}
while the remaining field equations are identically satisfied. Note that the coupling $\alpha$ of the $C^2+\tfrac16R^2$ invariant does not appear in (\ref{eq:vacuum}). Indeed, on configurations of the form (\ref{eq:background}), the Lagrangian (\ref{eq:LC2}) reduces to
\begin{equation}
e^{-1}\cLW=\tfrac{64}{3}\Big(\tfrac{1}{32}R-\tfrac14gN-\tfrac{1}{16}g^2L^2\Big)^2\,,
\label{eq:LWvac}
\end{equation}
which vanishes, together with its first variation, once (\ref{eq:susyconditions}) is imposed, since the expression in parentheses then vanishes for every $L_0$. On configurations of the form (\ref{eq:background}), the expression in parentheses is, up to a sign, the auxiliary field $D$ of the standard Weyl multiplet, which is a composite field in the dilaton Weyl multiplet \cite{Coomans:2012cf,Ozkan:2013uk}, and its vanishing on the supersymmetric background is the reason why the $C^2+\tfrac16R^2$ invariant does not contribute to the vacuum equations. Notice, however, that the vanishing of the invariant and of its first variation on the background does not imply the vanishing of its second variation. Therefore, the coupling $\alpha$ does not affect the vacuum, but it does affect the spectrum.

For $\tau<0$, the condition (\ref{eq:vacuum}) has a single positive root, whereas for $0<\tau<2$, it has two positive roots
\begin{equation}
L_0=\frac{2}{1+\sqrt{1-\tau/2}}\,,\qquad L_0^\prime=\frac{2}{1-\sqrt{1-\tau/2}}\,,
\label{eq:L0}
\end{equation}
which merge at $\tau=2$ into $L_0=L_0^\prime=2$. For $\tau>2$, there is no supersymmetric AdS$_5$ vacuum. The first root is continuously connected to the vacuum of the two-derivative theory, $L_0=1$, $N_0=-\tfrac32g$ and $\ell^2=2/g^2$, as $\tau\to0$, whereas the second root diverges as $\tau\to0^+$ and has no two-derivative limit. In this paper, we study the spectrum around the first root, which, for the individual models of Section~\ref{sec:theory}, is determined by
\begin{equation}
\begin{aligned}
R^2:&\ \tau=6\zeta g^2\,,&\qquad C^2+\tfrac16R^2:&\ \tau=0\,,\\
\mbox{Gauss-Bonnet}:&\ \tau=2\lambda g^2\,,&\mbox{LOG}:&\ \tau=\beta g^2\,.
\end{aligned}
\end{equation}
In particular, the $C^2+\tfrac16R^2$ extension shares the vacuum of the two-derivative theory, $L_0=1$, for all values of $\alpha$. As we shall see, the value $\tau=\tfrac32$ ($L_0=\tfrac43$) plays a special role in the spectra of the Ricci scalar squared and LOG extensions, and the value $\tau=2$ ($L_0=2$), where the two roots merge, in the spectra of the LOG and Gauss-Bonnet extensions.

\subsubsection{Linearized fluctuations and gauge symmetries}
\label{sec:fluctuations}

We parametrize the bosonic fluctuations around the vacuum (\ref{eq:background}), (\ref{eq:susyconditions}) as
\begin{align}
g_{\mu\nu}&=\bar g_{\mu\nu}+h_{\mu\nu}\,,&
L&=L_0+\sigma\,,&
N&=N_0+n\,,&
C_\mu&=c_\mu\,,\nonumber\\*
B_{\mu\nu}&=b_{\mu\nu}\,,&
E_{\mu\nu\rho}&=e_{\mu\nu\rho}\,,&
V_\mu&=v_\mu\,,&
Z_\mu&=z_\mu\,,
\label{eq:fluctuations}
\end{align}
and, since the fermions vanish in the background, we denote the fermionic fluctuations by $\psi_\mu^i$ and $\varphi^i$. In the background (\ref{eq:background}), (\ref{eq:susyconditions}), the linearized bosonic and fermionic gauge symmetries are given by
\begin{align}
\delta h_{\mu\nu}&=\bar\nabla_\mu\xi_\nu+\bar\nabla_\nu\xi_\mu\,,&
\delta c_\mu&=\partial_\mu\Lambda\,,&
\delta v_\mu&=\partial_\mu\theta\,,\nonumber\\*
\delta b_{\mu\nu}&=2\partial_{[\mu}\Lambda_{\nu]}-\tfrac12g\Lambda_{\mu\nu}\,,&
\delta e_{\mu\nu\rho}&=3\partial_{[\mu}\Lambda_{\nu\rho]}\,,&
\delta\psi_\mu^i&=\bar D_\mu\epsilon^i\,,
\label{eq:lingauge}
\end{align}
where $\bar D_\mu$ is defined in (\ref{eq:Killing}), and the remaining fluctuations $\sigma$, $n$, $z_\mu$ and $\varphi^i$ have no inhomogeneous shifts. Here, $\xi_\mu$, $\Lambda$, $\theta$, $\Lambda_\mu$, $\Lambda_{\mu\nu}$ and $\epsilon^i$ are the parameters of the diffeomorphisms, the gauge transformations of $C_\mu$, the $U(1)$ R-symmetry, the gauge transformations of the form fields (\ref{eq:formgauge}) and supersymmetry, respectively. Note that this does not mean that these fluctuations are inert. The fluctuation $z_\mu$ carries charge under the $U(1)$ R-symmetry of the background and transforms homogeneously, $\delta z_\mu=\rmi\theta z_\mu$. Similarly, the inhomogeneous term in the supersymmetry transformation of $\varphi^i$ is proportional to $N_0+\tfrac32gL_0^2$, which vanishes on the vacuum, but $\varphi^i$ still transforms into the bosonic fluctuations under the supersymmetries preserved by the vacuum. Furthermore, for $g\neq0$, the two-form parameter $\Lambda_{\mu\nu}$ acts as a Stueckelberg shift of $b_{\mu\nu}$. Choosing $\Lambda_{\mu\nu}=2b_{\mu\nu}/g$ sets $b_{\mu\nu}$ to zero, and the three-form then appears through the gauge invariant combination $e_{\mu\nu\rho}+\tfrac6g\partial_{[\mu}b_{\nu\rho]}$, in terms of which the linearized three-form field strength is $\tfrac12g\big(e_{\mu\nu\rho}+\tfrac6g\partial_{[\mu}b_{\nu\rho]}\big)$. Unlike in the two-derivative theory, where $E^a=0$ allows $E_{\mu\nu\rho}$ to be absorbed into $B_{\mu\nu}$, the three-form cannot in general be removed in this way in the presence of the curvature squared invariants, since $E^a$ need not vanish.

Since the background is bosonic, the quadratic action splits into a bosonic and a fermionic part. We derive it from the complete action (\ref{eq:general}), keeping all the fields, including those that are auxiliary in the two-derivative theory, and without expanding in the higher-derivative couplings. For the fermionic part, the two-fermion terms of the Lagrangians are sufficient \cite{repo}. In the two-derivative theory, the fields $n$, $v_\mu$, $z_\mu$ and $e_{\mu\nu\rho}$ are auxiliary, and the propagating degrees of freedom are those of $\mathcal N=2$ gauged supergravity coupled to a single vector multiplet. In the presence of the curvature squared invariants, some of these fields acquire kinetic terms, which give rise to additional propagating modes. In each of Sections~\ref{sec:R2}--\ref{sec:LOG}, we use the gauge symmetries (\ref{eq:lingauge}) and the constraints that follow from the linearized field equations to remove the unphysical modes, and we decompose the remaining fluctuations into the spin-2, spin-1 and spin-0 sectors for the bosons and the spin-3/2 and spin-1/2 sectors for the fermions. We report the masses of the propagating modes together with the associated roots of the mass-energy relations (\ref{eq:massenergy}) below. The energy branches used in the supermultiplet assignments are specified for each model, and where these assignments are compatible with the standard boundary conditions, they use the larger roots. On other ranges of the couplings, the larger root cannot be chosen independently for each field without spoiling the energy spacings within a supermultiplet, and we call the corresponding assignments formal: they continue the formulas valid elsewhere, but with the standard boundary conditions the modes do not form the multiplets that the assignment names. We do not discuss the choice of boundary data in detail, including the alternative boundary conditions that are available for scalars with $-4<m^2\ell^2<-3$ \cite{Klebanov:1999tb} and the boundary data for scalars at the BF bound, where the two roots coincide.

\subsubsection{Supermultiplets and stability}
\label{sec:multiplets}

\begin{table}[!htb]
\centering
\renewcommand{\arraystretch}{1.25}
\begin{tabular}{|c|c|c|c|}
\hline
Multiplet & Energy & $SO(4)\times U(1)_R$ & Field\\
\hline
\hline
$\mathcal G$ & $3$ & $(\tfrac12,\tfrac12)_0$ & massless vector\\
$8+8$ & $\tfrac72$ & $(1,\tfrac12)_{-1}\oplus(\tfrac12,1)_{+1}$ & massless spin-3/2\\
 & $4$ & $(1,1)_0$ & massless spin-2\\
\hline
$\mathcal V$ & $2$ & $(0,0)_0$ & scalar\\
$4+4$ & $\tfrac52$ & $(\tfrac12,0)_{-1}\oplus(0,\tfrac12)_{+1}$ & spin-1/2\\
 & $3$ & $(\tfrac12,\tfrac12)_0$ & massless vector\\
\hline
$\mathcal V_L(\Delta)$ & $\Delta$ & $(0,0)_0$ & scalar\\
$8+8$ & $\Delta+\tfrac12$ & $(\tfrac12,0)_{-1}\oplus(0,\tfrac12)_{+1}$ & spin-1/2\\
 & $\Delta+1$ & $(\tfrac12,\tfrac12)_0\oplus(0,0)_{+2}\oplus(0,0)_{-2}$ & massive vector, complex scalar\\
 & $\Delta+\tfrac32$ & $(\tfrac12,0)_{+1}\oplus(0,\tfrac12)_{-1}$ & spin-1/2\\
 & $\Delta+2$ & $(0,0)_0$ & scalar\\
\hline
$\mathcal L_2(\Delta)$ & $\Delta$ & $(\tfrac12,\tfrac12)_0$ & massive vector\\
$32+32$ & $\Delta+\tfrac12$ & $(1,\tfrac12)_{-1}\oplus(\tfrac12,1)_{+1}$ & spin-3/2\\ & & $\oplus\,(0,\tfrac12)_{-1}\oplus(\tfrac12,0)_{+1}$ & spin-1/2\\
 & $\Delta+1$ & $(1,1)_0\oplus(0,0)_0\oplus(1,0)_0\oplus(0,1)_0$ & massive spin-2, scalar,\\
 & & $\oplus\,(\tfrac12,\tfrac12)_{+2}\oplus(\tfrac12,\tfrac12)_{-2}$ & two-forms, complex vector\\
 & $\Delta+\tfrac32$ & $(1,\tfrac12)_{+1}\oplus(\tfrac12,1)_{-1}$ & spin-3/2\\ & & $\oplus\,(0,\tfrac12)_{+1}\oplus(\tfrac12,0)_{-1}$ & spin-1/2\\
 & $\Delta+2$ & $(\tfrac12,\tfrac12)_0$ & massive vector\\
\hline
\end{tabular}
\vspace{1.5mm}
\caption{The $SU(2,2|1)$ supermultiplets that appear in the spectra of the models we consider. The number of bosonic and fermionic states is given below the name of each multiplet. The fermions of each multiplet form symplectic Majorana doublets.}
\label{tab:multiplets}
\end{table}

In arranging the spectra into supermultiplets, it is useful to recall the unitary representations of $SU(2,2|1)$ \cite{Ceresole:1999zs}. We label the representations of its bosonic subalgebra $SO(4,2)\times U(1)_R$ by the lowest energy $\Delta$, in units of $\ell^{-1}$, and by $(j_L,j_R)_r$, where $(j_L,j_R)$ labels the representation of $SO(4)\simeq SU(2)_L\times SU(2)_R$ and $r$ is the R-charge. We fix the normalization of the R-charge such that the supercharges transform as $(\tfrac12,0)_{-1}\oplus(0,\tfrac12)_{+1}$. For a field of mass $m$, the lowest energy $\Delta$ is related to the mass by
\begin{equation}
\begin{aligned}
s&=0,2:&\quad m^2\ell^2&=\Delta(\Delta-4)\,,\\
s&=1:&m^2\ell^2&=(\Delta-1)(\Delta-3)\,,\\
s&=\tfrac12,\tfrac32:&|m|\ell&=\Delta-2\,,
\end{aligned}
\label{eq:massenergy}
\end{equation}
where $s$ denotes the spin. Here, the masses of the scalars and the vectors are defined by $(\bar\Box-m^2)\phi=0$ and $\bar\nabla^\mu f_{\mu\nu}-m^2a_\nu=0$ with $f_{\mu\nu}=2\partial_{[\mu}a_{\nu]}$, and the mass of a transverse traceless spin-2 mode by $(\bar\Box+2\ell^{-2}-m^2)h_{\mu\nu}=0$, so that the massless graviton has $m=0$. Real values of $\Delta$ require $m^2\ell^2\geq-4$ for the scalars and the spin-2 modes and $m^2\ell^2\geq-1$ for the vectors, which are the Breitenlohner-Freedman (BF) bounds \cite{Breitenlohner:1982jf,Breitenlohner:1982bm} in these conventions. For the fermions, $m$ is the mass parameter of the first-order Dirac or Rarita-Schwinger equation, and the relation $|m|\ell=\Delta-2$ gives the energy on the standard branch. In particular, the massless gravitino, which possesses the gauge symmetry $\delta\psi_\mu^i=\bar D_\mu\epsilon^i$, has $|m|\ell=\tfrac32$ and $\Delta=\tfrac72$. The last relation also applies to the two-forms that satisfy first-order field equations, which transform as $(1,0)$ or $(0,1)$ under $SO(4)$. The supermultiplets that appear in our analysis are the massless graviton multiplet $\mathcal G$, the massless vector multiplet $\mathcal V$, the long vector multiplet $\mathcal V_L(\Delta)$ and the long spin-2 multiplet $\mathcal L_2(\Delta)$, whose content is given in Table~\ref{tab:multiplets}. The long multiplets are labeled by the energy $\Delta$ of their lowest component, and they are obtained by acting with the four supercharges of positive energy on it. Unitarity requires $\Delta>2$ for $\mathcal V_L(\Delta)$ and $\Delta>3$ for $\mathcal L_2(\Delta)$, and at $\Delta=2$ and $\Delta=3$, respectively, the long multiplets reach the threshold for shortening.

The two-derivative theory propagates the massless graviton multiplet $\mathcal G$ and a massless vector multiplet $\mathcal V$, that is, $12+12$ states. The scalar of $\mathcal V$ arises from the fluctuation of $L$ and has $\Delta=2$, and its mass $m^2\ell^2=-4$ saturates the BF bound. As we show in Sections~\ref{sec:R2}--\ref{sec:LOG}, for generic values of the couplings, the curvature squared invariants add a long vector multiplet in the case of the Ricci scalar squared invariant and a long spin-2 multiplet in the case of the $C^2+\tfrac16R^2$ and the LOG invariants, while the Gauss-Bonnet combination propagates no additional states.

Finally, let us comment on the stability of the vacuum. As emphasized in \cite{Pang:2012xs}, the fact that the vacuum is supersymmetric does not guarantee its stability, since the positivity of energy in a supersymmetric theory relies on a positive-definite norm in the space of states, which is violated by the ghosts that typically arise from higher-derivative terms. Therefore, in addition to requiring that the masses be real and respect the BF bounds, we determine the sign of the kinetic term of each physical mode. For a mode $\Phi$ with a second-order field equation, we write its quadratic action in the form $-\tfrac12Z\big(\partial\Phi\cdot\partial\Phi+m^2\Phi^2\big)$, and the mode is a ghost if $Z<0$. We refer to the sign of $Z$ as the sign of the norm of the mode. Note that the unitarity of a representation, which constrains $\Delta$, and the sign of $Z$ are independent conditions. At special values of the couplings, the quadratic system may lose rank or develop repeated operator factors. We analyze these cases separately for each model, distinguishing degenerations of the quadratic action from logarithmic sectors associated with nontrivial Jordan blocks, as in critical gravity \cite{Lu:2011zk,Deser:2011xc}.

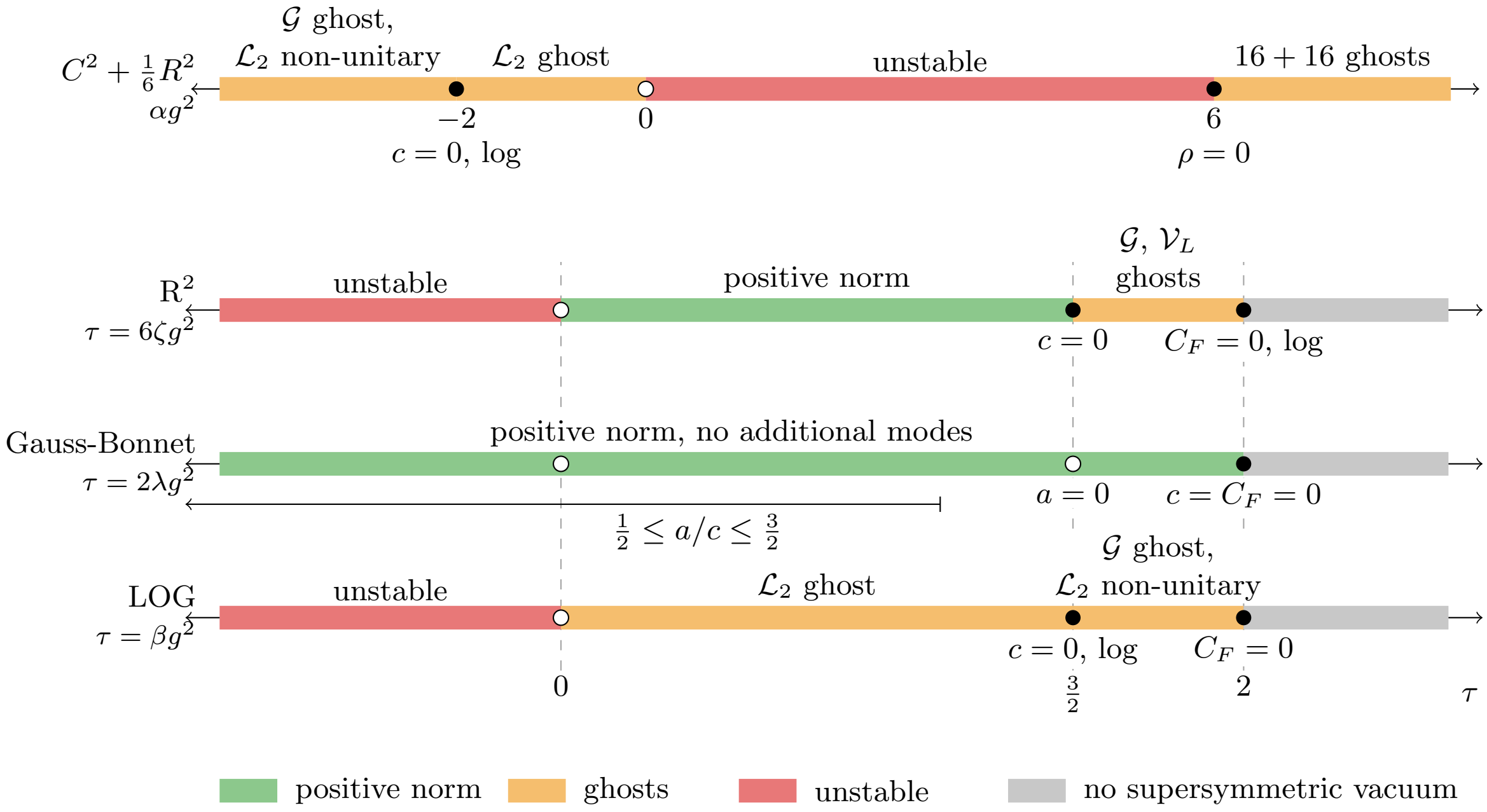
\begin{figure}[!htb]
\centering

\begin{tikzpicture}[x=1cm,y=1cm,font=\footnotesize]
\definecolor{okc}{RGB}{140,200,140}
\definecolor{ghc}{RGB}{245,190,110}
\definecolor{unc}{RGB}{232,120,120}
\definecolor{noc}{RGB}{200,200,200}
\newcommand{\tx}[1]{2.6+(#1+1)*3.55}
\newcommand{\qx}[1]{2.6+(#1+4.5)*0.985}
\tikzset{bar/.style={line width=7pt,line cap=butt},
         pt/.style={circle,fill=black,inner sep=0pt,minimum size=4.5pt},
         opt/.style={circle,fill=white,draw=black,inner sep=0pt,minimum size=4.5pt},
         seg/.style={anchor=south,inner sep=1pt,align=center,fill=white},
         blw/.style={anchor=north,inner sep=1.5pt,align=center,fill=white},
         rowlab/.style={anchor=east,inner sep=1pt,align=right}}
\def\yW{7.6}
\node[rowlab] at (2.4,\yW) {$C^2+\tfrac16R^2$\\[-2pt]{\scriptsize $\alpha g^2$}};
\draw[bar,ghc] ({\qx{-4.5}},\yW) -- ({\qx{-2}},\yW);
\draw[bar,ghc] ({\qx{-2}},\yW) -- ({\qx{0}},\yW);
\draw[bar,unc] ({\qx{0}},\yW) -- ({\qx{6}},\yW);
\draw[bar,ghc] ({\qx{6}},\yW) -- ({\qx{8.5}},\yW);
\draw[->,thin] ({\qx{8.5}},\yW) -- ({\qx{8.8}},\yW);
\draw[<-,thin] ({\qx{-4.8}},\yW) -- ({\qx{-4.5}},\yW);
\node[pt] at ({\qx{-2}},\yW) {};
\node[opt] at ({\qx{0}},\yW) {};
\node[pt] at ({\qx{6}},\yW) {};
\node[seg] at ({(\qx{-4.5}+\qx{-2})/2},\yW+0.14) {$\mathcal G$ ghost,\\[-2pt] $\mathcal L_2$ non-unitary};
\node[seg] at ({(\qx{-2}+\qx{0})/2},\yW+0.14) {$\mathcal L_2$ ghost};
\node[seg] at ({(\qx{0}+\qx{6})/2},\yW+0.14) {unstable};
\node[seg] at ({(\qx{6}+\qx{8.5})/2},\yW+0.14) {$16+16$ ghosts};
\node[blw] at ({\qx{-2}},\yW-0.14) {$-2$\\[-3pt] $c=0$, log};
\node[blw] at ({\qx{0}},\yW-0.14) {$0$};
\node[blw] at ({\qx{6}},\yW-0.14) {$6$\\[-3pt] $\rho=0$};
\def\yR{5.3}
\def\yG{3.7}
\def\yL{2.1}
\foreach \t in {0,1.5,2} {\draw[dashed,gray!70] ({\tx{\t}},\yL-0.55) -- ({\tx{\t}},\yR+0.5);}
\node[blw] at ({\tx{0}},\yL-0.55) {$0$};
\node[blw] at ({\tx{1.5}},\yL-0.55) {$\tfrac32$};
\node[blw] at ({\tx{2}},\yL-0.55) {$2$};
\node[anchor=west] at ({\tx{2.6}},\yL-0.8) {$\tau$};
\node[rowlab] at (2.4,\yR) {R$^2$\\[-2pt]{\scriptsize $\tau=6\zeta g^2$}};
\draw[bar,unc] ({\tx{-1}},\yR) -- ({\tx{0}},\yR);
\draw[bar,okc] ({\tx{0}},\yR) -- ({\tx{1.5}},\yR);
\draw[bar,ghc] ({\tx{1.5}},\yR) -- ({\tx{2}},\yR);
\draw[bar,noc] ({\tx{2}},\yR) -- ({\tx{2.6}},\yR);
\draw[->,thin] ({\tx{2.6}},\yR) -- ({\tx{2.7}},\yR);
\draw[<-,thin] ({\tx{-1.1}},\yR) -- ({\tx{-1}},\yR);
\node[pt] at ({\tx{1.5}},\yR) {};
\node[opt] at ({\tx{0}},\yR) {};
\node[pt] at ({\tx{2}},\yR) {};
\node[seg] at ({(\tx{-1}+\tx{0})/2},\yR+0.14) {unstable};
\node[seg] at ({(\tx{0}+\tx{1.5})/2},\yR+0.14) {positive norm};
\node[seg] at ({(\tx{1.5}+\tx{2})/2},\yR+0.14) {$\mathcal G$, $\mathcal V_L$\\[-2pt] ghosts};
\node[blw] at ({\tx{1.5}},\yR-0.14) {$c=0$};
\node[blw] at ({\tx{2}},\yR-0.14) {$C_F=0$, log};
\node[rowlab] at (2.4,\yG) {Gauss-Bonnet\\[-2pt]{\scriptsize $\tau=2\lambda g^2$}};
\draw[bar,okc] ({\tx{-1}},\yG) -- ({\tx{2}},\yG);
\draw[bar,noc] ({\tx{2}},\yG) -- ({\tx{2.6}},\yG);
\draw[->,thin] ({\tx{2.6}},\yG) -- ({\tx{2.7}},\yG);
\draw[<-,thin] ({\tx{-1.1}},\yG) -- ({\tx{-1}},\yG);
\node[pt] at ({\tx{2}},\yG) {};
\node[opt] at ({\tx{0}},\yG) {};
\node[opt] at ({\tx{1.5}},\yG) {};
\node[seg] at ({(\tx{-1}+\tx{2})/2},\yG+0.14) {positive norm, no additional modes};
\node[blw] at ({\tx{1.5}},\yG-0.14) {$a=0$};
\node[blw] at ({\tx{2}},\yG-0.14) {$c=C_F=0$};
\draw[thin,<-] ({\tx{-1.1}},\yG-0.42) -- ({\tx{1.111}},\yG-0.42);
\draw[thin] ({\tx{1.111}},\yG-0.34) -- ({\tx{1.111}},\yG-0.5);
\node[blw] at ({\tx{0.4}},\yG-0.46) {$\tfrac12\leq a/c\leq\tfrac32$};
\node[rowlab] at (2.4,\yL) {LOG\\[-2pt]{\scriptsize $\tau=\beta g^2$}};
\draw[bar,unc] ({\tx{-1}},\yL) -- ({\tx{0}},\yL);
\draw[bar,ghc] ({\tx{0}},\yL) -- ({\tx{1.5}},\yL);
\draw[bar,ghc] ({\tx{1.5}},\yL) -- ({\tx{2}},\yL);
\draw[bar,noc] ({\tx{2}},\yL) -- ({\tx{2.6}},\yL);
\draw[->,thin] ({\tx{2.6}},\yL) -- ({\tx{2.7}},\yL);
\draw[<-,thin] ({\tx{-1.1}},\yL) -- ({\tx{-1}},\yL);
\node[pt] at ({\tx{1.5}},\yL) {};
\node[opt] at ({\tx{0}},\yL) {};
\node[pt] at ({\tx{2}},\yL) {};
\node[seg] at ({(\tx{-1}+\tx{0})/2},\yL+0.14) {unstable};
\node[seg] at ({(\tx{0}+\tx{1.5})/2},\yL+0.14) {$\mathcal L_2$ ghost};
\node[seg] at ({(\tx{1.5}+\tx{2})/2},\yL+0.14) {$\mathcal G$ ghost,\\[-2pt] $\mathcal L_2$ non-unitary};
\node[blw] at ({\tx{1.5}},\yL-0.14) {$c=0$, log};
\node[blw] at ({\tx{2}},\yL-0.14) {$C_F=0$};
\def\yK{0.3}
\draw[bar,okc] (2.6,\yK) -- (3.2,\yK); \node[anchor=west] at (3.25,\yK) {positive norm};
\draw[bar,ghc] (5.6,\yK) -- (6.2,\yK); \node[anchor=west] at (6.25,\yK) {ghosts};
\draw[bar,unc] (8.0,\yK) -- (8.6,\yK); \node[anchor=west] at (8.65,\yK) {unstable};
\draw[bar,noc] (10.8,\yK) -- (11.4,\yK); \node[anchor=west] at (11.45,\yK) {no supersymmetric vacuum};
\end{tikzpicture}
\caption{The spectra of the four models around the supersymmetric AdS$_5$ vacuum as functions of their couplings. The filled circles are the points at which the quadratic action degenerates, which are zeros of $c$ or of $C_F$ except for $\alpha g^2=6$, where (\ref{eq:Wgravitino}) and (\ref{eq:WX}) have repeated factors, and ``log'' marks the critical points with logarithmic modes. The open circles are the two-derivative theory and the point $a=0$ of the Gauss-Bonnet extension. The bracket is the range $\tfrac12\leq a/c\leq\tfrac32$ of \cite{Hofman:2008ar}.}
\label{fig:phases}
\end{figure}

Before turning to the individual models, we summarize the results of Sections~\ref{sec:R2}--\ref{sec:LOG} in Figure~\ref{fig:phases} and Table~\ref{tab:summary}. For each model, Figure~\ref{fig:phases} shows the ranges of the coupling in which the spectrum has positive norm, in which it contains negative-norm modes, and in which the vacuum is unstable, together with the special values of the coupling. The R$^2$, Gauss-Bonnet and LOG extensions are displayed as functions of the same parameter $\tau$, which makes the values $\tau=\tfrac32$ and $\tau=2$ that they share visible. As we show in Section~\ref{sec:central}, the points at which the quadratic action degenerates are zeros of the central charge $c$ or of the flavor two-point coefficient $C_F$ of the dual theory, and they are labeled as such in the figure. Table~\ref{tab:summary} lists the multiplets that each invariant adds to the $12+12$ states of the two-derivative theory, in terms of the parameters $\nu$ and $\rho$ defined in (\ref{eq:R2ABnu}), (\ref{eq:Wrho}) and (\ref{eq:LOGnu}) below, and the case analysis behind the figure is tabulated model by model in Tables~\ref{tab:R2ranges}, \ref{tab:Wranges} and \ref{tab:LOGranges}.

\begin{table}[!htb]
\centering
\small
\renewcommand{\arraystretch}{1.3}
\begin{tabular}{|l|c|c|c|}
\hline
Model & Coupling & Additional multiplet & Positive-norm range\\
\hline\hline
R$^2$ & $\tau=6\zeta g^2$ & $\mathcal V_L(1+\nu)$, $8+8$ & $0<\tau<\tfrac32$\\
$C^2+\tfrac16R^2$ & $\alpha g^2$ & $\mathcal L_2(1+\rho)$, $32+32$ & none\\
Gauss-Bonnet & $\tau=2\lambda g^2$ & none & $\tau<2$\\
LOG & $\tau=\beta g^2$ & $\mathcal L_2(1+\nu)$, $32+32$ & none\\
\hline
\end{tabular}
\vspace{1.5mm}
\caption{The spectra of the four models around the supersymmetric AdS$_5$ vacuum for generic values of the couplings, with $\nu$ and $\rho$ defined in (\ref{eq:R2ABnu}), (\ref{eq:Wrho}) and (\ref{eq:LOGnu}). For $\alpha g^2>6$, the additional modes of the $C^2+\tfrac16R^2$ extension form $\mathcal L_2(1+\rho)$ only with non-standard boundary conditions, see Section~\ref{sec:Wmult}.}
\label{tab:summary}
\end{table}

\FloatBarrier
\subsection{Ricci Scalar Squared Extension}
\label{sec:R2}

In this subsection, we study the spectrum of the model $\cL_{LR}+\zeta\cL_{R^2}$ around the supersymmetric AdS$_5$ vacuum of Section~\ref{sec:vacuum}. The R$^2$ invariant adds to the $12+12$ states of the two-derivative theory a single long vector multiplet with $8+8$ states, and for $0<\tau<\tfrac32$, that is, for $0<\zeta g^2<\tfrac14$, all modes have real masses that respect the BF bounds and positive norm. Among the models that propagate additional modes, it is the only one with such a range, see Figure~\ref{fig:phases} and Table~\ref{tab:R2ranges}. For this model, $\tau=6\zeta g^2$, and the first root of (\ref{eq:vacuum}) is $L_0=2/\big(1+\sqrt{1-3\zeta g^2}\big)$. It is convenient to define
\begin{equation}
A\equiv L_0-3\zeta g^2L_0^2=4-3L_0\,,\qquad
B\equiv2-L_0\,,\qquad
\nu^2\equiv\frac{2}{\zeta g^2L_0}-2=\frac{4-L_0}{2(L_0-1)}\,,
\label{eq:R2ABnu}
\end{equation}
where the second equalities follow from (\ref{eq:vacuum}). The coefficient $A$ is the effective coefficient of the Ricci scalar on the vacuum. Indeed, the expression in the square of (\ref{eq:LR2}) takes the value $\tfrac{11}{4}g^2L_0^2$ on the vacuum, and together with the term $-\tfrac14g^2L^2R$ in (\ref{eq:LR2}), the R$^2$ invariant shifts the coefficient of $R$ from $L_0$ to $L_0-3\zeta g^2L_0^2$. The signs of these coefficients as functions of $\tau$ are the basis of the case analysis of Section~\ref{sec:R2mult}, which is summarized in Table~\ref{tab:R2ranges}; here we only record them. For $0<\tau<2$, we have $1<L_0<2$, and therefore $B>0$ and $\nu>1$, whereas $A$ is positive for $\tau<\tfrac32$ and negative for $\tfrac32<\tau<2$. For $\tau<0$, we have $L_0<1$ and $\nu^2<0$. In the limit $\zeta\to0$, $\nu$ diverges, and the modes that we find below in addition to those of the two-derivative theory decouple. In Sections~\ref{sec:R2bos} and \ref{sec:R2ferm}, we assume $A\neq0$, that is, $\tau\neq\tfrac32$; this point is discussed separately in Section~\ref{sec:R2mult}.

\subsubsection{Bosonic sector}
\label{sec:R2bos}

We start with the transverse traceless part of the metric fluctuation, which does not mix with the other fields. Since the R$^2$ invariant contributes to its quadratic action only through the effective coefficient $A$, it satisfies
\begin{equation}
A\big(\bar\Box+2\ell^{-2}\big)h_{\mu\nu}^{\rm TT}=0\,,
\label{eq:R2TT}
\end{equation}
and it describes a massless graviton whose kinetic term is proportional to $A$. In particular, the R$^2$ invariant does not give rise to a massive spin-2 mode.

The neutral scalar sector consists of the fluctuations of $L$, $N$ and the trace of the metric. To analyze it, we set the vectors and the tensors to zero and denote by $X$ the expression in the square of (\ref{eq:LR2}), which in this sector reduces to
\begin{equation}
X=-\tfrac12R-L^{-2}N^2+L^{-1}\Box L-\tfrac12L^{-2}\partial_aL\partial^aL\,.
\label{eq:R2X}
\end{equation}
Completing the square in the scalar sector introduces $\big(X+\tfrac14g^2L^2\big)^2$, up to a total derivative, with the remaining terms retained below. Introducing an auxiliary scalar $\Phi$ for this shifted square, the scalar part of the Lagrangian can be written, up to a total derivative, as
\begin{align}
e^{-1}\cL={}&\big(L-\zeta\Phi\big)R+L^{-1}\big(1+\zeta L^{-1}\Phi\big)\partial_aL\partial^aL\nonumber\\*
&-2\zeta L^{-1}\partial_a\Phi\partial^aL-2\zeta\,\partial_a\big(L^{-1}N\big)\partial^a\big(L^{-1}N\big)-U\,,\label{eq:R2scalar}
\end{align}
where the potential is given by
\begin{align}
U={}&2L^{-1}N^2+2gLN+4gN-2g^2L^2+\tfrac12g^2L^3\nonumber\\*
&+\zeta\Big(\Phi^2+2L^{-2}N^2\Phi-\tfrac12g^2L^2\Phi+4gL^{-2}N^3+2g^2N^2\Big)\,.\label{eq:R2pot}
\end{align}
The field equation of $\Phi$ sets $\Phi=X+\tfrac14g^2L^2$, restricted to the scalars, and on the vacuum $\Phi_0=3g^2L_0^2$, so that $L_0-\zeta\Phi_0=A$. Thus (\ref{eq:R2scalar}) is a scalar-tensor theory with three scalars. For $A\neq0$, after a Weyl rescaling of the metric by $|L-\zeta\Phi|^{2/3}$, the masses of the scalar fluctuations follow from the second derivatives of the potential and the kinetic matrix in the standard way, and we find
\begin{equation}
m_0^2\ell^2=-4\,,\qquad
m_\pm^2\ell^2=(\nu\pm1)^2-4\,.
\label{eq:R2scalarmasses}
\end{equation}
The first scalar is that of the two-derivative theory, which saturates the BF bound, whereas the other two arise from the R$^2$ invariant. We denote the corresponding eigenmodes by $\sigma_0$ and $\sigma_\pm$.

The R$^2$ invariant also provides a kinetic term for the longitudinal part of $Z_\mu$. From (\ref{eq:LLR}) and (\ref{eq:LR2}), the quadratic Lagrangian of its fluctuation $z_\mu$ is
\begin{equation}
e^{-1}\cL=4A\,z_\mu\bar z^\mu+4\zeta\,\big|\bar\nabla^\mu z_\mu\big|^2\,.
\label{eq:R2z}
\end{equation}
For $A\neq0$, its field equation sets the transverse part of $z_\mu$ to zero, and the divergence $\Pi\equiv\bar\nabla^\mu z_\mu$ satisfies $\big(\bar\Box-A/\zeta\big)\Pi=0$. Therefore, $\Pi$ is a complex scalar with R-charge $\pm2$ and mass
\begin{equation}
m_\Pi^2\ell^2=\frac{A}{\zeta}\,\ell^2=\nu^2-4\,.
\label{eq:R2Pi}
\end{equation}

Finally, the fluctuations $c_\mu$, $v_\mu$, $b_{\mu\nu}$ and $e_{\mu\nu\rho}$ form the vector sector. With a slight abuse of notation, we denote the linearized field strengths by the same symbols as the full ones, so that $G=dc$, where $(dX)_{\mu\nu}=2\partial_{[\mu}X_{\nu]}$, $H_{ab}$ satisfies $\bar\nabla^bH_{ab}=\tfrac12gE_a$, and $\cF=d\big(2L_0^{-1}E+\sqrt2v\big)$. Then, the quadratic Lagrangian of the vector sector reads
\begin{align}
e^{-1}\cL={}&-\tfrac12A\,G^2+4A\,H^2-2\big(L_0^{-1}+3\zeta g^2\big)E_aE^a-2\sqrt2E^av_a-4gE^ac_a\nonumber\\*
&+\zeta\Big(-\tfrac14\cF^2-\tfrac32gL_0\,\cF_{ab}\big(G^{ab}-4H^{ab}\big)-\tfrac94g^2L_0^2G^2+9g^2L_0^2G_{ab}H^{ab}\Big)\,.
\label{eq:R2vec}
\end{align}
In the two-derivative theory, $v_\mu$ is a Lagrange multiplier that imposes $E_a=0$. In the presence of the R$^2$ invariant, $E_a$ acquires a kinetic term through $\cF_{ab}$. To find the propagating modes, we perform a first-order dualization. In the gauge $b_{\mu\nu}=0$, we treat $H_{ab}$ and $E_a$ as independent fields and impose the constraint $\bar\nabla^bH_{ab}=\tfrac12gE_a$ by a Lagrange multiplier $u^a$. The two-form $H_{ab}$ then appears algebraically, and after eliminating the algebraic variables and redefining the remaining vectors, the vector sector is described in terms of $c_\mu$, $v_\mu$ and $E_\mu$ by the equivalent Lagrangian \cite{repo}
\begin{equation}
e^{-1}\cL=-\tfrac12A\,G^2-\frac{A}{4g^2L_0^2}\big(\sqrt2\,dv+2gL_0G\big)^2-\tfrac14\zeta\big(\cF+3gL_0G\big)^2-\frac{2A}{L_0^2}E_aE^a\,.
\label{eq:R2vec2}
\end{equation}
In terms of $\hat c_\mu\equiv c_\mu+\sqrt2v_\mu/(3gL_0)+g\zeta E_\mu/B$, the kinetic terms decouple, and we obtain
\begin{equation}
e^{-1}\cL=-\tfrac32B\,(d\hat c)^2-\frac{A}{6g^2L_0^2}(dv)^2-\frac{A\zeta}{L_0^2B}\Big((dE)^2+\frac{2B}{\zeta}E_aE^a\Big)\,.
\label{eq:R2vec3}
\end{equation}
Thus the vector sector contains two massless vectors, $\hat c_\mu$ and $v_\mu$, and a massive vector $E_\mu$ with
\begin{equation}
m_V^2\ell^2=\frac{B}{\zeta}\,\ell^2=\nu^2-1\,.
\label{eq:R2MV}
\end{equation}
Notice that the massive vector is the dual field strength of the three-form of the linear multiplet, which is non-dynamical in the two-derivative theory. Altogether, for $0<\tau<2$ and $\tau\neq\tfrac32$, where $A\neq0$, the bosonic sector contains $20$ propagating modes, namely the graviton (5), the massless vectors $\hat c_\mu$ and $v_\mu$ (3+3), the massive vector $E_\mu$ (4), the scalars $\sigma_0$ and $\sigma_\pm$ (1+2) and the complex scalar $\Pi$ (2), compared with the $12$ modes of the two-derivative theory. These counts are also listed in Table~\ref{tab:R2}.

\subsubsection{Fermionic sector}
\label{sec:R2ferm}

The fermionic fluctuations are $\psi_\mu^i$ and $\varphi^i$. Here, we take $g>0$; for $g<0$, the labels of the two additional spinors below are interchanged. It is convenient to write the operator in (\ref{eq:Killing}) as $\bar D_\mu=\bar\nabla_\mu-\bar m\gamma_\mu\mathbb J$, where
\begin{equation}
\bar m=\frac{gL_0}{2\sqrt2}=\frac{1}{2\ell}\,,\qquad
(\mathbb J\chi)^i=-\rmi\delta^{ij}\chi_j\,,\qquad
\mathbb J^2=1\,,
\label{eq:Jdef}
\end{equation}
and to define the gauge invariant Rarita-Schwinger operator $\mathscr R^\mu(\psi)=\gamma^{\mu\nu\rho}\bar D_\nu\psi_\rho$, which is invariant under $\delta\psi_\mu=\bar D_\mu\epsilon$ since $\bar D_\mu$ is flat on the vacuum. The quadratic fermionic action is the sum of a two-derivative part, whose coefficients are proportional to $A$, and a contribution of the R$^2$ invariant proportional to $\zeta\bar\Lambda\big(\bar{\slashed{\nabla}}-\bar m\mathbb J\big)\Lambda$, where $\Lambda$ is the linearized fermion of the composite vector multiplet (\ref{eq:compl}),
\begin{equation*}
\Lambda=-\frac{2\rmi}{L_0}\Big(\big(\bar{\slashed{\nabla}}+5\bar m\mathbb J\big)\varphi+\frac{L_0}{3\sqrt2}\,\mathbb J\gamma_\mu\mathscr R^\mu(\psi)\Big)\,,
\end{equation*}
which is invariant under $\delta\psi_\mu=\bar D_\mu\epsilon$, since $\mathscr R^\mu$ is and $\delta\varphi$ vanishes on the vacuum \cite{repo}. As a result, the fermionic sector contains a massless gravitino $\psi_\mu^{(0)}$, which is a combination of $\psi_\mu$ and $\varphi$ that satisfies $A\mathscr R^\mu(\psi^{(0)})=0$, a spinor $\varphi_0$ with $\big(\bar{\slashed{\nabla}}-\bar m\mathbb J\big)\varphi_0=0$, and two spinors whose wave operator is
\begin{equation}
\zeta\big(\mathbb J\bar{\slashed{\nabla}}+5\bar m\big)\big(\mathbb J\bar{\slashed{\nabla}}-3\bar m\big)-A
=\zeta\Big(\big(\mathbb J\bar{\slashed{\nabla}}+\bar m\big)^2-\nu^2\ell^{-2}\Big)\,.
\label{eq:R2fermop}
\end{equation}
Therefore, these spinors satisfy $\big(\bar{\slashed{\nabla}}-\mu_\pm\mathbb J\big)\chi_\pm=0$ with
\begin{equation}
\mu_\pm\ell=-\tfrac12\pm\nu\,.
\label{eq:R2mupm}
\end{equation}
Notice that the parameter $\nu$ that appears in (\ref{eq:R2fermop}) is the same as that of the bosonic sector, as required by supersymmetry. For $0<\tau<2$ and $\tau\neq\tfrac32$, where $A\neq0$, the fermionic sector contains $20$ propagating modes, namely the massless gravitino (8) and the spinors $\varphi_0$, $\chi_+$ and $\chi_-$ (4+4+4).

\subsubsection{Supermultiplets and stability}
\label{sec:R2mult}

We now arrange the modes found in Sections~\ref{sec:R2bos} and \ref{sec:R2ferm} into supermultiplets and determine the signs of their norms. Using (\ref{eq:massenergy}), these modes have the energies listed in Table~\ref{tab:R2}. Since $\nu>1$ for $0<\tau<2$, all of these energies are given by the larger roots of (\ref{eq:massenergy}). The massless graviton, the massless gravitino $\psi_\mu^{(0)}$ and $v_\mu$ form the massless graviton multiplet $\mathcal G$, while $\sigma_0$, $\varphi_0$ and $\hat c_\mu$ form the massless vector multiplet $\mathcal V$. The remaining modes form a long vector multiplet whose lowest component is $\sigma_-$, and therefore the spectrum is
\begin{equation}
\mathcal G\oplus\mathcal V\oplus\mathcal V_L(1+\nu)\,.
\label{eq:R2spectrum}
\end{equation}
For $0<\tau<2$, the energy of the lowest component of the long multiplet satisfies $1+\nu>2$, in accordance with the unitarity bound of $\mathcal V_L(\Delta)$.

\begin{table}[h!]
\centering
\renewcommand{\arraystretch}{1.3}
\setlength{\tabcolsep}{5pt}
\begin{tabular}{|c|c|c|c|c|c|}
\hline
Mode & States & Mass & Energy & Multiplet & Sign of the norm\\
\hline\hline
$h_{\mu\nu}^{\rm TT}$ & $5$ & $m^2\ell^2=0$ & $4$ & & \\
$\psi_\mu^{(0)}$ & $8$ & $|m|\ell=\tfrac32$ & $\tfrac72$ & $\mathcal G$ & ${\rm sgn}\,A$\\
$v_\mu$ & $3$ & $m^2\ell^2=0$ & $3$ & & \\
\hline
$\sigma_0$ & $1$ & $m^2\ell^2=-4$ & $2$ & & \\
$\varphi_0$ & $4$ & $|m|\ell=\tfrac12$ & $\tfrac52$ & $\mathcal V$ & $+$\\
$\hat c_\mu$ & $3$ & $m^2\ell^2=0$ & $3$ & & \\
\hline
$\sigma_-$ & $1$ & $m^2\ell^2=(\nu+1)(\nu-3)$ & $1+\nu$ & & \\
$\chi_+$ & $4$ & $|m|\ell=\nu-\tfrac12$ & $\tfrac32+\nu$ & & \\
$\Pi$, $E_\mu$ & $2$, $4$ & $m^2\ell^2=(\nu+2)(\nu-2)$, $(\nu+1)(\nu-1)$ & $2+\nu$ & $\mathcal V_L(1+\nu)$ & ${\rm sgn}\,A$\\
$\chi_-$ & $4$ & $|m|\ell=\nu+\tfrac12$ & $\tfrac52+\nu$ & & \\
$\sigma_+$ & $1$ & $m^2\ell^2=(\nu+3)(\nu-1)$ & $3+\nu$ & & \\
\hline
\end{tabular}
\vspace{1.5mm}
\caption{The spectrum of the R$^2$ extension around the supersymmetric AdS$_5$ vacuum for $0<\tau<2$ and $\tau\neq\tfrac32$, where $\nu$ is defined in (\ref{eq:R2ABnu}). The second column gives the number of states of each mode. For $g<0$, the labels of $\chi_\pm$ are interchanged.}
\label{tab:R2}
\end{table}

We now turn to the signs of the kinetic terms. The kinetic coefficients of all modes are given in \cite{repo}, and their signs are listed in Table~\ref{tab:R2}. For instance, the kinetic terms of the graviton and $v_\mu$ are proportional to $A$, as can be seen from (\ref{eq:R2TT}) and (\ref{eq:R2vec3}), those of $\sigma_0$ and $\hat c_\mu$ are proportional to $B$, and that of the massive vector $E_\mu$ is proportional to $A\zeta/B$. Therefore, the massless vector multiplet always has positive norm, whereas the norms of the graviton multiplet and of the long vector multiplet have the sign of $A$. Combining this with the masses gives the ranges listed in Table~\ref{tab:R2ranges} and shown in Figure~\ref{fig:phases}.

\begin{table}[!htb]
\centering
\small
\renewcommand{\arraystretch}{1.3}
\relpenalty=10000 \binoppenalty=10000
\begin{tabular}{|l|l|l|l|}
\hline
Range & Parameters & Masses & Norms\\
\hline\hline
$\tau<0$ & $\nu^2<0$, $A>0$ & $\sigma_\pm$ complex, $E_\mu$ below the BF bound & unstable vacuum\\
$0<\tau<\tfrac32$ & $\nu>2$, $A>0$ & real, bounds respected, $1+\nu>3$ & all positive\\
$\tau=\tfrac32$ & $\nu=2$, $A=0$ & real & kinetic terms of $\mathcal G$ vanish\\
$\tfrac32<\tau<2$ & $1<\nu<2$, $A<0$ & real, bounds respected, $2<1+\nu<3$ & $\mathcal G$ and $\mathcal V_L(1+\nu)$ negative\\
$\tau=2$ & $\nu=1$, $B=0$ & real, $\mathcal V_L(2)$ at the threshold & Jordan pair of $\mathcal V$ and $\mathcal V_L(2)$\\
\hline
\end{tabular}
\vspace{1.5mm}
\caption{The spectrum of the R$^2$ extension around the supersymmetric AdS$_5$ vacuum in the ranges of the coupling $\tau=6\zeta g^2$, where $A$, $B$ and $\nu$ are defined in (\ref{eq:R2ABnu}). For $\tau>2$, there is no supersymmetric AdS$_5$ vacuum.}
\label{tab:R2ranges}
\end{table}

A few features of Table~\ref{tab:R2ranges} deserve comment. First, for $\tfrac32<\tau<2$, the graviton multiplet and the long vector multiplet are ghosts even though $\mathcal V_L(1+\nu)$ with $2<1+\nu<3$ is a unitary representation of $SU(2,2|1)$, which provides an explicit example of the fact that the unitarity of the representation and the sign of the norm are independent conditions. The fermionic kinetic coefficients have the same signs as those of their bosonic partners: the light spinor has positive norm, whereas the gravitino and the two additional spinors have norms proportional in sign to $A$ \cite{repo}. Second, the point $\tau=\tfrac32$, where $L_0=\tfrac43$ and the kinetic terms of the graviton and the gravitino vanish identically, is a degeneration of the quadratic action rather than a logarithmic sector: since $A\big(\bar\Box+2\ell^{-2}\big)h_{\mu\nu}^{\rm TT}$ simply vanishes at this point, rather than turning into $\big(\bar\Box+2\ell^{-2}\big)^2h_{\mu\nu}^{\rm TT}$, no logarithmic partner of a repeated bulk kinetic operator arises.\footnote{Analyzing the original quadratic action directly at this point, without dividing by $A$, one finds that with zero-source boundary conditions its nonzero-norm bulk modes form a single massless vector multiplet, with positive norm, while the remaining directions belong to the kernel of the quadratic symplectic form \cite{repo}. For the scalar of this multiplet, which saturates the BF bound, zero source means that the coefficient of the logarithmic branch of its asymptotic falloff is set to zero \cite{Henneaux:2004zi}.}

Third, at the endpoint $\tau=2$, where the two supersymmetric vacua merge and $\mathcal V_L(2)$ is at the threshold for shortening, $B=0$ and the diagonalization (\ref{eq:R2vec3}) is singular, since $\hat c_\mu$ contains $E_\mu/B$. In the original fields $(c_\mu,v_\mu,E_\mu)$ of (\ref{eq:R2vec2}), the kinetic matrix $K$ is nondegenerate and, with $M$ the mass matrix, $K^{-1}M$ is nilpotent, with the field equations $\bar\nabla^\mu(dE)_{\mu\nu}=0$, $\bar\nabla^\mu(dc)_{\mu\nu}=-gE_\nu$ and $\bar\nabla^\mu(dv)_{\mu\nu}=0$. Thus the massless vector of $\mathcal V$ and the massive vector form a Jordan pair with a logarithmic partner, as the gravitons of critical gravity \cite{Lu:2011zk,Deser:2011xc}, and so do the spinors $\varphi_0$ and $\chi_+$, whereas the two scalars at the BF bound remain independent \cite{repo}. This analysis is local, and the admissible logarithmic modes depend on the boundary conditions, which we do not analyze.

In Section~\ref{sec:central}, the points $\tau=\tfrac32$ and $\tau=2$ are identified as the zeros of $c$ and of $C_F$, respectively. Finally, within the positive-norm range, at $\tau=\tfrac{114}{121}$, where $\nu=3$, the lowest component $\sigma_-$ of the long multiplet becomes massless and has $\Delta=4$.

\FloatBarrier
\subsection{\texorpdfstring{$C^2+\tfrac16R^2$ Extension}{C2 + R2/6 Extension}}
\label{sec:W}

In this subsection, we study the spectrum of the model $\cL_{LR}+\alpha\,\cLW$ around the supersymmetric AdS$_5$ vacuum of Section~\ref{sec:vacuum}. The $C^2+\tfrac16R^2$ invariant adds to the $12+12$ states of the two-derivative theory a long spin-2 multiplet with $32+32$ states, and there is no range of the coupling in which the spectrum is free of both ghosts and instabilities, see Figure~\ref{fig:phases} and Table~\ref{tab:Wranges}. The reason can be stated at once. The invariant makes the composite scalar $d$ associated with the auxiliary field $D$ of the standard Weyl multiplet dynamical, and, as we show in Section~\ref{sec:Wbos}, its kinetic term has the wrong sign for every finite nonzero $\alpha$. The remaining modes then determine whether the vacuum is in addition unstable, which is the case for $0<\alpha g^2<6$, and whether the massless graviton multiplet is itself a ghost, which is the case for $\alpha g^2<-2$. At $\alpha g^2=-2$, the model has a critical point of the kind familiar from critical gravity, at which the graviton multiplet acquires logarithmic partners while $d$ and other negative-norm modes remain. Since $\tau=0$ for this model, the vacuum coincides with that of the two-derivative theory for all values of $\alpha$,
\begin{equation}
L_0=1\,,\qquad N_0=-\tfrac32g\,,\qquad \ell^2=\frac{2}{g^2}\,,
\label{eq:Wvac}
\end{equation}
where we take $\alpha\neq0$ and $g>0$, and the parameter in (\ref{eq:Jdef}) is $\bar m=g/(2\sqrt2)=1/(2\ell)$. For $g<0$, the signs of all the fermion masses $\mu$ below are reversed. It is convenient to define
\begin{equation}
\rho^2\equiv1-\frac{6}{\alpha g^2}\,,
\label{eq:Wrho}
\end{equation}
in terms of which all masses below take a simple form. The range of $\rho$ as a function of $\alpha g^2$ is the basis of the case analysis of Section~\ref{sec:Wmult}, which is summarized in Table~\ref{tab:Wranges}; here we only record it. The parameter $\rho^2$ is positive for $\alpha g^2<0$, where $\rho>1$, and for $\alpha g^2>6$, where $0<\rho<1$, whereas $\rho^2<0$ for $0<\alpha g^2<6$. Whenever $\rho^2\geq0$, we take $\rho\geq0$. In the limit $\alpha\to0^-$, $\rho$ diverges, and the modes that we find below in addition to those of the two-derivative theory decouple, while the two-derivative theory itself, $\alpha=0$, is a separate limit. The values $\alpha g^2=-2$, where $\rho=2$, and $\alpha g^2=6$, where $\rho=0$, are special, and they are discussed in Section~\ref{sec:Wmult}. In Sections~\ref{sec:Wbos} and \ref{sec:Wferm}, we assume $\alpha g^2\neq-2$ and $\alpha g^2\neq6$.

Before turning to the individual sectors, let us comment on the structure of the quadratic action. As explained in Section~\ref{sec:weyl}, the $C^2+\tfrac16R^2$ invariant is the supersymmetric completion of the Weyl tensor squared, in which the fields of the standard Weyl multiplet are composite. On the supersymmetric vacuum, the Weyl tensor, the auxiliary fields $T_{ab}$ and $D$, the $SU(2)$ field strength and the fermions of the standard Weyl multiplet all vanish. As a result, only the terms of the invariant that are quadratic in these fields contribute to the quadratic action, which is therefore a quadratic form in their linearized expressions, with coefficients and covariant derivatives evaluated on the background \cite{repo}. In particular, this is why the invariant does not contribute to the vacuum equations, see (\ref{eq:LWvac}). For the bosonic fluctuations, we find, up to a total derivative,
\begin{align}
e^{-1}\cLW^{(2)}={}&\tfrac18C_{abcd}C^{abcd}+\tfrac{64}{3}\mathcal D^2-\tfrac16v_{ab}v^{ab}-\tfrac23z_{ab}\bar z^{ab}-\tfrac{4\sqrt2}{3}g\,v_{ab}T^{ab}\nonumber\\*
&-\tfrac{64}{9}(dT)_{abc}(dT)^{abc}+\tfrac{64}{3}\bar\nabla^bT_{ab}\bar\nabla_cT^{ac}-\tfrac{32}{3}g^2T_{ab}T^{ab}\,,
\label{eq:Wquad}
\end{align}
where $C_{abcd}$ is the linearized Weyl tensor, $v_{ab}=(dv)_{ab}$, $z_{ab}=(dz)_{ab}$, $(dT)_{abc}=3\partial_{[a}T_{bc]}$, and
\begin{equation}
\mathcal D=\tfrac1{32}R^{(1)}-\tfrac14gn-\tfrac18g^2\sigma\,,\qquad
T_{ab}=\tfrac18\big(G_{ab}-2H_{ab}\big)\,.
\label{eq:WDT}
\end{equation}
Here, $R^{(1)}$ is the linearized Ricci scalar, and, as in Section~\ref{sec:R2bos}, $G_{ab}$ and $H_{ab}$ denote the linearized field strengths. Up to a sign, $\mathcal D$ is the linearization of the combination that appears in (\ref{eq:LWvac}), that is, of the auxiliary field $D$, and $T_{ab}$ is the linearization of the auxiliary tensor of the standard Weyl multiplet. Notice that the Ricci scalar squared part of the invariant is contained in $\tfrac{64}{3}\mathcal D^2$, and that the term $-\tfrac{32}{3}g^2T_{ab}T^{ab}$ receives a contribution from the special conformal gauge field of the standard Weyl multiplet, which is nonzero on the background. Thus, the invariant makes the composite scalar associated with $D$ dynamical, and it supplies kinetic terms for $T_{ab}$ and for the $SU(2)$ gauge fields $v_\mu$ and $z_\mu$, which are auxiliary in the two-derivative theory. The Chern-Simons term $\tfrac1{16}\epsilon^{abcde}C_aR_{bc}{}^{fg}R_{defg}$ does not contribute at quadratic order. The bosonic quadratic action splits into the transverse traceless metric fluctuation, the neutral scalars, the complex vector $z_\mu$ and the vector-form sector, which we discuss in turn.

\subsubsection{Bosonic sector}
\label{sec:Wbos}

We start with the transverse traceless part of the metric fluctuation. Since $\mathcal D$ vanishes for $h_{\mu\nu}^{\rm TT}$, only the Weyl tensor contributes, and together with the two-derivative part, the quadratic Lagrangian of $h_{\mu\nu}^{\rm TT}$ is
\begin{equation}
e^{-1}\cL=\tfrac14h^{{\rm TT}\mu\nu}\big(\bar\Box+2\ell^{-2}\big)h_{\mu\nu}^{\rm TT}+\frac{\alpha}{12}h^{{\rm TT}\mu\nu}\big(\bar\Box+2\ell^{-2}\big)\big(\bar\Box+5\ell^{-2}\big)h_{\mu\nu}^{\rm TT}\,,
\end{equation}
in agreement with the general results for quadratic curvature gravity around AdS backgrounds \cite{Deser:2002jk}. The corresponding field equation reads
\begin{equation}
\big(\bar\Box+2\ell^{-2}\big)\Big(1+\tfrac12\alpha g^2+\frac{\alpha}{3}\big(\bar\Box+2\ell^{-2}\big)\Big)h_{\mu\nu}^{\rm TT}=0\,.
\label{eq:WTT}
\end{equation}
Therefore, the invariant changes the coefficient of the Einstein operator from $1$ to $1+\tfrac12\alpha g^2$ and gives rise to a massive spin-2 mode. For $\alpha g^2\neq-2$, we decompose $h_{\mu\nu}^{\rm TT}=h_{\mu\nu}^{(0)}+h_{\mu\nu}^{(1)}$, where $h_{\mu\nu}^{(0)}$ is a massless graviton and $h_{\mu\nu}^{(1)}$ is a massive spin-2 mode with
\begin{equation}
m^2\ell^2=-3-\frac{6}{\alpha g^2}=\rho^2-4\,.
\label{eq:WM2}
\end{equation}
Their kinetic coefficients are $1+\tfrac12\alpha g^2$ and $-\big(1+\tfrac12\alpha g^2\big)$, respectively, so that, as in quadratic curvature gravity \cite{Lu:2011zk,Deser:2011xc}, one of the two spin-2 modes is a ghost.

The neutral scalar sector consists of the fluctuations of $L$, $N$ and the trace of the metric. For these fluctuations, the linearized Weyl tensor vanishes, and the invariant contributes only through $\tfrac{64}{3}\mathcal D^2$. In fact, for configurations in which only the metric and the scalars are nontrivial, (\ref{eq:LC2}) reduces exactly to $\tfrac18C_{abcd}C^{abcd}+\tfrac{64}{3}\big(\tfrac1{32}R-\tfrac14gN-\tfrac1{16}g^2L^2\big)^2$. Introducing an auxiliary scalar $\Psi$ for the expression in parentheses, the scalar part of the Lagrangian can be written as
\begin{equation}
e^{-1}\cL=\big(L+\tfrac43\alpha\Psi\big)R+L^{-1}\partial_aL\partial^aL-U-\tfrac83\alpha\big(8\Psi^2+4gN\Psi+g^2L^2\Psi\big)\,,
\label{eq:Wscalar}
\end{equation}
where $U$ is the potential of the two-derivative theory, given by (\ref{eq:R2pot}) with $\zeta=0$. The field equation of $\Psi$ sets it equal to the expression in parentheses, so that $\Psi_0=0$ on the vacuum.  After eliminating $N$ and performing a Weyl rescaling of the metric by $F^{2/3}$ with $F=L+\tfrac43\alpha\Psi$, the kinetic terms of the scalars take the form $-\mathcal K_{ij}\partial_a\phi^i\partial^a\phi^j$, where $\phi^i=(L,\Psi)$ and $\mathcal K_{ij}=\tfrac43F^{-2}\partial_iF\partial_jF-F^{-1}K_{ij}$, with $K_{LL}=L^{-1}$ the only nonzero entry of the kinetic matrix in (\ref{eq:Wscalar}). On the vacuum, we find
\begin{equation}
\mathcal K=\begin{pmatrix}\tfrac13&\tfrac{16}{9}\alpha\\[2pt] \tfrac{16}{9}\alpha&\tfrac{64}{27}\alpha^2\end{pmatrix},\qquad
\det\mathcal K=-\tfrac{64}{27}\alpha^2\,,
\label{eq:WK}
\end{equation}
and the masses of the two scalar fluctuations are given by
\begin{equation}
m_0^2\ell^2=-4\,,\qquad m_d^2\ell^2=\rho^2-4\,.
\label{eq:Wscalarmasses}
\end{equation}
The first scalar is that of the two-derivative theory, which saturates the BF bound, and we denote it by $\sigma_0$. The second one is the fluctuation of the composite scalar associated with the auxiliary field $D$ of the standard Weyl multiplet, which the invariant makes dynamical, and we denote it by $d$. Its mass coincides with that of the massive spin-2 mode (\ref{eq:WM2}). Since $\det\mathcal K<0$ for all $\alpha\neq0$, the kinetic matrix is indefinite, and one of the two scalars is a ghost. We find that $\sigma_0$ has positive norm, and therefore the scalar $d$ is a ghost for every finite nonzero $\alpha$.

The complex vector $z_\mu$ appears in the two-derivative Lagrangian only through the term $4z_\mu\bar z^\mu$, see (\ref{eq:R2z}), and the invariant provides its kinetic term through (\ref{eq:Wquad}). Hence, its quadratic Lagrangian is
\begin{equation}
e^{-1}\cL=4z_\mu\bar z^\mu-\tfrac23\alpha\,z_{\mu\nu}\bar z^{\mu\nu}\,,
\label{eq:Wz}
\end{equation}
which describes a complex massive vector with R-charge $\pm2$, kinetic coefficient $\tfrac43\alpha$ and mass
\begin{equation}
m_z^2\ell^2=-\frac{3}{\alpha}\,\ell^2=\rho^2-1\,.
\label{eq:Wmz}
\end{equation}

Finally, the vector-form sector consists of the fluctuations $c_\mu$, $v_\mu$, $b_{\mu\nu}$ and $e_{\mu\nu\rho}$. Its two-derivative part is given by (\ref{eq:R2vec}) with $\zeta=0$ and $L_0=A=1$, and the contribution of the invariant follows from (\ref{eq:Wquad}). Since $G_{ab}$ is closed, the two-forms enter only through $T_{ab}$, and it is convenient to decompose
\begin{equation}
8T_{ab}=t_{ab}+(dw)_{ab}\,,\qquad \bar\nabla^bt_{ab}=0\,,
\label{eq:Wdecomp}
\end{equation}
into a coexact part $t_{ab}$ and an exact part, where $w_\mu$ is defined up to a gauge transformation. Since $\bar\nabla^bH_{ab}=\tfrac12gE_a$ at linear order, see (\ref{eq:Bianchi}), we have $gE_a=\bar\nabla^b(dc-dw)_{ab}$. In terms of these variables, the Lagrangian of the vector-form sector splits into two parts that do not mix. The first one,
\begin{equation}
e^{-1}\cL_t=-\frac{\alpha}{9}(dt)_{abc}(dt)^{abc}+\Big(1-\frac{\alpha g^2}{6}\Big)t_{ab}t^{ab}\,,
\label{eq:Wt}
\end{equation}
describes a massive two-form with kinetic coefficient $\tfrac43\alpha$ and mass
\begin{equation}
m_t^2\ell^2=\Big(\frac{g^2}{2}-\frac{3}{\alpha}\Big)\ell^2=\rho^2\,.
\label{eq:Wmt}
\end{equation}
Its second-order field equation $\bar\nabla^a(dt)_{abc}=m_t^2t_{bc}$ splits into two first-order equations, whose solutions transform as $(1,0)$ and $(0,1)$ under $SO(4)$ and have $|m|\ell=\rho$. The second part involves the transverse vectors $w_\mu$, $c_\mu$ and $v_\mu$,
\begin{align}
e^{-1}\cL_V={}&-\tfrac32(dc)^2-\frac{\sqrt2}{g}\,dc\cdot dv-\frac{\alpha}{6}(dv)^2+\Big(1-\frac{\alpha g^2}{6}\Big)\Big((dw)^2+\frac{\sqrt2}{g}\,dw\cdot dv\Big)\nonumber\\*
&+\frac{\alpha}{3}\,\bar\nabla^b(dw)_{ab}\bar\nabla_c(dw)^{ac}-2E_aE^a\,,\label{eq:WV}
\end{align}
where $X\cdot Y\equiv X_{ab}Y^{ab}$ and $X^2\equiv X\cdot X$. Here, we impose the field equation of the two-form before parametrizing its exact part by $w_\mu$. The field equation of $w_\mu$ alone gives only the divergence of the former and would introduce a spurious massless root \cite{repo}. For transverse modes with $\bar\nabla^\mu(dX)_{\mu\nu}=m^2X_\nu$, the resulting characteristic polynomial of the vector system is proportional to
\begin{equation}
m^4\Big(\big(m^2\ell^2-\rho^2\big)^2-4\rho^2\Big)\,.
\end{equation}
The massless modes are carried by $v_\mu$ and by
\begin{equation}
u_\mu\equiv c_\mu+\frac{\sqrt2}{3g}\,v_\mu\,.
\label{eq:Wu}
\end{equation}
On these modes, the field equations give $w_\mu=-v_\mu/(\sqrt2g)$ and $E_a=0$, and (\ref{eq:WV}) reduces to
\begin{equation}
e^{-1}\cL_V=-\tfrac32(du)^2-\frac{2+\alpha g^2}{12g^2}(dv)^2\,.
\label{eq:WVmassless}
\end{equation}
The remaining two modes are massive vectors $W_\mu^\pm$, which are linear combinations of $w_\mu$, $c_\mu$ and $v_\mu$, $(w_\mu,c_\mu,v_\mu)\propto\big(1\mp\rho,\mp\rho,-\tfrac{1}{\sqrt2}g\rho(\rho\mp1)\big)W_\mu^\pm$, with
\begin{equation}
m_\pm^2\ell^2=\rho(\rho\pm2)\,.
\label{eq:Wmpm}
\end{equation}
In the normalization given above, their kinetic coefficients are $-4\rho^2(\rho\pm2)/(\rho\pm1)$ \cite{repo}. Notice that, in the two-derivative limit, $t_{ab}$, $W_\mu^\pm$ and $z_\mu$ disappear from the spectrum, while the massless vectors $u_\mu$ and $v_\mu$ are those of the two-derivative theory. Altogether, for $\rho^2>0$, except at $\alpha g^2=-2$ and $\alpha g^2=6$, where $\rho=2$ and $\rho=0$, the bosonic sector contains $44$ propagating modes, namely the massless graviton (5), the massive spin-2 mode (9), the scalars $\sigma_0$ and $d$ (1+1), the complex vector $z_\mu$ (8), the two-forms $t_{\mu\nu}$ (6), the massless vectors $u_\mu$ and $v_\mu$ (3+3) and the massive vectors $W_\mu^\pm$ (4+4), compared with the $12$ modes of the two-derivative theory. These counts are also listed in Table~\ref{tab:W}.

\subsubsection{Fermionic sector}
\label{sec:Wferm}

The fermionic fluctuations are $\psi_\mu^i$ and $\varphi^i$. As for the bosons, the invariant contributes to the quadratic action through bilinears in the linearized fermions of the standard Weyl multiplet. These are the gamma-traceless part $\mathcal Q_{ab}$ of the linearized gravitino curvature and the linearized spinor $\mathcal X$ of the standard Weyl multiplet,
\begin{equation*}
\mathcal Q_{ab}=Q_{ab}+\tfrac23\gamma_{[a}\gamma^cQ_{b]c}-\tfrac1{12}\gamma_{ab}\gamma^{cd}Q_{cd}\,,\qquad\mathcal X=\tfrac1{32}\gamma^{ab}\bar\nabla_a\psi_b+\tfrac18\bar m\gamma^a\mathbb J\psi_a+\tfrac18g\varphi\,,
\end{equation*}
where $Q_{ab}=2\bar\nabla_{[a}\psi_{b]}$, so that $\gamma^a\mathcal Q_{ab}=0$. Both are invariant under $\delta\psi_\mu=\bar D_\mu\epsilon$, and the fermionic part of the quadratic Lagrangian of the invariant consists of the kinetic terms $\bar{\mathcal Q}_{ab}\gamma^c\bar\nabla_c\mathcal Q^{ab}$ and $\bar{\mathcal X}\bar{\slashed{\nabla}}\mathcal X$ together with mass terms proportional to $g$ \cite{repo}. Consequently, the spinor $\mathcal X$ propagates, and the gravitino acquires a kinetic operator of third order in derivatives. After separating the spin-3/2 and spin-1/2 parts of the field equations \cite{repo}, we find that the gamma-traceless and divergence-free part of a gravitino $\hat\psi_\mu$, which differs from $\psi_\mu$ by terms proportional to $\gamma_\mu$, satisfies
\begin{align}
&\Big(1+\tfrac12\alpha g^2\Big)\big(\mathbb J\bar{\slashed{\nabla}}-3\bar m\big)\hat\psi_\mu+\frac\alpha3\big(\mathbb J\bar{\slashed{\nabla}}-3\bar m\big)^2\big(\mathbb J\bar{\slashed{\nabla}}+5\bar m\big)\hat\psi_\mu\nonumber\\*
&\qquad=\frac\alpha3\big(\mathbb J\bar{\slashed{\nabla}}-3\bar m\big)\Big(\big(\mathbb J\bar{\slashed{\nabla}}+\bar m\big)^2-\rho^2\ell^{-2}\Big)\hat\psi_\mu=0\,,
\label{eq:Wgravitino}
\end{align}
which is the fermionic counterpart of (\ref{eq:WTT}). Its first factor describes a massless gravitino $\psi_\mu^{(0)}$, and the second factor describes two massive gravitini $\psi_\mu^\pm$ that satisfy $\big(\bar{\slashed{\nabla}}-\mu_\pm\mathbb J\big)\psi_\mu^\pm=0$ with
\begin{equation}
\mu_\pm\ell=-\tfrac12\mp\rho\,.
\end{equation}
In the spin-1/2 sector, a linear combination $\varphi_0$ of $\varphi$ and $\mathcal X$ satisfies $\big(\bar{\slashed{\nabla}}-\bar m\mathbb J\big)\varphi_0=0$, as in the two-derivative theory, while $\mathcal X$ satisfies
\begin{equation}
\Big(\big(\mathbb J\bar{\slashed{\nabla}}-\bar m\big)^2-\rho^2\ell^{-2}\Big)\mathcal X=0\,,
\label{eq:WX}
\end{equation}
and describes two spinors $\chi_\pm$ that satisfy $\big(\bar{\slashed{\nabla}}-\mu^\prime_\pm\mathbb J\big)\chi_\pm=0$ with
\begin{equation}
\mu^\prime_\pm\ell=\tfrac12\pm\rho\,.
\end{equation}
The labels are chosen such that $|\mu_\pm|\ell=|\mu^\prime_\pm|\ell=\rho\pm\tfrac12$ for $\rho>\tfrac12$. Notice that the parameter $\rho$ that appears in (\ref{eq:Wgravitino}) and (\ref{eq:WX}) is the same as that of the bosonic sector, as required by supersymmetry. The kinetic coefficients of $\psi_\mu^{(0)}$ and $\psi_\mu^\pm$ are proportional to $1+\tfrac12\alpha g^2$ and $\alpha\rho(\rho\pm2)$, respectively, the norm of $\varphi_0$ is positive, and those of $\chi_\pm$ are proportional to $-\rho(\rho\pm1)$ \cite{repo}. Here, the norms of the fermions refer to the symplectic Majorana reality condition, with the overall sign fixed such that the fermions of the two-derivative theory have positive norm. For $\rho^2>0$, except at $\alpha g^2=-2$ and $\alpha g^2=6$, the fermionic sector contains $44$ propagating modes, namely the massless gravitino (8), the massive gravitini $\psi_\mu^\pm$ (12+12), the spinor $\varphi_0$ (4) and the spinors $\chi_\pm$ (4+4).

\subsubsection{Supermultiplets and stability}
\label{sec:Wmult}

We now arrange the modes found in Sections~\ref{sec:Wbos} and \ref{sec:Wferm} into supermultiplets and determine the signs of their norms. Using (\ref{eq:massenergy}), these modes have the energies listed in Table~\ref{tab:W}. For $\alpha g^2<0$, where $\rho>1$, all of these energies are given by the larger roots of (\ref{eq:massenergy}). The massless graviton $h_{\mu\nu}^{(0)}$, the massless gravitino $\psi_\mu^{(0)}$ and $v_\mu$ form the massless graviton multiplet $\mathcal G$, while $\sigma_0$, $\varphi_0$ and $u_\mu$ form the massless vector multiplet $\mathcal V$. The remaining $32+32$ modes form a long spin-2 multiplet, whose lowest component is the massive vector $W_\mu^-$, and therefore the spectrum is
\begin{equation}
\mathcal G\oplus\mathcal V\oplus\mathcal L_2(1+\rho)\,.
\label{eq:Wspectrum}
\end{equation}
The masses (\ref{eq:WM2}), (\ref{eq:Wscalarmasses}), (\ref{eq:Wmz}), (\ref{eq:Wmt}) and (\ref{eq:Wmpm}), as well as those of the fermions, reproduce the energies $1+\rho,\ \tfrac32+\rho,\ \dots,\ 3+\rho$ of the components of $\mathcal L_2(1+\rho)$ in Table~\ref{tab:multiplets}. This is the AdS$_5$ counterpart of the massive $32+32$ multiplet that arises around the Minkowski$_5$ vacuum of the ungauged theory with curvature squared invariants \cite{Ozkan:2013uk}.

\begin{table}[!htb]
\centering
\renewcommand{\arraystretch}{1.3}
\begin{tabular}{|c|c|c|c|c|c|}
\hline
Mode & States & Mass & Energy & Multiplet & Sign of the norm\\
\hline\hline
$h_{\mu\nu}^{(0)}$ & $5$ & $m^2\ell^2=0$ & $4$ & & \\
$\psi_\mu^{(0)}$ & $8$ & $|m|\ell=\tfrac32$ & $\tfrac72$ & $\mathcal G$ & ${\rm sgn}(\rho-2)$\\
$v_\mu$ & $3$ & $m^2\ell^2=0$ & $3$ & & \\
\hline
$\sigma_0$ & $1$ & $m^2\ell^2=-4$ & $2$ & & \\
$\varphi_0$ & $4$ & $|m|\ell=\tfrac12$ & $\tfrac52$ & $\mathcal V$ & $+$\\
$u_\mu$ & $3$ & $m^2\ell^2=0$ & $3$ & & \\
\hline
$W_\mu^-$ & $4$ & $m^2\ell^2=\rho(\rho-2)$ & $1+\rho$ & & $-{\rm sgn}(\rho-2)$\\
$\psi_\mu^-$, $\chi_-$ & $12$, $4$ & $|m|\ell=\rho-\tfrac12$ & $\tfrac32+\rho$ & & $-{\rm sgn}(\rho-2)$, $-$\\
$h_{\mu\nu}^{(1)}$, $d$ & $9$, $1$ & $m^2\ell^2=(\rho+2)(\rho-2)$ & $2+\rho$ & & $-{\rm sgn}(\rho-2)$, $-$\\
$t_{\mu\nu}$ & $6$ & $|m|\ell=\rho$ & $2+\rho$ & $\mathcal L_2(1+\rho)$ & $-$\\
$z_\mu$ & $8$ & $m^2\ell^2=(\rho+1)(\rho-1)$ & $2+\rho$ & & $-$\\
$\psi_\mu^+$, $\chi_+$ & $12$, $4$ & $|m|\ell=\rho+\tfrac12$ & $\tfrac52+\rho$ & & $-$, $-$\\
$W_\mu^+$ & $4$ & $m^2\ell^2=\rho(\rho+2)$ & $3+\rho$ & & $-$\\
\hline
\end{tabular}
\vspace{1.5mm}
\caption{The spectrum of the $C^2+\tfrac16R^2$ extension around the supersymmetric AdS$_5$ vacuum for $\alpha g^2<0$ and $\alpha g^2\neq-2$, where $\rho$ is defined in (\ref{eq:Wrho}). The second column gives the number of states of each mode. In this range, ${\rm sgn}(\rho-2)={\rm sgn}(2+\alpha g^2)$. For the massive modes, the last column gives the sign of the norm of the components of highest helicity.}
\label{tab:W}
\end{table}

We now turn to the signs of the kinetic terms. The kinetic coefficients of the massless graviton multiplet are proportional to $1+\tfrac12\alpha g^2$, as can be seen from (\ref{eq:WTT}), (\ref{eq:WVmassless}) and (\ref{eq:Wgravitino}), and those of the massless vector multiplet are positive. For the long multiplet, the kinetic coefficients of the components of highest helicity are given in Sections~\ref{sec:Wbos} and \ref{sec:Wferm}. The norms of the remaining components have the signs of $Z\,m^2$ for the longitudinal components of the massive vectors and the helicity-one components of the massive spin-2 mode, of $Z\,m^2\big(m^2+3\ell^{-2}\big)$ for the helicity-zero component of the massive spin-2 mode, and of $Z\big(\mu^2-\tfrac94\ell^{-2}\big)$ for the helicity-1/2 components of the massive gravitini, where $Z$ is the corresponding kinetic coefficient, as in Section~\ref{sec:multiplets}, and $m$ and $\mu$ are the masses \cite{repo}. Since the scalar $d$ has negative norm for every finite nonzero $\alpha$, see (\ref{eq:WK}), the spectrum contains a ghost for every value of the coupling. What remains to be determined is whether the vacuum is in addition unstable, whether the massless multiplets have positive norm, and how many of the additional modes are ghosts. Combining the masses with the signs of the norms gives Table~\ref{tab:Wranges}, which is shown in Figure~\ref{fig:phases}.

\begin{table}[H]
\centering
\small
\renewcommand{\arraystretch}{1.15}
\relpenalty=10000 \binoppenalty=10000
\begin{tabular}{|l|l|P{5.3cm}|P{4.9cm}|}
\hline
Range & $\rho$ & Masses & Norms\\
\hline\hline
$\alpha g^2<-2$ & $1<\rho<2$ & real, bounds respected, $1+\rho<3$ & $8+8$ negative in $\mathcal G$, $24+24$ negative in $\mathcal L_2$\\
$\alpha g^2=-2$ & $\rho=2$ & $h^{(1)}_{\mu\nu}$, $\psi_\mu^-$, $W_\mu^-$ massless & kinetic terms of $\mathcal G$ vanish, logarithmic modes\\
$-2<\alpha g^2<0$ & $\rho>2$ & real, bounds respected, $1+\rho>3$ & $\mathcal G$, $\mathcal V$ positive, all $32+32$ of $\mathcal L_2$ negative\\
$0<\alpha g^2<6$ & $\rho^2<0$ & $h^{(1)}_{\mu\nu}$, $d$, $z_\mu$ below the BF bounds; $W_\mu^\pm$, $\psi_\mu^\pm$, $\chi_\pm$ complex & unstable vacuum\\
$\alpha g^2=6$ & $\rho=0$ & $W_\mu^\pm$, $\psi_\mu^\pm$, $\chi_\pm$ pairwise degenerate, $d$ at the BF bound & repeated factors in (\ref{eq:Wgravitino}), (\ref{eq:WX})\\
$\alpha g^2>6$ & $0<\rho<1$ & real, bounds respected; $\mathcal L_2(1+\rho)$ requires the smaller roots for $W_\mu^-$, and for $\psi_\mu^-$, $\chi_-$ if $\rho<\tfrac12$ & $\mathcal G$, $\mathcal V$ positive, $16+16$ negative\\
\hline
\end{tabular}
\vspace{1.5mm}
\caption{The spectrum of the $C^2+\tfrac16R^2$ extension around the supersymmetric AdS$_5$ vacuum in the ranges of the coupling, where $\rho$ is defined in (\ref{eq:Wrho}). The vacuum is the same for all $\alpha$, and the scalar $d$ has negative norm in every range.}
\label{tab:Wranges}
\end{table}

A few features of Table~\ref{tab:Wranges} deserve comment. For $\alpha g^2<-2$, the massless graviton multiplet is a ghost, since $1+\tfrac12\alpha g^2<0$, and the lowest energy $1+\rho<3$ of the long multiplet violates the unitarity bound of $\mathcal L_2(\Delta)$, so that the norms within $\mathcal L_2(1+\rho)$ have both signs. For $-2<\alpha g^2<0$, the long multiplet is a unitary representation of $SU(2,2|1)$ all of whose states have negative norm, which shows once more that the unitarity of the representation and the sign of the norm are independent conditions; for small $|\alpha|g^2$, where $\rho\simeq\big(6/(|\alpha|g^2)\big)^{1/2}$, the ghost multiplet decouples at energies small compared to $\rho/\ell$. For $\alpha g^2>6$, the larger roots of (\ref{eq:massenergy}) do not reproduce the energies of $\mathcal L_2(1+\rho)$: for $W_\mu^-$, the larger root is $3-\rho$ rather than $1+\rho$, and for $6<\alpha g^2<8$, where $\rho<\tfrac12$, the same happens for $\psi_\mu^-$ and $\chi_-$, whose larger roots are $\tfrac52-\rho$. These modes realize the energies of $\mathcal L_2(1+\rho)$ only if they are quantized with the smaller roots, that is, with non-standard boundary conditions, in the sense of Section~\ref{sec:fluctuations}. Therefore, the assignment (\ref{eq:Wspectrum}) is formal in this range: with the standard boundary conditions, the additional modes do not form a single representation of $SU(2,2|1)$ with lowest energy $1+\rho$, and the count of negative-norm states in Table~\ref{tab:Wranges} refers to the modes of the quadratic action. At $\alpha g^2=6$, which separates the last two ranges, (\ref{eq:Wgravitino}) and (\ref{eq:WX}) contain repeated factors, and the structure of the quadratic action at this point is described in \cite{repo}. The critical point $\alpha g^2=-2$ is discussed in the remainder of this subsection.

At $\alpha g^2=-2$, we have $\rho=2$, and the coefficient $1+\tfrac12\alpha g^2$ vanishes. In this case, the massive spin-2 mode becomes massless, and (\ref{eq:WTT}) becomes $\big(\bar\Box+2\ell^{-2}\big)^2h_{\mu\nu}^{\rm TT}=0$. As in critical gravity \cite{Lu:2011zk,Deser:2011xc}, the massive graviton is replaced by a logarithmic mode $h_{\mu\nu}^{\log}$ with $\big(\bar\Box+2\ell^{-2}\big)h_{\mu\nu}^{\log}\propto h_{\mu\nu}^{(0)}$. Supersymmetry extends this structure to the rest of the graviton multiplet. The spin-3/2 operator in (\ref{eq:Wgravitino}) becomes $\tfrac\alpha3\big(\mathbb J\bar{\slashed{\nabla}}-3\bar m\big)^2\big(\mathbb J\bar{\slashed{\nabla}}+5\bar m\big)$, and $\psi_\mu^-$ is replaced by a logarithmic partner of $\psi_\mu^{(0)}$, while $W_\mu^-$ becomes massless and forms a Jordan block in the vector sector, in which $u_\mu$ remains an ordinary Maxwell field \cite{repo}. Correspondingly, the long multiplet $\mathcal L_2(3)$ is at the threshold of unitarity, and at the level of characters, it decomposes into a massless graviton multiplet and two semishort multiplets with $12+12$ states each, the former combining with $\mathcal G$ into logarithmic pairs. The kinetic coefficients of $\mathcal G$ vanish at this point, and the inner product on each Jordan block is indefinite, as in critical gravity \cite{Porrati:2011ku}. This is the five-dimensional counterpart of critical Einstein-Weyl supergravity in four dimensions \cite{Lu:2011mw}. However, unlike in critical gravity, where the massive scalar mode is removed by a choice of the coefficient of the Ricci scalar squared, the scalar $d$ is part of the same supermultiplet as the massive spin-2 mode. The components of $\mathcal L_2(3)$ that do not combine with $\mathcal G$, such as $d$, $t_{\mu\nu}$ and $z_\mu$, retain negative norms at $\alpha g^2=-2$. In particular, $d$ becomes a massless scalar with $\Delta=4$ and kinetic coefficient $-512/(9g^4)$. Excluding only the logarithmic modes therefore leaves negative-norm modes.

\FloatBarrier
\subsection{Gauss-Bonnet Extension}
\label{sec:GB}

In this subsection, we study the spectrum of the model $\cL_{LR}+\lambda\,\cL_{\rm GB}$ around the supersymmetric AdS$_5$ vacuum of Section~\ref{sec:vacuum}. Around this vacuum, the Gauss-Bonnet extension propagates no modes beyond the $12+12$ states of the two-derivative theory, and all of them have positive norm for $\lambda g^2<1$, see Figure~\ref{fig:phases}. Since the Gauss-Bonnet invariant was introduced in Section~\ref{sec:general} as the combination $\cL_{\rm GB}=\cLW+2\cL_{\log}$, one might expect it to be treated after the LOG invariant, as a special case. We treat it first, because it is the key to the spectrum of the LOG extension, for the following reason.

A curvature squared invariant can change the spectrum around the supersymmetric vacuum in two ways. First, it can shift the vacuum through its contribution to the parameter $\tau$, which changes $L_0$ and the AdS radius. Second, it can propagate modes in addition to those of the two-derivative theory. The LOG invariant does both: it contributes $\tau=\beta g^2$, and it propagates an additional long spin-2 multiplet. The $C^2+\tfrac16R^2$ invariant does only the second: as explained in Section~\ref{sec:W}, the Weyl tensor and the other covariant quantities of the standard Weyl multiplet, from which it is built, vanish on every maximally supersymmetric AdS$_5$ vacuum, so that it does not contribute to $\tau$, and it contributes to the quadratic action only through bilinears in their linearized expressions. The Gauss-Bonnet invariant does only the first: as we show in this subsection, around its shifted vacuum the quadratic action of the Gauss-Bonnet extension is a rescaled copy of that of the two-derivative theory, and the extension propagates no additional modes. Since $\cL_{\log}=\tfrac12\cL_{\rm GB}-\tfrac12\cLW$, the LOG extension with the coupling $\beta$ is the Gauss-Bonnet extension with $\lambda=\tfrac12\beta$, which accounts for the entire shift of the vacuum, together with $-\tfrac12\beta$ times the $C^2+\tfrac16R^2$ invariant, which accounts for all the additional modes. As a result, in Section~\ref{sec:LOG}, the spectrum of the LOG extension is obtained from (\ref{eq:GBresult}) and the results of Section~\ref{sec:W} by a change of the coupling, without a separate analysis. Apart from this, the Gauss-Bonnet extension is of interest in its own right, since it propagates no ghosts for $\lambda g^2<1$. We compute the central charges of its dual theory in Section~\ref{sec:GBcc}, and we study it around supersymmetric AdS$_3\times\Sigma_2$ backgrounds in Section~\ref{sec:GBprod}.

Since the rescaling holds for the complete quadratic action, we do not analyze the bosonic and the fermionic sectors separately, as we did in Sections~\ref{sec:R2} and \ref{sec:W}. Instead, in Section~\ref{sec:GBquad} we state the result, show how it arises in each of the two sectors, and read off the spectrum, which consists of the supermultiplets of the two-derivative theory.

For this model, $\tau=2\lambda g^2$, and the vacuum is given by the first root $L_0$ in (\ref{eq:L0}), with $1<L_0<2$ for $0<\tau<2$ and $0<L_0<1$ for $\tau<0$. Here, we take $\tau<2$.

\subsubsection{Quadratic action and spectrum}
\label{sec:GBquad}

The main result of this subsection is that, at quadratic order around its supersymmetric vacuum, the Gauss-Bonnet extension is a rescaled copy of the two-derivative theory. To state this precisely, let $S^{(2)}_{LR}[g]$ denote the quadratic action of the two-derivative theory around its supersymmetric vacuum, $L_0=1$ and $\ell^2=2/g^2$, as a functional of the fluctuations (\ref{eq:fluctuations}), $\psi_\mu^i$ and $\varphi^i$. Then the quadratic action of the Gauss-Bonnet extension is given by
\begin{equation}
S^{(2)}_{LR+\lambda{\rm GB}}[g]=\big(2-L_0\big)\,S^{(2)}_{LR}[gL_0]\,,
\label{eq:GBresult}
\end{equation}
where the right-hand side is evaluated on $\big(\sigma/L_0,\,n/L_0,\,e_{\mu\nu\rho}/L_0,\,\varphi^i/L_0\big)$, while the other fluctuations are the same on both sides. Notice that the two-derivative theory with the gauge coupling $gL_0$ has the AdS radius $\ell^2=2/(g^2L_0^2)$, which coincides with that of the Gauss-Bonnet extension, see (\ref{eq:susyconditions}). Therefore, the two theories have the same masses, and in these variables their kinetic coefficients differ by the overall factor $2-L_0$.

Let us show how (\ref{eq:GBresult}) arises in the bosonic sector. For the transverse traceless metric fluctuation, the Gauss-Bonnet combination of the curvature squared terms does not change the order of the field equation and only contributes to the coefficient of the Einstein operator \cite{Deser:2002jk}. Together with the term $-\tfrac14g^2L^2R$ in (\ref{eq:GB}), it changes this coefficient from $L_0$ to $L_0-\lambda\ell^{-2}=2-L_0$, where we have used $\lambda\ell^{-2}=\tfrac14\tau L_0^2=2(L_0-1)$, which follows from (\ref{eq:vacuum}). In the vector-form sector, the higher-derivative terms of the forms cancel in the Gauss-Bonnet invariant. Using the linearized Bianchi identity (\ref{eq:Bianchi}) in the remaining derivative couplings, up to total derivatives, the terms $G_{ab}H^{ab}$ and $v_{ab}H^{ab}$ can be written as $gE^ac_a$ and $gE^av_a$, respectively, and we find
\begin{equation}
e^{-1}\cL=\big(2-L_0\big)\Big(-\tfrac12G^2+4H^2-\frac{2}{L_0^2}E_aE^a-\frac{2\sqrt2}{L_0}E^av_a-4gE^ac_a\Big)\,.
\label{eq:GBvec}
\end{equation}
This is $2-L_0$ times the two-derivative Lagrangian (\ref{eq:R2vec}) at $\zeta=0$ and $L_0=1$, with $g\to gL_0$ and $E_a\to E_a/L_0$, the latter following from the rescaling of $e_{\mu\nu\rho}$. In the same way, the quadratic Lagrangian of $z_\mu$ becomes $4(2-L_0)z_\mu\bar z^\mu$. Thus, as in the two-derivative theory, $v_\mu$ is a Lagrange multiplier that imposes $E_a=0$, and $z_\mu$ is auxiliary. In the scalar sector, we use the fact that, on AdS$_5$, the first and second bulk metric variations of the Gauss-Bonnet density $R_{abcd}R^{abcd}-4R_{ab}R^{ab}+R^2$, with a constant coefficient, coincide with those of $-4\ell^{-2}R-24\ell^{-4}$, although their background values differ \cite{Deser:2002jk}. We then find a single scalar with $m^2\ell^2=-4$, whose kinetic coefficient, like that of the metric, is rescaled by $2-L_0$. 

The fermionic part of (\ref{eq:GBresult}) arises in the same way. On their own, the $C^2+\tfrac16R^2$ and LOG invariants give the gravitino a kinetic operator of third order in derivatives and propagate additional spinors, see Sections~\ref{sec:Wferm} and \ref{sec:LOG}. In the Gauss-Bonnet combination, the terms with three and with two derivatives of the fermions cancel, and the remaining terms combine with those of the two-derivative theory into $2-L_0$ times the fermionic quadratic action of the two-derivative theory with $g\to gL_0$ and $\varphi^i\to\varphi^i/L_0$. We have obtained this result from the complete two-fermion Lagrangian of the Gauss-Bonnet invariant, keeping all the components of $\psi_\mu^i$ and $\varphi^i$ and without imposing any gauge condition \cite{repo}. Therefore, the fermionic sector contains a massless gravitino and a spinor with $|m|\ell=\tfrac12$. In terms of the original spinor $\varphi^i$, the kinetic coefficient of the latter is $2(2-L_0)/(3L_0^2)$, and all the coefficients of the fermionic quadratic action vanish at $\tau=2$, before any field is eliminated.

As a consequence of (\ref{eq:GBresult}), the spectrum of the Gauss-Bonnet extension consists of the massless graviton multiplet $\mathcal G$ and the massless vector multiplet $\mathcal V$, that is, of the $12+12$ states of the two-derivative theory, and the norms of all of these states are multiplied by $2-L_0$. Since $L_0<2$ for $\tau<2$, all kinetic coefficients are positive and the scalar saturates the BF bound. Thus the Gauss-Bonnet extension is free of ghosts and tachyons around the supersymmetric vacuum for all $\lambda g^2<1$, including negative values of $\lambda$, and it propagates no additional modes. This is the AdS$_5$ counterpart of the result of \cite{Ozkan:2013uk} that, around the Minkowski$_5$ vacuum of the ungauged theory, the Gauss-Bonnet combination propagates only the states of the two-derivative theory. At the endpoint $\tau=2$, where the two roots of (\ref{eq:vacuum}) merge, we have $L_0=2$, and the quadratic action vanishes identically. This is a degeneration of the quadratic action rather than a logarithmic sector, and the spectrum at this point is not determined by the linearized analysis.

\subsubsection{Central charges}
\label{sec:GBcc}

The central charges of the dual four-dimensional theory follow from the same data. The coefficient $a$ of the Euler density in the trace anomaly is determined by the value of the Lagrangian on the vacuum \cite{Henningson:1998gx,Imbimbo:1999bj,Myers:2010tj}, $a=-\pi^2\ell^5\,e^{-1}\cL\big|_{\rm vac}/(8\kappa^2)$ in the normalization (\ref{eq:general}), and the coefficient $c$ of the Weyl tensor squared is proportional to the coefficient $2-L_0$ of the Einstein operator for the transverse traceless fluctuations. We find
\begin{equation}
a=\frac{\pi^2\ell^3}{\kappa^2}\big(4-3L_0\big)\,,\qquad
c=\frac{\pi^2\ell^3}{\kappa^2}\big(2-L_0\big)\,,
\label{eq:ac}
\end{equation}
which reduce to $a=c=\pi^2\ell^3/\kappa^2$ in the two-derivative theory, and therefore
\begin{equation}
c-a=\frac{\pi^2\lambda\ell}{\kappa^2}\,,\qquad
\frac ac=2-\frac{1}{\sqrt{1-k}}\,,\qquad k\equiv\lambda g^2\,.
\label{eq:aoverc}
\end{equation}
Notice that the background value of the Gauss-Bonnet density contributes to $a$, which therefore cannot be read off from the quadratic action, and that $c-a$ is linear in $\lambda$ only at fixed radius, since $\ell$ depends on $\lambda$ through $L_0$. For the Gauss-Bonnet extension, $c$ vanishes at $\tau=2$, where the quadratic action degenerates. We give $a$ and $c$ for the general model (\ref{eq:general}) in Section~\ref{sec:central}, where we also compare them with earlier results at first order in the couplings.

The values (\ref{eq:ac}) coincide with those of Einstein-Gauss-Bonnet gravity \cite{Boulware:1985wk} with $\lambda_{\rm EGB}f_\infty=\tfrac12(L_0-1)$, in the conventions of \cite{Buchel:2009tt}. In this identification, $\tau=2$ corresponds to $\lambda_{\rm EGB}=\tfrac14$, where the two branches of vacua of Einstein-Gauss-Bonnet gravity merge and the theory becomes Chern-Simons gravity \cite{Crisostomo:2000bb}, and the linearized perturbations around AdS$_5$ have no kinetic terms \cite{Fan:2016zfs}. For $\mathcal N=1$ superconformal field theories, the positivity of the energy flux requires $\tfrac12\leq a/c\leq\tfrac32$ \cite{Hofman:2008ar}, and by (\ref{eq:aoverc}) this gives $-3\leq k\leq\tfrac59$. In the same identification, this window is the causality window $-\tfrac7{36}\leq\lambda_{\rm EGB}\leq\tfrac9{100}$ of Einstein-Gauss-Bonnet gravity \cite{Brigante:2008gz,Buchel:2009tt,Hofman:2009ug,Camanho:2009vw}. It lies strictly inside the range $k<1$ in which the vacuum exists and the spectrum is free of ghosts, and the spectrum (\ref{eq:GBresult}) is regular at its endpoints, so the two windows do not coincide. We note, however, that a finite Gauss-Bonnet coupling leads to causality violation at high energies unless an infinite tower of higher-spin states is added \cite{Camanho:2014apa}, so that statements at finite $k$ refer to the truncated theory.

\subsection{LOG Extension}
\label{sec:LOG}

In this subsection, we study the spectrum of the model $\cL_{LR}+\beta\,\cL_{\log}$ around the supersymmetric AdS$_5$ vacuum. For this model, $\tau=\beta g^2$, and as in Section~\ref{sec:GB}, we take $\tau<2$. The result is that, up to an overall factor, the quadratic action of the LOG extension, in both the bosonic and the fermionic sectors, is that of the $C^2+\tfrac16R^2$ extension with a coupling determined by $\tau$. Consequently, for $0<\tau<2$, the LOG extension propagates a long spin-2 multiplet in addition to the $12+12$ states of the two-derivative theory, and its spectrum contains $32+32$ negative-norm states, for $\tau<0$ the vacuum is unstable, and $\tau=\tfrac32$ is a critical point, see Figure~\ref{fig:phases}. To derive this, we separate the two effects of the LOG invariant in the way explained in Section~\ref{sec:GB}, so that the sector-by-sector analysis of Section~\ref{sec:W} need not be repeated. Since $\cL_{\rm GB}=\cLW+2\cL_{\log}$, the model can be written as
\begin{equation}
\cL_{LR}+\beta\,\cL_{\log}=\Big(\cL_{LR}+\tfrac12\beta\,\cL_{\rm GB}\Big)-\tfrac12\beta\,\cLW\,.
\label{eq:LOGsplit}
\end{equation}
The terms in parentheses form the Gauss-Bonnet extension with $\lambda=\tfrac12\beta$, which has the same value of $\tau$ and therefore the same vacuum, and whose quadratic action is given by (\ref{eq:GBresult}), while the second term accounts for all the modes that are not present in the two-derivative theory. The latter contributes through the quadratic form (\ref{eq:Wquad}) and its fermionic counterpart. Indeed, the Weyl tensor, the auxiliary fields $T_{ab}$ and $D$, the $SU(2)$ field strength and the fermions of the standard Weyl multiplet vanish on every supersymmetric AdS$_5$ vacuum (\ref{eq:background}), (\ref{eq:susyconditions}), and the argument of Section~\ref{sec:W} applies. For $L_0\neq1$, these expressions hold with $g$ replaced by $gL_0$, in terms of the same variables as in (\ref{eq:GBresult}), including the contribution of the background special conformal gauge field to the $T_{ab}T^{ab}$ term. For instance, the linearization $\mathcal D$ of the composite scalar $D$, (\ref{eq:WDT}), becomes $\tfrac1{32}R^{(1)}-\tfrac14gn-\tfrac18g^2L_0\sigma$. Therefore, denoting by $S^{(2)}_W[g]$ the quadratic action of the $C^2+\tfrac16R^2$ invariant around the vacuum with $L_0=1$, the quadratic action of the LOG extension is given by
\begin{equation}
S^{(2)}_{LR+\beta\log}[g]=\big(2-L_0\big)\,S^{(2)}_{LR}[gL_0]-\tfrac12\beta\,S^{(2)}_W[gL_0]\,,
\label{eq:LOGundivided}
\end{equation}
where both terms on the right-hand side are evaluated on the variables of (\ref{eq:GBresult}). For $L_0\neq2$, this can be written as
\begin{equation}
S^{(2)}_{LR+\beta\log}[g]=\big(2-L_0\big)\,S^{(2)}_{LR+\hat\alpha W}[gL_0]\,,\qquad
\hat\alpha=-\frac{\beta}{2(2-L_0)}\,,
\label{eq:LOGresult}
\end{equation}
where $S^{(2)}_{LR+\hat\alpha W}[gL_0]$ denotes the quadratic action of the $C^2+\tfrac16R^2$ extension of Section~\ref{sec:W} with the coupling $\hat\alpha$ and the gauge coupling $gL_0$. In other words, for $L_0\neq2$, the spectrum of the LOG extension is that of the $C^2+\tfrac16R^2$ extension with
\begin{equation}
\alpha g^2\ \to\ \hat q\equiv\hat\alpha g^2L_0^2=-\frac{4(L_0-1)}{2-L_0}\,,
\label{eq:LOGmap}
\end{equation}
where we have used (\ref{eq:vacuum}), and, in these variables, all kinetic coefficients are multiplied by $2-L_0>0$. We have verified (\ref{eq:LOGundivided}) in the bosonic sector by a direct computation of the quadratic action of the LOG extension from (\ref{eq:LLR}) and (\ref{eq:Llog}). The fermionic quadratic action of the LOG extension has also been computed directly \cite{repo}, with the result described below.

Under (\ref{eq:LOGmap}), the parameter $\rho$ of Section~\ref{sec:W} becomes
\begin{equation}
\nu^2=\frac{4-L_0}{2(L_0-1)}\,,
\label{eq:LOGnu}
\end{equation}
which is positive for $0<\tau<2$, where $\nu>1$. For $0<\tau<2$ and $\tau\neq\tfrac32$, the spectrum of the LOG extension is
\begin{equation}
\mathcal G\oplus\mathcal V\oplus\mathcal L_2(1+\nu)\,.
\label{eq:LOGspectrum}
\end{equation}
Notice that, as a function of $L_0$, $\nu$ coincides with the parameter (\ref{eq:R2ABnu}) of the R$^2$ extension. On this domain, $\hat q<0$, and the masses, the energies and the multiplet assignments are those of Table~\ref{tab:W} with $\rho$ replaced by $\nu$. In terms of the coupling $\beta$, the massive spin-2 mode and the scalar $d$, the complex vector $z_\mu$ and the two-form $t_{\mu\nu}$ have
\begin{equation}
m_{h^{(1)}}^2=m_d^2=\frac{6(4-3L_0)}{\beta}\,,\qquad
m_z^2=\frac{6(2-L_0)}{\beta}\,,\qquad
m_t^2=\frac{2(4-L_0)}{\beta}\,,
\label{eq:LOGmasses}
\end{equation}
while the massive vectors have $m_\pm^2\ell^2=\nu(\nu\pm2)$, and the fermions have $\mu_\pm\ell=-\tfrac12\mp\nu$ and $\mu^\prime_\pm\ell=\tfrac12\pm\nu$ for $g>0$. The direct computation of the fermionic quadratic action \cite{repo} gives the same result: after the spin-3/2 and spin-1/2 parts of the field equations are separated, the gravitino and the spinor satisfy equations of third order in derivatives, whose roots give the massless gravitino, the gravitini with the masses $\mu_\pm$, the spinor $\varphi_0$ with $|m|\ell=\tfrac12$ and the spinors with the masses $\mu^\prime_\pm$. Thus the fermionic sector contains $44$ propagating modes, namely the massless gravitino (8), the gravitini with the masses $\mu_\pm$ (12+12), the spinor $\varphi_0$ (4) and the spinors with the masses $\mu^\prime_\pm$ (4+4), as for the $C^2+\tfrac16R^2$ extension. The kinetic coefficients are those of Sections~\ref{sec:Wbos} and \ref{sec:Wferm} multiplied by $2-L_0$. In particular, those of the massless graviton multiplet are proportional to $(2-L_0)\big(1+\tfrac12\hat q\big)=4-3L_0$, and those of the massless vector multiplet are proportional to $2-L_0$. Therefore, the signs of the norms are obtained from the last column of Table~\ref{tab:W} by replacing ${\rm sgn}(\rho-2)$ with ${\rm sgn}(4-3L_0)$.

The map (\ref{eq:LOGmap}) sends the ranges of $\tau$ to the ranges of Table~\ref{tab:Wranges} as listed in Table~\ref{tab:LOGranges} and shown in Figure~\ref{fig:phases}; the range $\alpha g^2>6$ of Table~\ref{tab:Wranges} is not realized by the LOG extension.

\begin{table}[!htb]
\centering
\small
\renewcommand{\arraystretch}{1.3}
\relpenalty=10000 \binoppenalty=10000
\begin{tabular}{|l|P{3.4cm}|P{4.2cm}|P{4.6cm}|}
\hline
Range & Parameters & Masses & Norms\\
\hline\hline
$\tau<0$ & $0<L_0<1$, $0<\hat q<2$, $\nu^2<0$ & $d$ below the BF bound & unstable vacuum\\
$0<\tau<\tfrac32$ & $1<L_0<\tfrac43$, $-2<\hat q<0$, $\nu>2$ & real, bounds respected, $1+\nu>3$ & $\mathcal G$, $\mathcal V$ positive, all $32+32$ of $\mathcal L_2$ negative\\
$\tau=\tfrac32$ & $L_0=\tfrac43$, $\hat q=-2$, $\nu=2$ & $h^{(1)}_{\mu\nu}$, $\psi_\mu^-$, $W_\mu^-$ massless & kinetic terms of $\mathcal G$ vanish, logarithmic modes\\
$\tfrac32<\tau<2$ & $\tfrac43<L_0<2$, $\hat q<-2$, $1<\nu<2$ & real, bounds respected, $1+\nu<3$ & $8+8$ negative in $\mathcal G$, $24+24$ negative in $\mathcal L_2$\\
$\tau=2$ & $L_0=2$, $\nu=1$ & quadratic action of the $C^2+\tfrac16R^2$ invariant alone, spin-2 mode partially massless & kinetic terms of $\mathcal V$ vanish\\
\hline
\end{tabular}
\vspace{1.5mm}
\caption{The spectrum of the LOG extension around the supersymmetric AdS$_5$ vacuum in the ranges of the coupling $\tau=\beta g^2$, obtained from Table~\ref{tab:Wranges} through the map (\ref{eq:LOGmap}), where $\nu$ is defined in (\ref{eq:LOGnu}). The conventions are those of Table~\ref{tab:Wranges}. For $\tau>2$, there is no supersymmetric AdS$_5$ vacuum.}
\label{tab:LOGranges}
\end{table}

At $\tau=\tfrac32$, where $L_0=\tfrac43$ and $\nu=2$, (\ref{eq:LOGmap}) gives $\hat q=-2$, so that the LOG extension has the critical point of Section~\ref{sec:Wmult}: the massive spin-2 mode, $\psi_\mu^-$ and $W_\mu^-$ become massless and are replaced by logarithmic partners of the members of $\mathcal G$, and the components of $\mathcal L_2(3)$ that do not combine with $\mathcal G$, such as $d$, $t_{\mu\nu}$ and $z_\mu$, retain negative norms. Notice that the same value of $\tau$ is also special for the R$^2$ extension, where, however, the quadratic action degenerates without logarithmic modes, see Section~\ref{sec:R2mult}.

At $\tau=2$, where $L_0=2$ and $\nu=1$, the first term in (\ref{eq:LOGundivided}) vanishes, and the quadratic action of the LOG extension reduces to $-\tfrac12\beta$ times that of the $C^2+\tfrac16R^2$ invariant alone, given by (\ref{eq:Wquad}) and its fermionic counterpart with $g\to2g$. In particular, the transverse traceless metric fluctuation satisfies $\big(\bar\Box+2\ell^{-2}\big)\big(\bar\Box+5\ell^{-2}\big)h_{\mu\nu}^{\rm TT}=0$, so that the massive spin-2 mode has $m^2\ell^2=-3$, which is the partially massless value, and the quadratic action has additional gauge symmetries.\footnote{For instance, since the background is conformally flat, the linearized Weyl tensor is invariant under $\delta h_{\mu\nu}=\omega\,\bar g_{\mu\nu}$, and $\mathcal D$ is invariant if this is accompanied by $\delta n=\delta R^{(1)}/(8g)$. Combined with a diffeomorphism, this gives the partially massless gauge transformation $\delta h_{\mu\nu}=\big(\bar\nabla_\mu\bar\nabla_\nu-\ell^{-2}\bar g_{\mu\nu}\big)\epsilon$.} We do not analyze this point further.

\FloatBarrier
\subsection{Central Charges and Special Points}
\label{sec:central}

The special values of the couplings found in Sections~\ref{sec:R2}--\ref{sec:LOG} can be understood in terms of the central charges of the putative dual four-dimensional $\mathcal N=1$ superconformal field theory. We first compute the central charges and relate them to the kinetic terms of the massless multiplets, then show that the vacuum equation is $a$-maximization, and finally identify the special points as zeros of the central charges.

\subsubsection{Central charges and the kinetic terms of the massless multiplets}

For the general model (\ref{eq:general}), the value of the Lagrangian on the vacuum and the coefficient of the Einstein operator for the transverse traceless fluctuations \cite{Bueno:2016ypa} give, as in Section~\ref{sec:GB},
\begin{equation}
a=\frac{\pi^2\ell^3}{\kappa^2}\big(4-3L_0\big)\,,\qquad
c=\frac{\pi^2\ell^3}{\kappa^2}\Big(4-3L_0+\tfrac12\alpha g^2L_0^2\Big)\,,
\label{eq:acgen}
\end{equation}
where $L_0$ is determined by $\tau$ through (\ref{eq:vacuum}). In the vector sector, the two massless vectors are $v_\mu$ and $u_\mu=c_\mu+\sqrt2v_\mu/(3gL_0)$, with the kinetic terms
\begin{equation}
e^{-1}\cL=-\tfrac32\big(2-L_0\big)(du)^2-\frac{1}{6g^2L_0^2}\Big(4-3L_0+\tfrac12\alpha g^2L_0^2\Big)(dv)^2\,.
\label{eq:uvgen}
\end{equation}
We have computed (\ref{eq:acgen}) and (\ref{eq:uvgen}) directly from the Lagrangians of Section~\ref{sec:theory} with all three couplings switched on \cite{repo}. They reproduce the results of the individual models: the coefficients $A=4-3L_0$ and $B=2-L_0$ in (\ref{eq:R2vec3}), the coefficient $1+\tfrac12\alpha g^2$ in (\ref{eq:WVmassless}), the factor $2-L_0$ in (\ref{eq:GBresult}), and the map (\ref{eq:LOGmap}). The four-derivative corrections to $a$ and $c$ in minimal gauged supergravity were determined at first order in the couplings in \cite{Bobev:2021qxx,Cassani:2022lrk}, and for the three invariants considered here, after a truncation to minimal gauged supergravity, in \cite{Gold:2023ymc,Gold:2025ttt}. At first order in the couplings, (\ref{eq:acgen}) agrees with the results of \cite{Gold:2023ymc}. Beyond first order, the vector multiplet cannot be truncated as in the two-derivative theory, since $L_0$ depends on the couplings and, as we show below, so does the mixing of the R-symmetry with the flavor symmetry. The vector $v_\mu$ belongs to the massless graviton multiplet $\mathcal G$, and its kinetic coefficient is proportional to $c$, as is that of the graviton. The vector $u_\mu$ belongs to the massless vector multiplet $\mathcal V$ and is dual to a flavor current, whose two-point coefficient we denote by $C_F$, so that $C_F\propto2-L_0$. Thus $a$ and $C_F$ depend on the couplings only through $\tau$, whereas $c$ also depends on $\alpha$, which multiplies the only invariant that contains the Weyl tensor squared. In particular, $a=c$ for the R$^2$ and LOG extensions.

\subsubsection{The trial \texorpdfstring{$a$}{a}-function and \texorpdfstring{$a$}{a}-maximization}

The functions $a$ and $C_F$ are related in the way required by $a$-maximization \cite{Intriligator:2003jj}. On the configurations (\ref{eq:background}) that satisfy (\ref{eq:susyconditions}) and $\delta\varphi^i=0$, but with $L$ not fixed by its field equation, the Lagrangian gives the cubic function of $y=L^{-1}$
\begin{equation}
\hat a(y)=\frac{2\sqrt2\,\pi^2}{g^3\kappa^2}\Big(3y^2-2y^3-\tfrac34\tau y\Big)
\label{eq:atrial}
\end{equation}
for all values of $\alpha$, $\beta$ and $\zeta$. Its stationary points are the roots of (\ref{eq:vacuum}), and its value at $y=L_0^{-1}$ is $a$. Its second derivative at $y=L_0^{-1}$ is $-12\sqrt2\,\pi^2(2-L_0)/(g^3\kappa^2L_0)$, so that the vacuum we study is a maximum of $\hat a$, while for $0<\tau<2$ the second root $L_0^\prime>2$ is a minimum, and the two coincide at $\tau=2$. In a four-dimensional superconformal field theory, the second derivative of the trial $a$-function with respect to the mixing of the R-symmetry with a flavor symmetry is $-\tfrac9{16}$ times the two-point coefficient of the flavor current \cite{Intriligator:2003jj}. Accordingly, the second derivative of (\ref{eq:atrial}) depends on the couplings through $2-L_0$, like $C_F$ and unlike $c$, in all four models.

In fact, (\ref{eq:atrial}) is the trial $a$-function of the dual theory. Its 't~Hooft anomalies follow from the Chern-Simons couplings of the model. By the Bianchi identity (\ref{eq:Bianchi}), $E_a=\tfrac2g\nabla^bH_{ab}+\tfrac1{8g}\epsilon_{abcde}G^{bc}G^{de}$. The second term turns the couplings $-2\sqrt2E^aV_a-4gE^aC_a$ of (\ref{eq:LLR}) into Chern-Simons terms, while the first one, after the dualization of $B_{\mu\nu}$, gives the kinetic terms of $V_\mu$, which on shell is the vector dual to $B_{\mu\nu}$. Together with the Chern-Simons terms of the invariants, we find
\begin{equation}
e^{-1}\cL_{\rm CS}=-\tfrac12\epsilon CGG-\frac{\sqrt2}{4g}\,\epsilon VGG-\tfrac1{24}\big(\alpha+\beta+6\zeta\big)\epsilon C\cV\cV+\tfrac1{16}\alpha\,\epsilon CRR\,,
\label{eq:LCS}
\end{equation}
where $\epsilon XYZ\equiv\epsilon^{abcde}X_aY_{bc}Z_{de}$, $\epsilon CRR\equiv\epsilon^{abcde}C_aR_{bc}{}^{fg}R_{defg}$, and we have used $\cF_{ab}\simeq\sqrt2\cV_{ab}$ at this order in (\ref{eq:LR2}). The other terms of the invariants that contain the Levi-Civita tensor are gauge invariant and do not contribute. Reduced on $\Sigma_2$ with the fluxes of Section~\ref{sec:GBprodbg}, (\ref{eq:LCS}) gives the three-dimensional Chern-Simons levels used in Section~\ref{sec:GBprodspecial}. The gravitino is charged only under $V_\mu$, so that the background gauge field $A_R$ of a trial R-symmetry, normalized such that the gravitino has charge one, corresponds to $V_\mu=-2A_{R\mu}$ and $C_\mu=\varepsilon A_{R\mu}$, where $\varepsilon$ parametrizes the mixing with the flavor symmetry. Anomaly inflow \cite{Tachikawa:2005tq,Hristov:2025ygn} then gives
\begin{equation}
{\rm Tr}R^3=-\frac{8\pi^2\varepsilon}{\kappa^2}\Big(3\varepsilon^2-\frac{3\sqrt2}{g}\,\varepsilon+\alpha+\beta+6\zeta\Big)\,,\qquad
{\rm Tr}R=-\frac{24\pi^2\alpha\,\varepsilon}{\kappa^2}\,,
\label{eq:TrR}
\end{equation}
where the overall sign, which is common to both, is fixed by requiring that $a$ reduce to its two-derivative value $a=c=\pi^2\ell^3/\kappa^2$ for $\alpha=\beta=\zeta=0$. With these anomalies, we find the following.

First, the trial function $a(\varepsilon)=\tfrac3{32}\big(3{\rm Tr}R^3-{\rm Tr}R\big)$ does not depend on $\alpha$, since the contribution of the mixed Chern-Simons term to ${\rm Tr}R$ cancels that of $\alpha$ to ${\rm Tr}R^3$, and it coincides with (\ref{eq:atrial}) for $\varepsilon=2\sqrt2y/(3g)$, for all values of the couplings. Thus $y=L^{-1}$ is proportional to the mixing parameter, and the vacuum equation (\ref{eq:vacuum}) is $a$-maximization. In the presence of the $C^2+\tfrac16R^2$ invariant, this was first shown in \cite{Hanaki:2006pj}. At the maximum $\varepsilon_*=2\sqrt2/(3gL_0)$, the anomalies (\ref{eq:TrR}) give $c=\tfrac1{32}\big(9{\rm Tr}R^3-5{\rm Tr}R\big)$ in agreement with (\ref{eq:acgen}).

Second, the superconformal R-symmetry, $C_\mu/V_\mu=-\sqrt2/(3gL_0)$, is the direction of the graviphoton $v_\mu$ in (\ref{eq:uvgen}), on which $u_\mu=0$. Third, the second derivative of $a(\varepsilon)$ at $\varepsilon_*$ equals $-\tfrac9{16}$ times the two-point coefficient of the flavor current obtained from the kinetic terms (\ref{eq:uvgen}), normalized such that the two-point coefficient of the R-current is $\tfrac{16}3c$, as required by \cite{Intriligator:2003jj}. For the flavor symmetry whose mixing is parametrized by $\varepsilon$, that is, for the background $C_\mu=A_{F\mu}$ and $V_\mu=0$, this coefficient is $C_F=24\pi^2\ell\,(2-L_0)/\kappa^2$, which has the dimension of an inverse length squared, since $\varepsilon$ has the dimension of a length.

Finally, on the supersymmetric AdS$_3\times\Sigma_2$ backgrounds of the Gauss-Bonnet extension, the central charges of the dual two-dimensional theory are reproduced for all values of the fluxes by $c$-extremization \cite{Benini:2013cda} with the anomalies obtained from (\ref{eq:LCS}), see Section~\ref{sec:GBprodspecial}.

This is an explicit instance of the general analysis of \cite{Hristov:2025ygn}, in which the trial function is the action evaluated on supersymmetric configurations with the scalars of the vector multiplets left free, and in which the Weyl tensor squared invariant drops out in the same way. Here the trial function is exact in the couplings and has two extrema that merge at $\tau=2$, since the model contains a physical vector multiplet; in minimal gauged supergravity, the scalars are fixed completely, and $a$ is linear in the couplings.

\subsubsection{The special points as zeros of \texorpdfstring{$c$}{c} and \texorpdfstring{$C_F$}{C\_F}}

The special points of Sections~\ref{sec:R2}--\ref{sec:LOG} at which the quadratic action degenerates, marked by filled circles in Figure~\ref{fig:phases}, can now be characterized as follows. Since the kinetic terms of $\mathcal G$ are proportional to $c$ and those of $\mathcal V$ to $C_F$, the points at which these kinetic terms vanish are zeros of $c$ or $C_F$. The coefficient $c$ vanishes at $\alpha g^2=-2$ for the $C^2+\tfrac16R^2$ extension and at $\tau=\tfrac32$ for the LOG extension, which are their critical points, as in critical gravity \cite{Lu:2011zk,Deser:2011xc}, and at $\tau=\tfrac32$ for the R$^2$ extension, where the kinetic terms of $\mathcal G$ vanish. For the R$^2$ and LOG extensions, $a$ vanishes at the same point, since $a=c$. 

The coefficient $C_F$ vanishes at $\tau=2$, where the two supersymmetric vacua merge. For the R$^2$ extension, this is where $B=0$ and the long vector multiplet, at the threshold for shortening, merges with the massless vector multiplet into a logarithmic pair, see Section~\ref{sec:R2mult}. This is parallel to $c=0$, where the long spin-2 multiplet merges with the graviton multiplet at the critical points, and unlike the R$^2$ point $\tau=\tfrac32$, where the kinetic terms of $\mathcal G$ vanish without logarithmic modes. For the LOG extension, $C_F=0$ is where the first term of (\ref{eq:LOGundivided}) vanishes, and for the Gauss-Bonnet extension, where the entire quadratic action vanishes, since for the latter $c$ and $C_F$ are both proportional to $2-L_0$. Since $C_F$ is proportional to the second derivative of (\ref{eq:atrial}), the merger of the vacua is the point at which the maximum and the minimum of $\hat a$ coincide, where $\hat a$ has a stationary inflection point, and the behavior of the quadratic action there reflects the vanishing of the flavor two-point function. The vanishing of $a$ alone, in contrast, has no effect: for the Gauss-Bonnet extension, $a=0$ at $\tau=\tfrac32$, where the spectrum is regular. 

The remaining special values are of a different kind: at $\tau=\tfrac{114}{121}$ for the R$^2$ extension, a scalar becomes marginal, and at $\alpha g^2=6$ for the $C^2+\tfrac16R^2$ extension, where $\rho=0$, the masses of $W_\mu^+$ and $W_\mu^-$, of $\psi_\mu^+$ and $\psi_\mu^-$, and of $\chi_+$ and $\chi_-$ coincide, see Section~\ref{sec:Wmult}. There, $W_\mu^\pm$ become massless, although $c$ and $C_F$ do not vanish.

\section{Gauss-Bonnet Extension on Supersymmetric AdS\texorpdfstring{$_3\times\Sigma_2$}{3 x Sigma2}}
\label{sec:GBprod}

In Section~\ref{sec:GB}, we have seen that around the supersymmetric AdS$_5$ vacuum the Gauss-Bonnet extension is a rescaled copy of the two-derivative theory. This simplicity is partly due to the vacuum itself: all vector and tensor fields vanish, and the Weyl tensor, the auxiliary fields of the standard Weyl multiplet and the $SU(2)$ field strength vanish as well, so that most of the couplings of (\ref{eq:GB}) do not contribute to the quadratic action. In this section, we study the Gauss-Bonnet extension around supersymmetric backgrounds of the form AdS$_3\times\Sigma_2$, where $\Sigma_2=S^2$, $T^2$ or $H^2/\Gamma$, supported by the fluxes of $C_\mu$ and $V_\mu$ through $\Sigma_2$ and by a three-form flux along AdS$_3$. These backgrounds preserve four real supersymmetries, and they are candidate near-horizon geometries of supersymmetric black strings \cite{Maldacena:2000mw,Benini:2013cda}, although a flow from the AdS$_5$ vacuum to them has not been constructed in this theory. On such backgrounds, all the terms of (\ref{eq:GB}) contribute, including the Chern-Simons terms, and as we shall see, the spectrum of the Gauss-Bonnet extension is very different from that of the two-derivative theory: at a generic Kaluza-Klein level it contains ten long multiplets of the superalgebra $su(1,1|1)$, among them a spin-2 multiplet that originates from a gravitational Chern-Simons term, as in topologically massive gravity \cite{Deser:1981wh,Deser:1982vy}, and, if all modes are retained, it contains negative-norm states for generic nonzero values of the coupling. Moreover, at large levels, towers of states accumulate at a finite energy, so that the spectrum on a compact $\Sigma_2$ has no finite density of states, and for $kL<2$ and $k(L-1)<2$, the tower of spin-2 multiplets has positive norm. Throughout this section, we use the notation $k\equiv\lambda g^2$. The family of backgrounds and its special loci are summarized in Figure~\ref{fig:l0} and Table~\ref{tab:loci}, the spectrum at a generic level in Table~\ref{tab:generic}, and the spectrum at the special levels, together with the signs of the norms, in Table~\ref{tab:light}.

\subsection{Supersymmetric backgrounds}
\label{sec:GBprodbg}

We consider configurations of the form
\begin{align}
&ds^2=ds^2({\rm AdS}_3)+ds^2(\Sigma_2)\,,\qquad
\bar R_{\alpha\beta}=-2\ell^{-2}\bar g_{\alpha\beta}\,,\qquad
\bar R_{mn}=v\,\bar g_{mn}\,,\nonumber\\*
&G_{34}=p\,,\qquad H_{34}=q\,,\qquad \cV_{34}=r\,,\qquad L=L_0\,,\qquad N=N_0\,,\qquad E^a=Z_\mu=0\,,
\label{eq:prodbg}
\end{align}
where $\alpha,\beta$ and $m,n$ denote the directions of AdS$_3$ and $\Sigma_2$, respectively, $\ell$ is the radius of AdS$_3$, $v$ is the Gaussian curvature of $\Sigma_2$, and the indices $3,4$ refer to an oriented orthonormal frame on $\Sigma_2$. The constants $p$ and $r$ are the fluxes of $G_{\mu\nu}$ and of the $U(1)$ R-symmetry field strength $\cV_{\mu\nu}=2\partial_{[\mu}V_{\nu]}$ through $\Sigma_2$, and $q$ is the component of the dual field strength $H_{ab}$ along $\Sigma_2$, which corresponds to a three-form flux along AdS$_3$. The three-form flux can be carried either by $B_{\mu\nu}$ or by $E_{\mu\nu\rho}$, see (\ref{eq:H3}), and $E^a=0$ holds in both cases. In this section, we write $L$ and $N$ for the background values $L_0$ and $N_0$. Depending on the sign of $v$, the internal space is a sphere, a torus or a compact quotient $H^2/\Gamma$ of the hyperbolic plane, and as we shall see, all three cases belong to a single family of backgrounds.

\begin{figure}[!htb]
\centering
\includegraphics[width=0.8\textwidth]{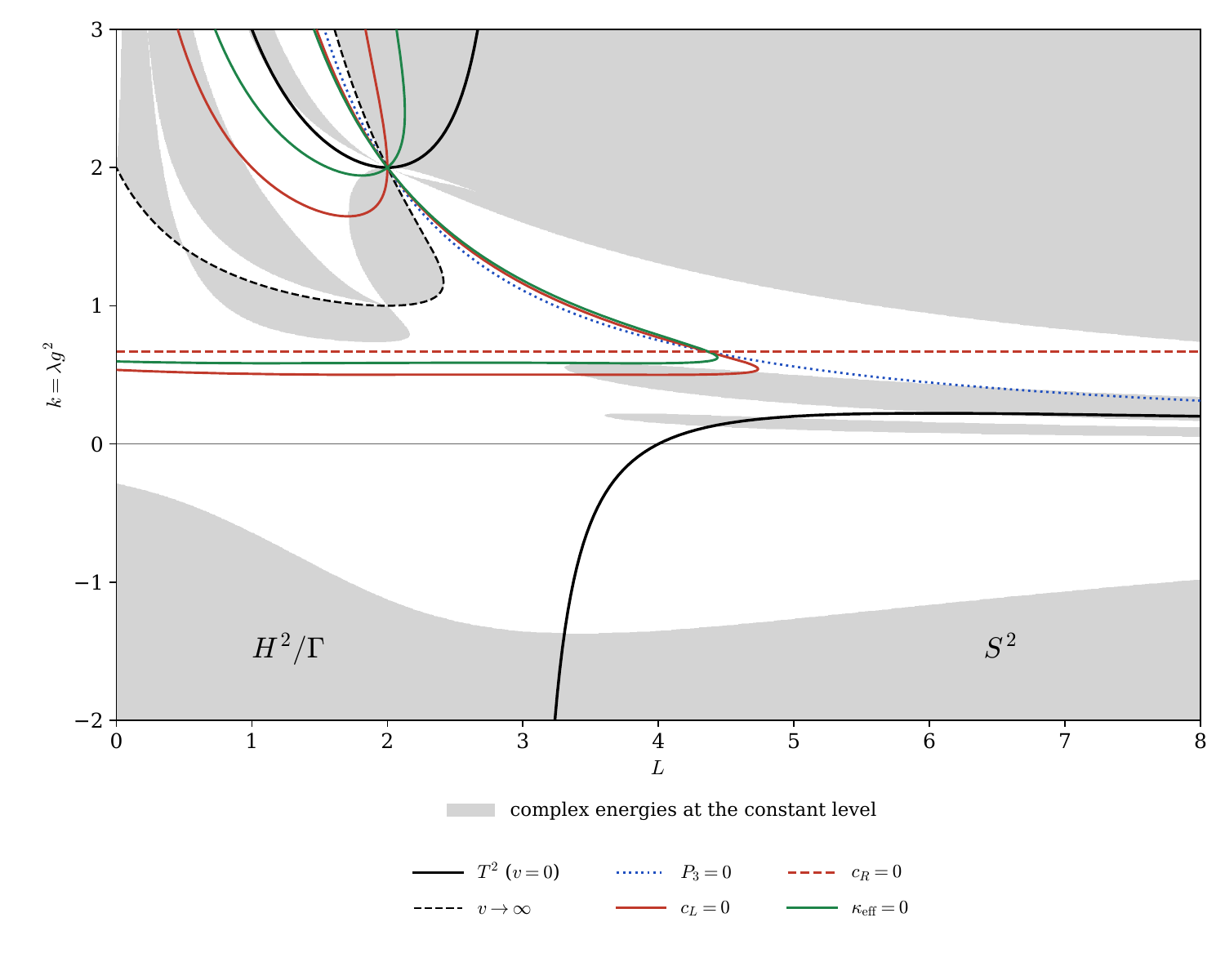}
\caption{The supersymmetric AdS$_3\times\Sigma_2$ backgrounds (\ref{eq:prodsol}), with the upper signs in (\ref{eq:prodsusy}), in the $(L,k)$ plane. The curves are the loci of Table~\ref{tab:loci}; both values of the chirality $\chi$ occur, separated by the curve $P_3=0$. In the shaded region, the constant level contains a pair of $(1,0)$ multiplets with complex energies.}
\label{fig:l0}
\end{figure}

On (\ref{eq:prodbg}), the supersymmetry transformation rules (\ref{eq:susy}) reduce to
\begin{align}
\delta\psi_\alpha^i={}&\bar\nabla_\alpha\epsilon^i-\rmi q\,\gamma_\alpha\gamma^{34}\epsilon^i-\frac{gL}{2\sqrt2}\,\gamma_\alpha(\mathbb J\epsilon)^i\,,\nonumber\\*
\delta\psi_m^i={}&\partial_m\epsilon^i+\tfrac12\omega_m{}^{34}\gamma^{34}\epsilon^i-\tfrac12\rmi V_m(\mathbb J\epsilon)^i-\tfrac12\rmi p\,\gamma_m\gamma^{34}\epsilon^i-\frac{gL}{2\sqrt2}\,\gamma_m(\mathbb J\epsilon)^i\,,\nonumber\\*
\delta\varphi^i={}&\tfrac12\Big(N+\tfrac32gL^2\Big)\epsilon^i+\frac{\rmi L(p+4q)}{2\sqrt2}\,\gamma^{34}(\mathbb J\epsilon)^i\,,
\label{eq:prodKilling}
\end{align}
with $\mathbb J$ defined in (\ref{eq:Jdef}), where we used $\gamma_\alpha{}^{34}=\gamma_\alpha\gamma^{34}$, $\gamma_m{}^{34}=0$ and $G_{ma}\gamma^a=p\,\gamma_m\gamma^{34}$. They are solved by spinors that satisfy the projection condition $\rmi\gamma^{34}\mathbb J\epsilon=\mp\epsilon$, on which $\gamma^{34}\epsilon=\pm\rmi\,\mathbb J\epsilon$, provided that
\begin{equation}
p=\pm\frac{gL}{\sqrt2}\,,\qquad
r=\pm v\,,\qquad
\ell^{-2}=4\Big(q\mp\frac{gL}{2\sqrt2}\Big)^2\,,\qquad
N=-\tfrac32gL^2\pm\frac{L(p+4q)}{\sqrt2}\,.
\label{eq:prodsusy}
\end{equation}
The first two conditions are the vanishing of the last four terms of $\delta\psi_m^i$ in (\ref{eq:prodKilling}): the two terms proportional to $\gamma_m\mathbb J\epsilon$ cancel each other for $p=\pm gL/\sqrt2$, and the R-symmetry gauge field cancels the spin connection of $\Sigma_2$ for $V_m=\pm\omega_m{}^{34}$, which is the topological twist of \cite{Maldacena:2000mw} and gives $r=\pm v$. The third condition is the integrability condition of the Killing spinor equation $\bar\nabla_\alpha\epsilon=\mp\big(q\mp gL/(2\sqrt2)\big)\gamma_\alpha\mathbb J\epsilon$ on AdS$_3$ that remains in $\delta\psi_\alpha^i$, and the last one is $\delta\varphi^i=0$. The Killing spinors are then constant on $\Sigma_2$, and they comprise four real supercharges. As in Section~\ref{sec:vacuum}, the conditions (\ref{eq:prodsusy}) follow from supersymmetry alone and hold for all values of the couplings. In the following, we take the upper signs. Substituting (\ref{eq:prodbg}) and (\ref{eq:prodsusy}) into the field equations of $\cL_{LR}+\lambda\cL_{\rm GB}$, we find that all field equations are satisfied provided that
\begin{equation}
v=\frac{2g^2L\big(2L-8+3kL-kL^2\big)}{P_4}\,,\quad
\ell_s=\frac{\sqrt2P_2}{gP_3}\,,\quad
q=\frac{g(L-2)}{\sqrt2P_2}\,,\quad
N=-\frac{gLP_3}{P_2}\,,
\label{eq:prodsol}
\end{equation}
where $\ell=|\ell_s|$, and we introduced the polynomials
\begin{align}
&P_1=2-kL\,,\qquad
P_2=4-kL\,,\qquad
P_3=2L+4-kL^2\,,\nonumber\\*
&P_4=8-4k-4kL+k^2L^2\,,\qquad
P_5=8+4k-8kL+k^2L^2\,.
\label{eq:Pdef}
\end{align}
The polynomials $P_1$ and $P_5$ will appear in the spectrum below. Thus, for each value of $k$, there is a one-parameter family of supersymmetric backgrounds labeled by $L$, and $\Sigma_2$ is a sphere for $v>0$, a torus for $v=0$ and a hyperbolic surface for $v<0$. In the two-derivative theory, (\ref{eq:prodsol}) reduces to
\begin{equation}
v=\tfrac12g^2L(L-4)\,,\qquad
\ell=\frac{2\sqrt2}{g(L+2)}\,,\qquad
q=\frac{g(L-2)}{4\sqrt2}\,,\qquad
N=-\tfrac12gL(L+2)\,,
\end{equation}
so that $\Sigma_2=H^2/\Gamma$ for $L<4$, $T^2$ for $L=4$ and $S^2$ for $L>4$, and $N$ takes the value that follows from its two-derivative field equation. For $k\neq0$, the torus is realized on the curve
\begin{equation}
k=\frac{2(L-4)}{L(L-3)}\,,
\label{eq:T2curve}
\end{equation}
and both the sphere and the hyperbolic surfaces occur for positive and for negative values of $k$, see Figure~\ref{fig:l0}.

The sign of $\ell_s$, $\chi\equiv{\rm sgn}(P_3/P_2)$, determines the chirality of the preserved supersymmetries. The isometry superalgebra of the background is $su(1,1|1)\oplus sl(2,\mathbb R)$, extended by the $U(1)$ symmetries, and for $\chi=+1$, which includes the two-derivative theory, the supercharges belong to the left-moving factor $su(1,1|1)$. We give our results for $\chi=+1$. For $\chi=-1$, the polynomials that determine the spectrum below are replaced by their images under $D\to2-D$, which exchanges the left- and right-moving weights. The background (\ref{eq:prodsol}) degenerates where $P_2$, $P_3$ or $P_4$ vanishes. Away from the point $(L,k)=(2,2)$, where they vanish simultaneously and which we discuss at the end of this subsection, for $P_3=0$ the radius of AdS$_3$ diverges, and the background becomes Minkowski$_3\times H^2$, for $P_4=0$ the curvature of $\Sigma_2$ diverges, and at $P_2=0$ the solution (\ref{eq:prodsol}) is singular. The background is regular where $P_1$ or $P_5$ vanishes. On these two loci, however, some of the ranks of the quadratic action change. For $P_1=0$, the field $N$ enters the quadratic action as a Lagrange multiplier, and for $P_5=0$, the quadratic action of the modes of boost weight zero, which we define in Section~\ref{sec:GBprodfluc}, acquires additional null directions, both for the neutral fields and for $z_\mu$. We exclude the loci $P_i=0$, $i=1,\dots,5$, from the analysis below. Table~\ref{tab:loci} collects these loci together with the other special curves of the $(L,k)$ plane that appear in the course of this section.

\begin{table}[!htb]
\centering
\small
\renewcommand{\arraystretch}{1.3}
\relpenalty=10000 \binoppenalty=10000
\begin{tabular}{|l|l|P{6.95cm}|l|}
\hline
Locus & Equation & Meaning & Figure~\ref{fig:l0}\\
\hline\hline
$T^2$ & (\ref{eq:T2curve}) & $v=0$; separates the spheres ($v>0$) from the hyperbolic surfaces ($v<0$) & solid black\\
$P_4=0$ & $8-4k-4kL+k^2L^2=0$ & curvature of $\Sigma_2$ diverges, for $L\neq2$ & dashed black\\
$P_3=0$ & $2L+4-kL^2=0$ & $\ell\to\infty$, background becomes Minkowski$_3\times H^2$, for $L\neq2$; $\chi$ changes sign & dotted blue\\
$P_2=0$ & $kL=4$ & solution (\ref{eq:prodsol}) singular & not shown\\
$L=k=2$ & $P_2=P_3=P_4=P_5=0$ & (\ref{eq:prodsol}) is singular; one-parameter family of backgrounds labeled by $q$, excluded from the analysis & not shown\\
$P_1=0$ & $kL=2$ & $N$ enters the quadratic action as a Lagrange multiplier; the flat multiplets of Section~\ref{sec:GBprodgeneric} reach $\Delta=1$ & not shown\\
$P_5=0$ & $8+4k-8kL+k^2L^2=0$ & additional null directions in the sectors of boost weight zero; (\ref{eq:quintic}) has a double root at $D=0$ at every level & not shown\\
$c_L=0$ & $X=0$, (\ref{eq:Xpoly}) & chiral point; the massive graviton and its partner degenerate with $\{J_R,T\}$ & solid red\\
$c_R=0$ & $k=\tfrac23$ & chiral point; $c$-extremization does not fix the R-symmetry & dashed red\\
$\kappa_{\rm eff}=0$ & $Y=0$, (\ref{eq:Ypoly}) & coefficient of the Einstein term in the reduction to AdS$_3$ vanishes & green\\
${\rm Disc}\,R=0$ & Appendix~\ref{app:quintic} & boundary of the region in which two $(1,0)$ multiplets of the constant level, roots of $R(D)$, have complex energies & shaded\\
\hline
\end{tabular}
\vspace{1.5mm}
\caption{The special loci of the family (\ref{eq:prodsol}) of supersymmetric AdS$_3\times\Sigma_2$ backgrounds in the $(L,k)$ plane, where the polynomials $P_i$ are defined in (\ref{eq:Pdef}). The last column gives the corresponding curve in Figure~\ref{fig:l0}.}
\label{tab:loci}
\end{table}

The parametrization (\ref{eq:prodsol}) is singular at $(L,k)=(2,2)$, where $P_2$, $P_3$, $P_4$ and $P_5$ vanish simultaneously, and its limit there depends on the direction of approach. At this point, the conditions (\ref{eq:prodsusy}), with the upper signs, and the field equations, before any division by the $P_i$, are solved by the one-parameter family
\begin{equation*}
p=\sqrt2\,g\,,\qquad r=v=\frac{2g^2\big(g+2\sqrt2\,q\big)}{g-2\sqrt2\,q}\,,\qquad \ell^{-2}=4\Big(q-\frac{g}{\sqrt2}\Big)^2\,,\qquad N=-4g+4\sqrt2\,q\,,
\end{equation*}
with $q\neq g/(2\sqrt2)$ and $q\neq g/\sqrt2$, in which $\Sigma_2$ is a sphere for $|q|<g/(2\sqrt2)$, a torus with $\ell=\sqrt2/(3g)$ for $q=-g/(2\sqrt2)$, and a hyperbolic surface otherwise. This family is excluded from the spectral analysis below, and we have not computed its spectrum.
\FloatBarrier

\subsection{Linearized fluctuations and supermultiplets}
\label{sec:GBprodfluc}

We parametrize the fluctuations around (\ref{eq:prodbg}) as in (\ref{eq:fluctuations}), and we derive the quadratic action from the complete action $\cL_{LR}+\lambda\cL_{\rm GB}$, keeping all fields off-shell and without expanding in $k$. On $\Sigma_2$, the neutral fluctuations are expanded in the eigenfunctions $\mathcal Y$ of the Laplacian, $-\nabla^2_\Sigma\mathcal Y=\lambda_\Sigma\mathcal Y$, together with the vector and tensor harmonics built from them, and we label a Kaluza-Klein level by
\begin{equation}
M^2\equiv\lambda_\Sigma\ell^2=\left\{\begin{array}{ll}
j(j+1)\,v\ell^2\,, & S^2\,,\\[2pt]
|\vec q\,|^2\ell^2\,, & T^2\,,\\[2pt]
\big(\tfrac14+\nu_H^2\big)|v|\ell^2\,, & H^2\,,
\end{array}\right.
\label{eq:levels}
\end{equation}
where $\vec q$ is a momentum on $T^2$, and for $H^2$ we have written the eigenvalues of the principal series, $\nu_H\in\mathbb R$. A compact quotient $H^2/\Gamma$ can also have finitely many eigenvalues $0<\lambda_\Sigma<\tfrac14|v|$, for which $\nu_H$ is imaginary. The results below depend on the level only through $M^2$, and they apply to all eigenvalues of the quotient, including these. The charged field $z_\mu$ couples to the R-symmetry flux, and its harmonics are sections of a line bundle on $\Sigma_2$. We call a level generic if all scalar, vector and tensor harmonics exist at that level. This excludes the constant level of $\Sigma_2$, the level $j=1$ of $S^2$, the harmonic one-forms of $T^2$, and the harmonic one-forms and holomorphic quadratic differentials of $H^2/\Gamma$, which we discuss separately in Section~\ref{sec:GBprodspecial}.

Instead of deriving the wave operators of each sector, we determine the spectrum from the quadratic action evaluated on modes of fixed scaling in AdS$_3$. In Poincar\'e coordinates, a field dual to a primary operator with weights $(h,\bar h)$, that is, with energy $\Delta=h+\bar h$ and spin $s=h-\bar h$, has a solution of the linearized field equations that is independent of the boundary coordinates and whose frame components behave as $z^\Delta$, with a polarization of boost weight $s$. We therefore evaluate the quadratic action on fluctuations whose frame components behave as $z^D$, multiplied by a harmonic on $\Sigma_2$, paired with fluctuations that behave as $z^{2-D}$; the exponent $D$ is unrelated to the auxiliary field $D$ of Section~\ref{sec:theory}, which does not appear in this section. Since the background has constant frame components, total derivatives do not contribute, and in each sector of boost weight $n$ the quadratic action becomes
\begin{equation}
S^{(2)}_n=\mathbf b^T K_n(D)\,\mathbf a\,,\qquad
(s,\Delta)=\begin{cases}(n,D)\,,&{\rm Re}\,D>1\,,\\ (-n,2-D)\,,&{\rm Re}\,D<1\,,\end{cases}
\label{eq:Kform}
\end{equation}
where $\mathbf a$ and $\mathbf b$ are the constant polarizations and $K_n(D)$ is a polynomial matrix, and the second relation gives the spin and the energy of the mode associated with a root $D$ of $K_n(D)$, as we now explain. The linearized gauge symmetries, which are the analogs of (\ref{eq:lingauge}) on this background, give rise to null directions of $K_n(D)$ at every value of $D$. Their number does not depend on $D$, while the directions themselves do, and the physical modes are the roots of the greatest common divisor of the maximal minors of $K_n(D)$, at which the rank of $K_n(D)$ drops below its generic value, and the two cases in (\ref{eq:Kform}) correspond to the two scalings $z^D$ and $z^{2-D}$ of a mode of energy $\Delta$.

The sign of the norm of each mode is given by the sign of the residue of a gauge invariant two-point function at the corresponding root, normalized such that all modes of the two-derivative theory have positive norm. We illustrate this in Section~\ref{sec:GBprodnorms} for the spin-2 sector at the constant level. These signs refer to the primary of each mode. When one of its weights is negative, the representation also contains descendants of the opposite norm, unless they are removed by the boundary conditions, which we do not discuss here. All quadratic actions, the resulting polynomials and the norms are given in \cite{repo}.

For $\chi=+1$, the long multiplets of $su(1,1|1)$ have the weights $(h,\bar h)$, $2\times(h+\tfrac12,\bar h)$ and $(h+1,\bar h)$. If the lowest component is a boson, a long multiplet contains two bosons with $(s,\Delta)$ and $(s+1,\Delta+1)$, which have the same $U(1)$ charges, and two fermions with $(s+\tfrac12,\Delta+\tfrac12)$. If the lowest component is a fermion, the two bosons have the same $(s,\Delta)$ and their R-charges differ by two units, in the normalization in which the supercharges have charges $\pm1$, so that one of them is neutral and the other one is a mode of the charged field $z_\mu$. In terms of the roots, the two bosons of a multiplet with a bosonic lowest component correspond to roots $D$ and $D+1$ in the sectors $n$ and $n+1$. When these two roots lie on opposite sides of ${\rm Re}\,D=1$, the multiplet requires the alternative quantization of one of its members, and we include such multiplets below. The bosonic spectrum of the Gauss-Bonnet extension organizes into such multiplets at every level, and the relations between the norms of the members of a multiplet that follow from the superalgebra are satisfied in all cases. This requires the terms of (\ref{eq:GB}) that contain $E_a$, which arise from the gauged Bianchi identity (\ref{eq:Bianchi}), see footnote~\ref{fn:Eterms}. They vanish on the backgrounds (\ref{eq:prodbg}), but in the presence of the fluxes they couple the fluctuation of $E_a$ to those of $G_{\mu\nu}$ and $H_{\mu\nu\rho}$. If they are omitted, then at $(L,k)=(10,\tfrac1{10})$ and the level $j=2$ of $S^2$, for example, the roots $D\simeq-2.108$, $3.550$ and $10.061$ of the spin-2 sector are unchanged, but the sector $n=1$ no longer contains their partners at $D-1$ and has a pair of complex roots $D\simeq2.643\pm1.866\,\rmi$ instead, and none of the multiplets with a bosonic lowest component is formed. We have also computed the bosonic spectra of the $C^2+\tfrac16R^2$, LOG and R$^2$ extensions on their own supersymmetric AdS$_3\times H^2$ backgrounds, which we have determined numerically, at two points for each model, and they organize into $su(1,1|1)$ multiplets in the same way. This tests the terms of (\ref{eq:LC2}), (\ref{eq:Llog}) and (\ref{eq:LR2}) that depend on the fluxes and do not contribute to the quadratic action around the AdS$_5$ vacuum.

\subsection{The spectrum at generic levels}
\label{sec:GBprodgeneric}

At a generic level, the bosonic modes form ten long multiplets of $su(1,1|1)$: three spin-2 multiplets, whose bosons have spins $1$ and $2$, five multiplets whose bosons have spins $0$ and $1$, which we call $(1,0)$ multiplets, where the label refers to the spins of their bosons and not to conformal weights, and two multiplets whose lowest component is a fermion. The neutral bosonic states per level are thus three spin-2 states, ten vectors and five scalars, and in addition there are two modes of $z_\mu$. In the two-derivative theory, the same analysis gives six long multiplets, two spin-2 and four $(1,0)$ multiplets, and the Gauss-Bonnet invariant adds one spin-2 multiplet, one $(1,0)$ multiplet and the two multiplets with a fermionic lowest component, whose energies diverge as $k\to0$. Table~\ref{tab:generic} summarizes the generic level; the remainder of this subsection derives its entries.

\begin{table}[!htb]
\centering
\small
\renewcommand{\arraystretch}{1.3}
\relpenalty=10000 \binoppenalty=10000
\begin{tabular}{|l|c|P{3.2cm}|P{3.6cm}|P{4.2cm}|}
\hline
Multiplet & Number & Bosons & Determined by & Large levels\\
\hline\hline
spin-2 & $3$ & spins $2$ and $1$ & roots $D$ of the cubic (\ref{eq:spin2cubic}), partner $D-1$ in the sector $n=1$ & two roots $D\simeq\pm M+\tfrac12(2+\mu\ell_s-\nu_f)$, one root $\to1+\nu_f$\\
$(1,0)$ & $5$ & spins $0$ and $1$ & roots $D$ of the quintic (\ref{eq:quintic}), partner $D+1$ in the sector $n=1$ & roots $D\simeq\pm M$ and $\pm\sqrt{P_4P_5}\,M/|P_1P_2|$, one root $\to\nu_f-1$\\
flat & $2$ & neutral vector and a mode of $z_\mu$ or $\bar z_\mu$, lowest component a fermion & root $D=\nu_f$ in the sector $n=1$, independent of the level & $(s,\Delta)=(1,\nu_f)$ for $\nu_f>1$, $(-1,2-\nu_f)$ for $\nu_f<1$\\
\hline
\end{tabular}
\vspace{1.5mm}
\caption{The bosonic spectrum of the Gauss-Bonnet extension at a generic Kaluza-Klein level $M$, defined in (\ref{eq:levels}), for $\chi=+1$, which consists of three spin-2 states, ten vectors, five scalars and two modes of $z_\mu$ in ten long multiplets of $su(1,1|1)$. The flat multiplets are defined in the text below.}
\label{tab:generic}
\end{table}

The spin-2 sector at each level is described by a single cubic polynomial in the scaling exponent $D$, with the level entering through the Kaluza-Klein mass $M$ of (\ref{eq:levels}),
\begin{equation}
K_2(D)\propto D(D-2)\big(D-1-\mu\ell_s\big)-M^2\big(D-1-\nu_f\big)\,,
\label{eq:spin2cubic}
\end{equation}
where
\begin{equation}
\mu\ell_s=\frac{Y}{kP_3P_4}\,,\qquad
\nu_f=\frac{16+4k-10kL+k^2L^2}{kP_3}\,,
\label{eq:munuf}
\end{equation}
where the polynomial $Y$, of degree four in $k$ and $L$, is given in (\ref{eq:Ypoly}) of Appendix~\ref{app:quintic}.
Each root $D$ of (\ref{eq:spin2cubic}) is paired with the root $D-1$ in the sector $n=1$, and together they give the bosons of one spin-2 multiplet. At $M=0$, (\ref{eq:spin2cubic}) is the field equation of topologically massive gravity, which we discuss in Section~\ref{sec:GBprodspecial}. At generic levels, the three roots describe the two helicities of a massive Kaluza-Klein graviton and a third spin-2 mode, which arises from the gravitational Chern-Simons term $\tfrac1{16}\epsilon^{abcde}C_aR_{bc}{}^{fg}R_{defg}$ in (\ref{eq:GB}) in the presence of the flux $p$. At the two isolated points where $\mu\ell_s$ and $\nu_f$ vanish simultaneously, $(L,k)=\big(\tfrac18(25\mp\sqrt{65}),\tfrac1{49}(69\pm5\sqrt{65})\big)$, (\ref{eq:spin2cubic}) has the root $D=1$ at every level, for which the prescription (\ref{eq:Kform}) does not apply, and we exclude these points.

The scalars of the five $(1,0)$ multiplets are the roots of a single polynomial of degree five, which is irreducible at a generic level,
\begin{equation}
F(D,M^2)=D(D-2)R(D)+M^2B(D)+M^4\,\frac{P_4P_5}{P_1^2P_2^2}\Big(D-\nu_f+1\Big)\,,
\label{eq:quintic}
\end{equation}
where $R(D)$ and $B(D)$ are polynomials of degree three, whose coefficients are given in Appendix~\ref{app:quintic}. Each root $D$ of (\ref{eq:quintic}) is paired with the root $D+1$ in the sector $n=1$, and together they give the scalar and the vector of one $(1,0)$ multiplet.

The two remaining multiplets do not depend on the level. Their bosons are a neutral vector and a mode of $z_\mu$ or $\bar z_\mu$, both corresponding to the root $D=\nu_f$ in the sector $n=1$, that is, with $(s,\Delta)=(1,\nu_f)$ for $\nu_f>1$ and $(-1,2-\nu_f)$ for $\nu_f<1$, and their lowest components are fermions. The field $z_\mu$ has no Maxwell term in (\ref{eq:GB}), and its quadratic action is of first order in derivatives, with a coefficient proportional to $kP_3\big(D-\nu_f\big)$. We call these two multiplets the flat multiplets.

\begin{figure}[t]
\centering
\includegraphics[width=0.8\textwidth]{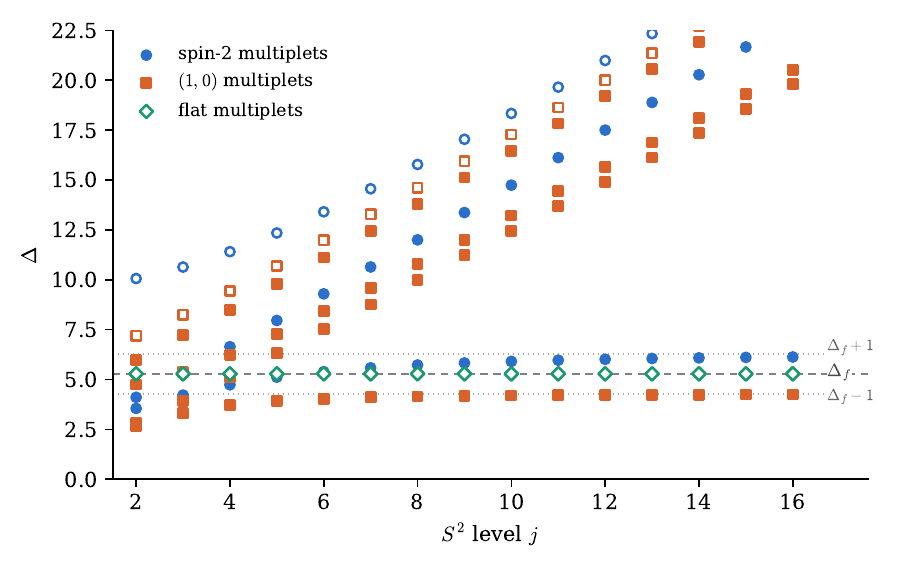}
\caption{The energies of the bosonic modes as a function of the $S^2$ level $j$, at $(L,k)=(10,\tfrac1{10})$, where $\chi=+1$ and $\Delta_f=\tfrac{37}7$. For each spin-2 multiplet the spin-2 member is shown, and for each $(1,0)$ multiplet the scalar. Filled symbols denote positive norm and open symbols negative norm. The towers grow linearly in $j$, except for one tower of spin-2 multiplets and one tower of $(1,0)$ multiplets, whose members shown here, of helicity $+2$ and $0$, approach $\Delta_f+1$ and $\Delta_f-1$ (dotted lines), so that their vector members approach $\Delta_f$ (dashed line). These two towers have positive norm, whereas the heavy spin-2 and $(1,0)$ towers and the flat multiplets have negative norm.}
\label{fig:towers}
\end{figure}

The large-level behaviour of the spectrum shows a feature that is absent in the two-derivative theory. As $M\to\infty$, two roots of (\ref{eq:spin2cubic}) grow as $D\simeq\pm M+\tfrac12(2+\mu\ell_s-\nu_f)$, while the third one tends to the finite value $1+\nu_f$. Similarly, four roots of (\ref{eq:quintic}) grow as $D\simeq\pm M$ and $D\simeq\pm\sqrt{P_4P_5}\,M/|P_1P_2|$, while the fifth one tends to $\nu_f-1$. Therefore, at every level, the vector members of one spin-2 multiplet and of one $(1,0)$ multiplet lie at, or approach, the energy of the flat multiplets, $\Delta_f=1+|\nu_f-1|$. Since the levels of $S^2$ have degeneracy $2j+1$, and those of $T^2$ and $H^2/\Gamma$ are discrete, the spectrum on a compact $\Sigma_2$ thus contains infinitely many states in a bounded range of energies. We illustrate this in Figure~\ref{fig:towers}, and we discuss the norms of these states in Section~\ref{sec:GBprodnorms}. Furthermore, where $P_4P_5<0$, two of the $(1,0)$ multiplets have complex energies at large levels, whose imaginary parts grow linearly in $M$. The case $L=2$, in which the three-form flux vanishes and $P_4=P_5$, is special. There, the leading order does not decide, and the next order gives $D\simeq\pm M+b\sqrt M$ with $b^2=\pm k/\big(4(k-1)\big)$, where the signs are correlated, so that for all $k\neq0,1$ one pair of $(1,0)$ multiplets has complex energies at every large level. These asymptotics assume $P_5\neq0$. On the locus $P_5=0$, which we have excluded in Section~\ref{sec:GBprodbg}, the term of order $M^4$ in (\ref{eq:quintic}) vanishes, and $F$ has a double root at $D=0$ at every level.

\subsection{Special levels}
\label{sec:GBprodspecial}

\subsubsection{The constant level}

On $S^2$ and $H^2/\Gamma$, the modes that are constant on $\Sigma_2$ give the $M\to0$ limit of the generic spectrum, except for the flat multiplets, since $z_\mu$ has no constant mode in the presence of the R-symmetry flux. At $M=0$, (\ref{eq:spin2cubic}) is the field equation of topologically massive gravity \cite{Deser:1981wh,Deser:1982vy}. Indeed, the reduction of the action to AdS$_3$ gives the Einstein-Hilbert term with the effective coefficient
\begin{equation}
\kappa_{\rm eff}=L+\lambda\big(\tfrac12v+pq-\tfrac14g^2L^2\big)=\frac{LY}{4P_2P_4}\,,
\label{eq:kappaeff}
\end{equation}
and a gravitational Chern-Simons term proportional to $\lambda p$, which arises from the mixed Chern-Simons term of (\ref{eq:GB}), so that the spin-2 sector at this level is topologically massive gravity with $\mu=2\kappa_{\rm eff}/(\lambda p)$. The roots $D=2$ and $D=0$ of (\ref{eq:spin2cubic}) give the massless graviton, that is, the stress tensors $T$ and $\bar T$ with weights $(2,0)$ and $(0,2)$, and the third root gives the massive graviton of topologically massive gravity with $\Delta=1+|\mu\ell|$ and helicity $2\,{\rm sgn}(\mu\ell_s)$.

\begingroup
\emergencystretch=1em
The central charges of the dual two-dimensional theory are obtained from the Brown-Henneaux central charge \cite{Brown:1986nw} and the gravitational Chern-Simons term \cite{Kraus:2005zm}. For $\chi=+1$, we find\par
\endgroup
\begin{equation}
c_L=\frac{12\pi A_\Sigma}{\kappa^2}\,\frac{LX\ell}{2P_2P_4}\,,\qquad
c_R=\frac{12\pi A_\Sigma}{\kappa^2}\,\frac{L(2-3k)P_2\,\ell}{P_4}\,,
\label{eq:cLcR}
\end{equation}
where $A_\Sigma$ is the area of $\Sigma_2$, and the polynomial $X$, of degree four in $k$ and $L$, is given in (\ref{eq:Xpoly}) of Appendix~\ref{app:quintic}. Here $c_L$ is the central charge of the supersymmetric side. In the two-derivative theory, $c_L=c_R=12\pi A_\Sigma L\ell/\kappa^2$. The difference $c_R-c_L=12\pi A_\Sigma\lambda p/\kappa^2$ is fixed by the mixed Chern-Simons term alone, as appropriate for a gravitational anomaly. Central charges of supersymmetric AdS$_3\times\Sigma_2$ solutions of four-derivative supergravity were obtained for general twists in the presence of vector multiplets and hypermultiplets in \cite{Baggio:2014hua}, and for minimal gauged supergravity in \cite{Bobev:2021qxx}. These central charges are those given by $c$-extremization \cite{Benini:2013cda}.

The reduction of the Chern-Simons terms (\ref{eq:LCS}) on $\Sigma_2$ gives the 't~Hooft anomalies of the dual theory as functions of the fluxes $p$ and $r$. Extremizing the trial central charge over the mixing of the two-dimensional R-symmetry with the flavor symmetry reproduces $c_L$ in (\ref{eq:cLcR}) identically in $(L,k)$, and the gravitational anomaly reproduces $c_R-c_L$, on the whole family, that is, on $S^2$, $T^2$ and $H^2/\Gamma$. At $k=\tfrac23$, where $c_R=0$, the two-dimensional 't~Hooft anomalies that involve the flavor symmetry vanish, so that the trial central charge does not depend on the mixing. Its value still reproduces $c_L$, but $c$-extremization does not determine the R-symmetry there.

The universal twist \cite{Maldacena:2000mw,Benini:2015bwz,Bobev:2017uzs} is the special case in which the fluxes of $C_\mu$ and $V_\mu$ are in the ratio of the superconformal R-symmetry of Section~\ref{sec:central}, $p/r=-\sqrt2/(3gL_0)$. In the two-derivative theory, it is $L=1$, where $\Sigma_2=H^2/\Gamma$ and $L$ takes its AdS$_5$ value. The values $c_L=\tfrac{16}3(\mathsf g-1)(5a-3c)$ and $c_R=\tfrac{32}3(\mathsf g-1)\,a$ for a surface of genus $\mathsf g$ \cite{Benini:2015bwz,Bobev:2017uzs}, with $a$ and $c$ given by (\ref{eq:ac}), assume that the two-dimensional R-symmetry does not mix with the flavor symmetry, which by $a$-maximization is the case if ${\rm Tr}F=0$, where $F$ generates the flavor symmetry. Here ${\rm Tr}F\neq0$, and $c$-extremization, as in \cite{Benini:2015bwz}, gives
\begin{equation}
c_L=\tfrac{16}3(\mathsf g-1)(5a-3c)+\delta c\,,\qquad
c_R=\tfrac{32}3(\mathsf g-1)\,a+\delta c\,,
\label{eq:UTc}
\end{equation}
with
\begin{equation}
\delta c=-3(\mathsf g-1)\frac{\big({\rm Tr}R^2F\big)^2}{{\rm Tr}RF^2}=\frac{8(\mathsf g-1)(c-a)^2}{3c}\,.
\label{eq:deltac}
\end{equation}
Since $c-a$ is of order $k$, the values of \cite{Benini:2015bwz,Bobev:2017uzs} hold at first order in $k$. We have also verified that the Chern-Simons level of the R-symmetry current obtained from the reduction satisfies $c_L=3k_R$, as required by the $\mathcal N=2$ superconformal algebra.

The chiral points of topologically massive gravity \cite{Li:2008dq} are therefore $c_R=0$, which happens for $k=\tfrac23$ for every $L$, and $c_L=0$, which happens on the curve $X=0$. At the latter, the massive graviton and its multiplet partner become degenerate with the stress tensor multiplet $\{J_R,T\}$, which is a supersymmetric version of the degeneration at the chiral point of topologically massive gravity \cite{Grumiller:2008qz}.

At $M=0$, the polynomial (\ref{eq:quintic}) reduces to $D(D-2)R(D)$, and the roots $D=0,2$ are not physical at this level. The cubic $R(D)$ gives three $(1,0)$ multiplets. Its three roots are real outside the shaded region of Figure~\ref{fig:l0}, whose boundary is the locus on which the discriminant of $R(D)$ vanishes; inside it, two of the roots are complex, so that the background is unstable already at the constant level. The region with real roots contains the two-derivative line $k=0$ together with a range of negative $k$ around it, and the band of spheres between the curves $v=0$ and $P_3=0$. Two of the three multiplets are continuously connected to the two multiplets of the two-derivative theory at this level, which consist of a scalar with $\Delta=x_+$ and a vector with $\Delta=1+x_+$, and of a vector with $\Delta=1-x_-$ and a scalar with $\Delta=2-x_-$, with $x_\pm=\big(L-2\pm\sqrt{9L^2-4L+4}\big)/(L+2)$. The third multiplet is new, and its energy diverges as $k\to0$. Its scalar is the fluctuation of the three-form flux $q$, which multiplies the three-dimensional curvature in the reduced action and therefore propagates, like a Brans-Dicke scalar. The constant level also contains the R-symmetry current $J_R$, which belongs to the stress tensor multiplet, and a flavor current.

\subsubsection{The sphere at \texorpdfstring{$j=1$}{j=1}}

At the level $j=1$ of $S^2$, the tensor harmonics vanish. The spectrum is that of a generic level with $M^2=2v\ell^2$, with one exception. At this value, the polynomial (\ref{eq:quintic}) has the root $D=0$ for all $(L,k)$, and the corresponding multiplet loses its scalar member, which would be a tensor harmonic. Its vector member at $D=1$ remains, and it is the Kaluza-Klein gauge field of the $SU(2)$ isometries of $S^2$. Thus the level $j=1$ consists of nine long multiplets and an $SU(2)$ current. The residue at $D=1$ shows that this current is right-moving, that is, it belongs to the non-supersymmetric side, and that its level is positive at all points we have examined, including points where $c_L<0$.

\subsubsection{The torus}

On $T^2$, the background lies on the curve (\ref{eq:T2curve}), along which $\chi=+1$, $\ell=\sqrt2/(3g)$, and all quantities above are rational functions of $L$,
\begin{equation}
\begin{aligned}
\kappa_{\rm eff}&=\frac{2L^2-9L+12}{2(L-3)}\,,&\qquad
\mu\ell&=\frac{2L^2-9L+12}{3(L-4)}\,,\\
\nu_f&=\frac{5L-12}{3(L-4)}\,,&
\frac{c_L}{c_R}&=\frac{L^2-6L+12}{L(L-3)}\,.
\end{aligned}
\end{equation}
The two-derivative point is $L=4$, and $L=3$ corresponds to $k\to\infty$. At the constant level, the massive spectrum is the $M\to0$ limit of the generic spectrum: the topologically massive multiplet, three $(1,0)$ multiplets and the two flat multiplets. The $(1,0)$ multiplets are given by the roots of the cubic $R(D)$, which on the torus reduces to (\ref{eq:T2cubic}) of Appendix~\ref{app:quintic}. The flat multiplets exist here, since there is no R-symmetry flux and $z_\mu$ has a constant mode, and their neutral members come from the harmonic one-forms of $T^2$. The light sector consists of the stress tensor multiplet $\{J_R,T\}$, the stress tensor $\bar T$, a flavor current with weights $(0,1)$, two long multiplets made of a Wilson line of $V_\mu$ with weights $(1,1)$ and a vector with weights $(2,1)$, the vectors with weights $(1,2)$ and the two moduli of the complex structure of $T^2$, the last two in short multiplets. The volume modulus and the Wilson lines of $C_\mu$ are massive. In the two-derivative theory, all these states have positive norm, and the topologically massive and the flat multiplets decouple.

\subsubsection{Hyperbolic surfaces}

On $H^2/\Gamma$ of genus $\mathsf g\geq2$, each of the $\mathsf g$ holomorphic one-forms gives two vectors with $\Delta=3$, each with the two helicities, that is, with weights $(2,1)$ and $(1,2)$, four moduli with $\Delta=2$, which are the Wilson lines of $C_\mu$ and $V_\mu$, and the two flat multiplets, whose charged members are zero modes of $z_\mu$. Each of the $3\mathsf g-3$ holomorphic quadratic differentials gives a complex structure modulus with $\Delta=2$. Remarkably, apart from those of the flat multiplets, these energies do not depend on $(L,k)$, and they coincide with those of the corresponding states on $T^2$, except for the Wilson lines of $C_\mu$, which are massive on $T^2$. The states found at the special levels in this subsection are collected in Table~\ref{tab:light}, together with the signs of their norms, which we derive in Section~\ref{sec:GBprodnorms}.

\begin{table}[p]
\centering
\renewcommand{\baselinestretch}{1.15}
\footnotesize
\renewcommand{\arraystretch}{1.0}
\relpenalty=10000 \binoppenalty=10000
\begin{tabular}{|P{2.7cm}|P{4.5cm}|P{2.9cm}|P{4.3cm}|}
\hline
Level & States & Weights or $(s,\Delta)$ & Norm of the primary\\
\hline\hline
constant level, all $\Sigma_2$ & stress tensor multiplet $\{J_R,T\}$ & $(1,0)$, $(2,0)$ & ${\rm sgn}\,c_L$ for both, since $c_L=3k_R$\\
 & stress tensor $\bar T$ & $(0,2)$ & ${\rm sgn}\,c_R$\\
 & massive graviton of topologically massive gravity & $\Delta=1+|\mu\ell|$, helicity $2\,{\rm sgn}(\mu\ell_s)$ & ${\rm sgn}\big(\kappa_{\rm eff}(1-\mu^2\ell^2)\big)$, with negative-norm descendants for $|\mu\ell|<1$\\
 & two $(1,0)$ multiplets connected to those of the two-derivative theory, roots of $R(D)$ & $\Delta=x_+$, $1+x_+$ and $1-x_-$, $2-x_-$ at $k=0$ & for $\kappa_{\rm eff}<0$, one scalar of the three $(1,0)$ multiplets negative\\
 & a new $(1,0)$ multiplet, root of $R(D)$, its scalar the fluctuation of $q$ & $\Delta\to\infty$ as $k\to0$ & its scalar negative as $k\to0$\\
 & flavor current & $(0,1)$ on $T^2$; --- on $S^2$ and $H^2/\Gamma$ & ---\\
\hline
$S^2$, $j=1$ & nine long multiplets & --- & ---\\
 & $SU(2)$ current of the isometries & $(0,1)$, right-moving & level positive at all points examined\\
\hline
$T^2$, constant level & two flat multiplets, neutral members from the harmonic one-forms & $(s,\Delta)=(1,\nu_f)$ for $\nu_f>1$, $(-1,2-\nu_f)$ for $\nu_f<1$ & charged member negative iff $kL<2$; neutral member the same norm if $h_f>1$, the opposite if $0<h_f<1$\\
 & Wilson line of $V_\mu$ and its partner vector, in two long multiplets & $(1,1)$, $(2,1)$ & vector negative iff $kL>2$\\
 & vectors & $(1,2)$ & negative iff $P_2\big(2-k(L-1)\big)<0$\\
 & two complex structure moduli, in short multiplets & $(1,1)$ & negative iff $P_2P_4<0$\\
 & volume modulus and Wilson lines of $C_\mu$ & massive & ---\\
\hline
$H^2/\Gamma$, per harmonic one-form & a vector & $(2,1)$ & negative iff $kL>2$\\
 & a vector & $(1,2)$ & negative iff $P_2\big(2-k(L-1)\big)<0$\\
 & four moduli, Wilson lines of $C_\mu$ and $V_\mu$ & $(1,1)$ & ---\\
 & two flat multiplets, charged members zero modes of $z_\mu$ & as on $T^2$ & as on $T^2$\\
\hline
$H^2/\Gamma$, per quadratic differential & a complex structure modulus & $(1,1)$ & negative iff $P_2P_4<0$\\
\hline
\end{tabular}
\vspace{1.5mm}
\caption{The spectrum of the Gauss-Bonnet extension at the special levels of the supersymmetric AdS$_3\times\Sigma_2$ backgrounds for $\chi=+1$ and $L>0$, where $x_\pm$ are defined in Section~\ref{sec:GBprodspecial} and $h_f$ is the left-moving weight of the flat multiplets. A surface of genus $\mathsf g$ has $\mathsf g$ holomorphic one-forms and $3\mathsf g-3$ quadratic differentials. The weights are $(h,\bar h)$ unless marked as $(s,\Delta)$, the norms are those of the primaries of Section~\ref{sec:GBprodnorms}, and a dash denotes an entry that is not determined in this paper.}
\label{tab:light}
\end{table}

\subsection{Norms and unitarity}
\label{sec:GBprodnorms}

In all cases we have examined, on $S^2$, $T^2$ and $H^2/\Gamma$, the spectrum contains negative-norm or complex modes. Thus, in contrast with the AdS$_5$ vacuum of Section~\ref{sec:GB}, where the Gauss-Bonnet extension is free of ghosts, the supersymmetric AdS$_3\times\Sigma_2$ backgrounds of the Gauss-Bonnet extension are generically not unitary. Away from the loci excluded in Section~\ref{sec:GBprodbg}, which include $kL=2$, the closed-form results below establish this for every $k\neq0$. On $S^2$ in the range $kL>2$, where the arguments based on the harmonic forms do not apply, the spin-2 tower of Section~\ref{sec:GBprodgeneric} has, at large levels, either negative norm or a negative weight, as we show at the end of this subsection, and this includes the curves $\kappa_{\rm eff}=0$ and $|\mu\ell|=1$. The origin of the difference is the flux: in its presence, the Chern-Simons terms of (\ref{eq:GB}) give rise to a gravitational Chern-Simons term and to first-order kinetic terms for $V_\mu$ and $Z_\mu$, and the coupling of the three-form flux to the curvature gives a propagating scalar. These statements refer to the primaries of the Kaluza-Klein modes, with all modes retained. We have not analyzed the boundary conditions, which may remove some of the negative-norm descendants, nor the global multiplicities of the charged harmonics.

The norms of the modes follow from the residues described in Section~\ref{sec:GBprodfluc}, and several of them can be given in closed form. As an example, consider the spin-2 sector at the constant level, where the quadratic action of the component $h_{++}$ of the metric fluctuation is $\mathbf b^TK_2(D)\,\mathbf a$ with a single function $K_2(D)$. Up to a positive factor, it is the sum of the Einstein term and of the gravitational Chern-Simons term,
\begin{equation*}
K_2(D)\propto\kappa_{\rm eff}\,D(D-2)\Big(1-\frac{D-1}{\mu\ell_s}\Big)\,,
\end{equation*}
and the norm of the mode at a real root $D_i$ has the sign of ${\rm sgn}(D_i-1)/K_2'(D_i)$. This is the sign of the residue of $1/K_2$ with respect to the mass variable, which depends on $D$ through $(D-1)^2$, and a root with $D_i<1$ describes the mode with $\Delta=2-D_i$. For $D=2$ and $D=0$, which correspond to the stress tensors $T$ and $\bar T$, this gives the signs of $\kappa_{\rm eff}\big(1-1/(\mu\ell_s)\big)$ and $\kappa_{\rm eff}\big(1+1/(\mu\ell_s)\big)$, which are the signs of $c_L$ and $c_R$ in (\ref{eq:cLcR}), whose ratio is $\big(1-1/(\mu\ell_s)\big)/\big(1+1/(\mu\ell_s)\big)$. In the two-derivative limit, where $\kappa_{\rm eff}=L>0$ and $|\mu\ell|\to\infty$, both are positive, as required by the normalization of Section~\ref{sec:GBprodfluc}. For the massive graviton, $D=1+\mu\ell_s$, the norm of the primary has the sign of $\kappa_{\rm eff}(1-\mu^2\ell^2)$, as in topologically massive gravity \cite{Li:2008dq}, and this mode decouples in the two-derivative limit. Thus, for $\kappa_{\rm eff}>0$ the massive graviton is a ghost if $|\mu\ell|>1$. If $|\mu\ell|<1$, its primary has positive norm, but one of its weights is negative, and the representation contains descendants of negative norm. For $\kappa_{\rm eff}<0$, one of the scalars has negative norm. Therefore, if all bulk modes and their descendants are retained, the constant level contains negative-norm states for every $k\neq0$, apart from the loci $\kappa_{\rm eff}=0$ and $|\mu\ell|=1$, and in the limit $k\to0$ the negative-norm states are the massive graviton and the scalar of the new $(1,0)$ multiplet, whose energies diverge. This is the usual picture of the higher-derivative modes as ghosts whose energies lie beyond the regime of validity of the four-derivative theory \cite{Pang:2012xs}.

At the other levels, the norms of the flat multiplets and of the modes associated with harmonic forms can also be given in closed form. For $\chi=+1$ and $L>0$, the rules are listed in the last column of Table~\ref{tab:light}. For $\chi=-1$, the rules for the two helicities of the vectors are exchanged, and both helicities have negative norm on the whole domain with $\chi=-1$. As a consequence, on $T^2$ and on $H^2/\Gamma$ every nonzero value of $k$ leads to negative-norm states, apart from the locus $kL=2$, where the flat multiplets reach $\Delta=1$: for $kL<2$ the flat multiplets have negative norm at every level, while for $kL>2$ the vectors with weights $(2,1)$ do. On the torus, for $L>3$, which includes small positive values of $k$, the light sector has positive norm and $c_L,c_R>0$, but the topologically massive multiplet and the two flat multiplets are ghosts. For $0<L<3$, the entire left-moving light sector, including the stress tensor $T$ and the R-symmetry current, has negative norm, and $c_L<0$. By (\ref{eq:T2curve}), this range corresponds to $k\geq2$, where the theory has no supersymmetric AdS$_5$ vacuum, see Section~\ref{sec:vacuumsolution}, so that these backgrounds cannot arise as near-horizon geometries of black strings in an asymptotically AdS$_5$ spacetime.

On the sphere, we have determined the norms numerically, level by level. At $(L,k)=(10,\tfrac1{10})$, the point of Figure~\ref{fig:towers}, each level contains four multiplets of negative norm, namely the heavy spin-2 and $(1,0)$ multiplets and the two flat multiplets, while the remaining six multiplets have positive norm. At the point with $k<0$ of Table~\ref{tab:GBbos}, $(L,k)=(5,-1)$, where the constant level is real, pairs of multiplets with complex energies appear from $j=1$ on.

\begin{table}[!htb]
\centering
\renewcommand{\arraystretch}{1.2}
\begin{tabular}{|c|c|c|c|}
\hline
$\Sigma_2$ & $(L,k)$ & $\chi$ & levels\\
\hline
$S^2$ & $(10,\tfrac1{10})$ & $+1$ & $j=2,\dots,16$\\
$S^2$ & $(5,-1)$ & $+1$ & $j=2,3$\\
$H^2$ & $(6,2)$ & $+1$ & $\nu_H=0,1,3$\\
$H^2$ & $(3,\tfrac65)$ & $-1$ & $\nu_H=\tfrac12,2$\\
\hline
$S^2$ & $(10,\tfrac1{10})$ & $+1$ & $j=10,30,100$\\
$S^2$ & $(12,\tfrac1{20})$ & $+1$ & $j=10,30,100$\\
$S^2$ & $(5,-1)$ & $+1$ & $j=10,30,100$\\
$H^2$ & $(3,\tfrac1{10})$ & $+1$ & $M=10,100,1000$\\
$H^2$ & $(\tfrac32,\tfrac12)$ & $+1$ & $M=10,100,1000$\\
$T^2$ & $(5,\tfrac15)$ & $+1$ & $M=10,100,1000$\\
$T^2$ & $(8,\tfrac15)$ & $+1$ & $M=10,100,1000$\\
$H^2$ & $(5,1)$ & $+1$ & $M=100,1000,10^4$\\
$H^2$ & $(3,2)$ & $+1$ & $M=100,1000,10^4$\\
$H^2$ & $(\tfrac72,1)$ & $-1$ & $M=100,1000,10^4$\\
$H^2$ & $(\tfrac74,\tfrac{12}5)$ & $-1$ & $M=100,1000,10^4$\\
$H^2$ & $(3,\tfrac12)$ & $+1$ & $M=100,1000,10^4$\\
$H^2$ & $(5,\tfrac3{10})$ & $+1$ & $M=100,1000,10^4$\\
$H^2$ & $(4,\tfrac9{20})$ & $+1$ & $M=100,1000,10^4$\\
$H^2$ & $(4,\tfrac{11}{20})$ & $+1$ & $M=100,1000,10^4$\\
\hline
\end{tabular}
\vspace{1.5mm}
\caption{The points and levels at which the norms of the bosonic modes of the Gauss-Bonnet extension were determined numerically, with $j$, $\nu_H$ and $M$ of (\ref{eq:levels}). In the upper part, the norms of all modes with real energies were computed, and in the lower part, those of the towers that approach the flat multiplets.}
\label{tab:GBbos}
\end{table}

The accumulation found in Section~\ref{sec:GBprodgeneric} does not fit the picture of heavy ghosts. At the points of the lower part of Table~\ref{tab:GBbos}, on $S^2$, $T^2$ and $H^2$, we have determined the norm of the spin-2 tower that approaches the flat multiplets numerically, and on $S^2$ and $H^2$ also that of the $(1,0)$ tower, which has the same sign at all these points; at $(L,k)=(3,\tfrac1{10})$, the positive norm of the spin-2 tower sets in at $M=17$, after a range of levels in which its two roots form a complex pair. In closed form, for $L>0$, the spin-2 tower consists, at large levels, of states of positive norm with positive weights if and only if $kL<2$ and $k(L-1)<2$. Indeed, for $\chi=+1$, $K_2(D)$ is $-\lambda p/(2\ell_s)$ times (\ref{eq:spin2cubic}), up to a positive factor, and the derivative of (\ref{eq:spin2cubic}) at the root that tends to $1+\nu_f$ is dominated by the term $-M^2$, so that for $\nu_f\neq0$ the norm of the primary has the sign of $k\nu_f$, and its weights are positive for $|\nu_f|>1$, where $\nu_f-1=2P_1P_2/(kP_3)$ and $\nu_f+1=8(P_1+k)/(kP_3)$; for $\nu_f=0$, the root tends to $1$ and the norm has the sign of $\kappa_{\rm eff}$, which is negative at $(L,k)=(3,2)$. For $\chi=-1$, the norm is negative, and for $k>0$ the two conditions reduce to $kL<2$. This includes points such as that of Figure~\ref{fig:towers}, where $c_L,c_R>0$ and $-3\leq k\leq\tfrac59$. Therefore, in this range, the quadratic spectrum on a compact $\Sigma_2$ contains infinitely many positive-norm primaries with bounded weights. It has no finite density of states, and a thermal sum over these states diverges at every temperature, if they are all admitted as physical states. This admits two readings. In terms of an effective theory, the towers bend over only at levels $M\sim\nu_f\sim1/k$, whereas the four-derivative truncation can be trusted only for energies and levels small compared to $|k|^{-1/2}$, so that the accumulation lies outside its regime of validity. Taken as a theory in its own right, however, the Gauss-Bonnet extension has no finite density of states on these backgrounds.

\subsection{Fermionic sector}
\label{sec:GBprodferm}

At every point at which we have computed it directly, the fermionic spectrum consists of exactly the twenty modes that $su(1,1|1)$ predicts from the bosonic multiplets, with no missing and no additional modes, see Table~\ref{tab:GBferm}. The fermionic spectrum follows from the bosonic one by $su(1,1|1)$. Each multiplet with a bosonic lowest component $(s,\Delta)$ contains two fermions with $(s+\tfrac12,\Delta+\tfrac12)$, and each flat multiplet contains two fermions with $(s\mp\tfrac12,\Delta\mp\tfrac12)$. At a generic level, the ten multiplets of Section~\ref{sec:GBprodgeneric} thus predict twenty fermionic modes.

We have also computed the fermionic spectrum directly, with the method of Section~\ref{sec:GBprodfluc}. The quadratic action is obtained from the complete terms of $\cL_{LR}+\lambda\cL_{\rm GB}$ that are quadratic in the fermions \cite{repo}, evaluated on (\ref{eq:prodbg}) with all supercovariant derivatives in full. The frame components of the gravitino $\psi_a^i$ and of the spinor $\varphi^i$ are expanded in modes that behave as $z^D$ in AdS$_3$, multiplied by a harmonic on $\Sigma_2$, and paired with modes that behave as $z^{2-D}$. Since the R-symmetry flux cancels the spin connection of $\Sigma_2$, the Killing spinors are constant on $\Sigma_2$, and the fermions can be compared with the bosons at the same level. The quadratic action splits into sectors labeled by the boost weight $n=\pm\tfrac12,\pm\tfrac32$ and by the eigenvalue $\pm1$ of $\mathbb J$, where the symplectic Majorana condition relates each sector to the one with the opposite labels, and in each sector it takes the form $\mathbf b^TK_n(D)\,\mathbf a$ with a polynomial matrix $K_n(D)$. The physical modes are the common roots of its maximal minors, each confirmed by a drop of the rank of $K_n(D)$, and the assignment of $(s,\Delta)$ is the same as for the bosons. The operators $K_n(D)$, their roots and the comparison with the bosonic multiplets are given in \cite{repo}.

In every sector of spin $\tfrac12$, the kernel of $K_n(D)$ is two-dimensional at generic $D$. These two null directions exist at every value of $D$ and correspond to the local supersymmetry of the background, while the kernel is larger only at the roots that give the physical modes. The two null directions depend on $D$, since the supersymmetry parameters are expanded in the same modes as the fluctuations. They exist only for the relative normalization of the fermionic and the bosonic terms of the Gauss-Bonnet invariant, and they disappear if the fermionic terms are rescaled. Since most of the two-fermion couplings of the invariant do not contribute to the quadratic action around the AdS$_5$ vacuum, this tests terms that the analysis of Section~\ref{sec:GB} does not probe.

\begin{table}[!htb]
\centering
\renewcommand{\arraystretch}{1.2}
\begin{tabular}{|c|c|c|c|c|c|}
\hline
$\Sigma_2$ & $(L,k)$ & $\chi$ & $M^2$ & fermionic modes & $\max|\Delta_{\rm direct}-\Delta_{\rm pred}|$\\
\hline
$H^2$ & $(3,2)$ & $+1$ & $0.0278$ & $20$ & $3.4\times10^{-13}$\\
$H^2$ & $(3,2)$ & $+1$ & $0.0234$ & $20$ & $1.7\times10^{-13}$\\
$H^2$ & $(5,1)$ & $+1$ & $-0.882$ & $20$ & $5.3\times10^{-12}$\\
$H^2$ & $(\tfrac72,\tfrac12)$ & $+1$ & $0.678$ & $20$ & $2.5\times10^{-10}$\\
$H^2$ & $(3,\tfrac1{10})$ & $+1$ & $0.136$ & $20$ & $4.0\times10^{-12}$\\
$H^2$ & $(3,\tfrac1{100})$ & $+1$ & $0.109$ & $20$ & $5.7\times10^{-11}$\\
$H^2$ & $(3,\tfrac1{1000})$ & $+1$ & $0.107$ & $20$ & $5.2\times10^{-10}$\\
$H^2$ & $(3,\tfrac1{10})$ & $+1$ & $1.19$ & $20$ & $5.7\times10^{-12}$\\
$H^2$ & $(\tfrac72,1)$ & $-1$ & $5.31$ & $20$ & $7.6\times10^{-12}$\\
$H^2$ & $(\tfrac74,\tfrac{12}5)$ & $-1$ & $23.8$ & $20$ & $8.2\times10^{-12}$\\
$S^2$ & $(10,\tfrac1{10})$ & $+1$ & $12.0$ & $20$ & $1.3\times10^{-10}$\\
$S^2$ & $(5,-1)$ & $+1$ & $1.35$ & $20$ & $3.4\times10^{-13}$\\
\hline
\end{tabular}
\vspace{1.5mm}
\caption{The points at which the fermionic spectrum of the Gauss-Bonnet extension was computed directly, with the level $M^2$ of (\ref{eq:levels}). At each point, the twenty predicted fermionic modes are found, and no others, with a one-to-one matching of spin, R-charge and multiplicity. The last column gives the largest difference between the direct and the predicted energies.}
\label{tab:GBferm}
\end{table}

Table~\ref{tab:GBferm} lists the points at which we have compared the direct spectrum with the prediction.\footnote{On $S^2$, the level in Table~\ref{tab:GBferm} is $j=2$. The last three points on $H^2$ have $\lambda_\Sigma=w^2(\tfrac14+1.3^2)$ with $w^2=-v$, in the principal series of $H^2$. At $(L,k)=(5,1)$, $M^2<0$ is not a level of $\Sigma_2$, and the comparison there is an algebraic check of the operators. The matching of spin, R-charge and multiplicity holds within $10^{-6}$, and the differences in the last column are computed in double precision and depend on the machine at the level of $10^{-9}$.} They include three points at a level in the principal series of $H^2$, two of which have $\chi=-1$. At all of them, the twenty predicted fermions are found, with no missing and no additional modes, including the pairs with complex energies that occur at several of these points. In the limit $k\to0^+$ at fixed $L$, twelve of the twenty modes approach the fermions of the two-derivative theory, with deviations of order $k$, while the remaining eight, which belong to the four multiplets added by the Gauss-Bonnet invariant, have ${\rm Re}\,\Delta\simeq8/\big((L+2)k\big)$, that is $1.6/k$ for the points of Table~\ref{tab:GBferm} with $L=3$ and $k=\tfrac1{10},\tfrac1{100},\tfrac1{1000}$. In the two-derivative theory, the same comparison gives the twelve fermions of the six multiplets at generic levels of $H^2$ and $S^2$, and the fourteen fermions predicted at the constant level of $T^2$ in Section~\ref{sec:GBprodspecial}, whereas at $k\neq0$ we have not analyzed $T^2$. Thus, at all the levels we have examined, every bosonic multiplet is completed by its fermions, and no further fermionic modes arise. These comparisons are local: they concern the modes at the levels examined, at regular points of the family. We have not computed the norms of the fermionic modes, the multiplicities of the harmonics on $\Sigma_2$, or the fermions on the loci excluded in Section~\ref{sec:GBprodbg}.

\FloatBarrier

\section{Conclusions}
\label{sec:conclusions}

In this paper, we have studied the spectra of off-shell gauged $D=5$, $\mathcal N=2$ supergravity extended by the $C^2+\tfrac16R^2$, LOG and Ricci scalar squared invariants around its maximally supersymmetric AdS$_5$ vacuum, and the spectrum of its Gauss-Bonnet extension around supersymmetric AdS$_3\times\Sigma_2$ backgrounds. In Section~\ref{sec:theory}, we have presented the bosonic Lagrangians of the three gauged invariants in terms of the independent fields, including the terms that arise from the gauged Bianchi identity. Throughout the paper, we have treated the couplings as finite parameters and kept all the auxiliary fields, and we have derived both the bosonic and the fermionic quadratic actions from the complete off-shell Lagrangians, which we have made available in \cite{repo}.

We have shown that the supersymmetric AdS$_5$ vacuum depends on the couplings only through $\tau=(\beta+6\zeta)g^2$, see (\ref{eq:vacuum}), since the $C^2+\tfrac16R^2$ invariant does not contribute to the vacuum equations. The spectra of the four models are given by (\ref{eq:R2spectrum}), (\ref{eq:Wspectrum}), (\ref{eq:GBresult}) and (\ref{eq:LOGspectrum}), the masses and the signs of the norms of all modes are listed in Tables~\ref{tab:R2} and \ref{tab:W}, and the ranges of the couplings in which the spectra have positive norm are summarized in Figure~\ref{fig:phases} and Table~\ref{tab:summary}. In addition to the $12+12$ states of the two-derivative theory, the Ricci scalar squared extension propagates a long vector multiplet, and it has a stable vacuum with a positive-norm spectrum for $0<\zeta g^2<\tfrac14$. The $C^2+\tfrac16R^2$ and LOG extensions propagate a long spin-2 multiplet with $32+32$ states, and there is no range of their couplings in which the spectrum is free of both ghosts and instabilities. At their critical points, $\alpha g^2=-2$ and $\tau=\tfrac32$, the graviton multiplet acquires logarithmic partners, but the components of the long multiplet that do not combine with the graviton multiplet, such as the scalar $d$, retain negative norm, so that the critical points do not remove the ghosts. The quadratic action of the Gauss-Bonnet extension, on the other hand, is a rescaled copy of that of the two-derivative theory, see (\ref{eq:GBresult}), and the extension propagates the $12+12$ states of the two-derivative theory with positive norm for all $\lambda g^2<1$. We obtained this result in both the bosonic and the fermionic sectors, and it implies that, up to an overall factor, the LOG extension is the $C^2+\tfrac16R^2$ extension with the coupling (\ref{eq:LOGmap}). In Section~\ref{sec:central}, we have shown that these results are controlled by the central charges of the dual theory. The kinetic coefficients of the massless graviton and vector multiplets are proportional to $c$ and to the flavor central charge $C_F$, respectively, the critical points are zeros of $c$, the merger of the two supersymmetric vacua at $\tau=2$ is a zero of $C_F$, and the vacuum equation is $a$-maximization for all values of the couplings. The superconformal R-symmetry mixes with the flavor symmetry by an amount that depends on the couplings, so that, beyond first order, the vector multiplet cannot be truncated as in the two-derivative theory.

In Section~\ref{sec:GBprod}, we have studied the Gauss-Bonnet extension around its supersymmetric AdS$_3\times\Sigma_2$ backgrounds, which form a one-parameter family for each value of $k=\lambda g^2$ and include $S^2$, $T^2$ and $H^2/\Gamma$. At a generic Kaluza-Klein level, the bosonic spectrum consists of three spin-2 multiplets, determined by the cubic polynomial (\ref{eq:spin2cubic}), five $(1,0)$ multiplets, determined by the quintic polynomial (\ref{eq:quintic}), and two flat multiplets whose energy does not depend on the level. The central charges (\ref{eq:cLcR}) are reproduced by $c$-extremization on the whole family, and for the universal twist, they differ from the values of \cite{Benini:2015bwz,Bobev:2017uzs} by the shift (\ref{eq:deltac}), which is of second order in $k$ and arises from the mixing of the two-dimensional R-symmetry with the flavor symmetry. Unlike around the AdS$_5$ vacuum, the spectrum is generically not unitary: closed-form arguments give negative-norm modes for every $k\neq0$ away from a few exceptional loci, and the spectrum contains negative-norm or complex modes in all cases we have examined. The reason is the flux: in its presence, the Chern-Simons terms of the invariant give rise to a gravitational Chern-Simons term and to first-order kinetic terms for the vector fields, and the coupling of the three-form flux to the curvature gives a propagating scalar. Moreover, at large levels, a tower of spin-2 multiplets and a tower of $(1,0)$ multiplets approach the energy of the flat multiplets. The limiting energies follow in closed form, so that, if all modes are retained, the spectrum on a compact $\Sigma_2$ has no finite density of states, and since the spin-2 tower has positive norm for $kL<2$ and $k(L-1)<2$, in this range it contains infinitely many positive-norm states with bounded energies. We have also computed the fermionic spectrum directly at the points of Table~\ref{tab:GBferm} and found, at each of them, exactly the twenty fermions predicted by $su(1,1|1)$. Together with the multiplet structure of the bosonic spectrum, which requires the terms of the invariant that arise from the gauged Bianchi identity, this is a nontrivial test of the gauged Gauss-Bonnet invariant, including its two-fermion terms.

From the point of view of effective field theory, the curvature squared invariants are the leading corrections to the two-derivative action, and their couplings are small in units of the AdS radius. In this regime, the additional modes around the AdS$_5$ vacuum have masses of order $\ell^{-1}$ times the inverse square root of $|\alpha|g^2$, $|\beta|g^2$ or $|\zeta|g^2$, and those around the AdS$_3\times\Sigma_2$ backgrounds have energies of order $k^{-1}$. These modes lie beyond the regime of validity of the four-derivative truncation, and their negative norms are not a problem of the effective theory, as in \cite{Pang:2012xs}. The same applies to the accumulation of states on AdS$_3\times\Sigma_2$, which sets in at levels of order $k^{-1}$, whereas the truncation can be trusted only at energies and levels small compared to $|k|^{-1/2}$. Taken as theories in their own right, with finite couplings, the off-shell models we have studied are free of ghosts and instabilities around the AdS$_5$ vacuum only for the Ricci scalar squared extension in the range $0<\zeta g^2<\tfrac14$ and for the Gauss-Bonnet extension, and the latter loses this property around the supersymmetric AdS$_3\times\Sigma_2$ backgrounds. Thus the absence of ghosts around one supersymmetric vacuum does not imply their absence around another.

There are a number of directions to pursue following our work. First of all, we have not analyzed the boundary conditions. Around the AdS$_5$ vacuum, they determine whether the logarithmic modes at the critical points can be removed, as has been discussed for critical gravity \cite{Lu:2011zk,Porrati:2011ku}, although in our case negative-norm modes that are not logarithmic partners remain. Around the AdS$_3\times\Sigma_2$ backgrounds, they may remove some of the negative-norm descendants, and together with the global multiplicities of the charged harmonics, they determine the spectrum beyond the primaries we have considered. Furthermore, at $\tau=2$, where the two supersymmetric vacua merge, the quadratic action of the Gauss-Bonnet extension vanishes identically, as for Einstein-Gauss-Bonnet gravity at the corresponding point \cite{Fan:2016zfs}, and that of the LOG extension reduces to the $C^2+\tfrac16R^2$ invariant alone, with a partially massless spin-2 mode and additional gauge symmetries. For the Gauss-Bonnet extension, the spectrum at this point is not determined by the linearized analysis, and both cases deserve future investigation. It would also be interesting to construct the black string solutions that interpolate between the AdS$_5$ vacuum and the AdS$_3\times\Sigma_2$ backgrounds of Section~\ref{sec:GBprod}, since such solutions have not been constructed in this theory. Finally, the models we have considered contain a single physical vector multiplet. Coupling further vector multiplets and hypermultiplets, as in \cite{Baggio:2014hua}, would allow a comparison with specific superconformal field theories, and the spectra of the resulting models also remain to be investigated.

\section*{Acknowledgments and Use of AI}
\addcontentsline{toc}{section}{Acknowledgments and Use of AI}

We are grateful to Saurish Khandelwal, Yi Pang and Gabriele Tartaglino-Mazzucchelli for communication on the $g$-dependent terms of the gauged curvature squared invariants. C.O. is supported in part by the Istanbul Technical University Research Fund under grant number TDK-2026-48977. C.B.S. is supported in part by the Istanbul Technical University Research Fund under grant number TDK-2025-47479. M.O. is supported in part by T\"UB\.ITAK grant 125F471 and by the Istanbul Technical University Research Fund under grant number PTA-2026-47860, and acknowledges support from the T\"UB\.ITAK Incentive Award.

The question addressed in this paper and the plan of the analysis are ours, but most of the explicit work was carried out with artificial intelligence. The public repository \cite{repo}, which contains the component Lagrangians up to quadratic order in fermions, their quadratic actions around the supersymmetric backgrounds and the resulting spectra, was built with Claude Opus 5.5 (Anthropic) and GPT-6 Astra (OpenAI). The manuscript was written with Claude Opus 5.5 and Claude Fable 5.1, section by section. Each section was then independently derived and audited with GPT-6 Astra, including the fermionic sectors, and all authors repeated these checks with further models. The corrections to the gauged invariants pointed out in footnote~1 were confirmed by the authors of \cite{Gold:2023ymc,Gold:2025ttt}. All checks are included as scripts in the repository. Several results also provide checks that do not depend on these tools: the two-derivative limits reproduce the known spectra, the central charges of the Gauss-Bonnet extension coincide with those of Einstein-Gauss-Bonnet gravity, the AdS$_3$ central charges agree with $c$-extremization and, at first order in the coupling, with the universal-twist values of \cite{Bobev:2017uzs}, and the fermionic spectra complete the supermultiplets predicted by the bosonic ones. The authors take full responsibility for the entire paper.

\clearpage %
\appendix

\section{Conventions} \label{app:conventions}
In this appendix, we summarize our conventions, which follow those of \cite{Bergshoeff:2001hc,Ozkan:2013uk}. We use the mostly plus signature, $\eta_{ab}={\rm diag}(-,+,+,+,+)$. Curved indices are denoted by $\mu,\nu,\ldots=0,\ldots,4$, flat indices by $a,b,\ldots=0,\ldots,4$, and $SU(2)$ indices by $i,j,\ldots=1,2$. The vielbein satisfies $g_{\mu\nu}=e_\mu{}^ae_\nu{}^b\eta_{ab}$, and $e=\det(e_\mu{}^a)=\sqrt{-\det g_{\mu\nu}}$. We write $e^{-1}\cL$ for the scalar Lagrangians, whose overall normalization is given in (\ref{eq:general}), and all Lagrangians are given up to total derivatives. Symmetrization and antisymmetrization are with unit weight, for instance $X_{[ab]}=\tfrac12(X_{ab}-X_{ba})$, and indices enclosed by vertical bars are excluded.

The Levi-Civita symbol with flat indices is constant, with $\epsilon^{01234}=1$ and therefore $\epsilon_{01234}=-1$. With curved indices, $\epsilon^{\mu\nu\rho\sigma\lambda}=e\,e_a{}^\mu e_b{}^\nu e_c{}^\rho e_d{}^\sigma e_f{}^\lambda\epsilon^{abcdf}$ denotes the corresponding tensor density, whose components are those of the symbol, so that $e^{-1}\epsilon^{\mu\nu\rho\sigma\lambda}$ in (\ref{eq:strengths}) is a tensor. The Levi-Civita symbol satisfies
\begin{equation}
\epsilon_{a_1\cdots a_nb_1\cdots b_p}\epsilon^{a_1\cdots a_nc_1\cdots c_p}=-n!\,p!\,\delta_{[b_1}{}^{c_1}\cdots\delta_{b_p]}{}^{c_p}\,,\qquad n+p=5\,.
\end{equation}
The gamma matrices satisfy
\begin{equation}
\{\gamma^a,\gamma^b\}=2\eta^{ab}\,,\qquad
\gamma^{a_1\cdots a_n}=\gamma^{[a_1}\cdots\gamma^{a_n]}\,,\qquad
\gamma^{abcdf}=\rmi\epsilon^{abcdf}\,,
\end{equation}
with $\gamma_\mu=e_\mu{}^a\gamma_a$, where $\gamma_0$ is anti-Hermitian and $\gamma_1,\ldots,\gamma_4$ are Hermitian. As a consequence,
\begin{equation}
\gamma^{a_1\cdots a_n}=\frac{\rmi}{(5-n)!}\,\epsilon^{a_1\cdots a_nb_1\cdots b_{5-n}}\gamma_{b_{5-n}\cdots b_1}\,.
\end{equation}
We use the notation $\slashed X=\gamma^aX_a$ and, for an antisymmetric tensor, $\gamma\cdot X=\gamma^{a_1\cdots a_n}X_{a_1\cdots a_n}$, without a factor $1/n!$.

The $SU(2)$ indices are raised and lowered with $\varepsilon^{ij}$ and $\varepsilon_{ij}$ according to
\begin{equation}
X^i=\varepsilon^{ij}X_j\,,\qquad X_i=X^j\varepsilon_{ji}\,,\qquad \varepsilon_{12}=\varepsilon^{12}=1\,.
\end{equation}
The gauging selects the direction $\delta^{ij}$ in $SU(2)$, which is real and symmetric, and $\delta_{ij}=\varepsilon_{ik}\varepsilon_{jl}\delta^{kl}$ has the same components. The spinors $\psi_\mu^i$, $\varphi^i$ and the supersymmetry parameter $\epsilon^i$ are anticommuting symplectic Majorana spinors. With the charge conjugation matrix $\mathsf C$, which satisfies $\mathsf C^T=-\mathsf C$, $(\mathsf C\gamma_a)^T=-\mathsf C\gamma_a$ and $\mathsf C^\dagger\mathsf C=1$, the Majorana conjugate and the symplectic Majorana condition are
\begin{equation}
\bar\zeta^i=(\zeta^i)^T\mathsf C\,,\qquad
\zeta^i=\mathsf B^{-1}\varepsilon^{ij}(\zeta^j)^*\,,\qquad
\mathsf B=\mathsf C\gamma_0\,,
\end{equation}
where $\mathsf B^*\mathsf B=-1$. In bilinears with suppressed $SU(2)$ indices, the indices are contracted as $\bar\lambda\chi=\bar\lambda^i\chi_i$, and the bilinears satisfy
\begin{equation}
\bar\lambda^i\gamma_{a_1\cdots a_n}\chi^j=t_n\,\bar\chi^j\gamma_{a_1\cdots a_n}\lambda^i\,,\qquad
\bar\lambda\gamma_{a_1\cdots a_n}\chi=-t_n\,\bar\chi\gamma_{a_1\cdots a_n}\lambda\,,
\end{equation}
with $t_0=t_1=1$ and $t_2=t_3=-1$.

The covariant derivative of a spinor is $\nabla_\mu\zeta^i=\big(\partial_\mu+\tfrac14\omega_\mu{}^{ab}\gamma_{ab}\big)\zeta^i$, where the spin connection $\omega_\mu{}^{ab}$ also contains terms bilinear in the gravitino in the complete Lagrangians \cite{repo}, and the $SU(2)$ connection enters as in (\ref{eq:susy}). The curvatures are given by
\begin{align}
&R_{\mu\nu}{}^{ab}=2\partial_{[\mu}\omega_{\nu]}{}^{ab}+2\omega_{[\mu}{}^{ac}\omega_{\nu]c}{}^b\,,\qquad
R_{\mu\nu\rho\sigma}=R_{\mu\nu}{}^{ab}e_{\rho a}e_{\sigma b}\,,\nonumber\\*
&R_{\mu\nu}=g^{\rho\sigma}R_{\rho\mu\sigma\nu}\,,\qquad
R=g^{\mu\nu}R_{\mu\nu}\,,
\end{align}
so that $[\nabla_\mu,\nabla_\nu]\zeta=\tfrac14R_{\mu\nu}{}^{ab}\gamma_{ab}\zeta$ and $[\nabla_\mu,\nabla_\nu]X^\rho=R_{\mu\nu}{}^\rho{}_\sigma X^\sigma$. In these conventions, a space of constant curvature with $R_{\mu\nu\rho\sigma}=k\big(g_{\mu\rho}g_{\nu\sigma}-g_{\mu\sigma}g_{\nu\rho}\big)$ has $R_{\mu\nu}=4kg_{\mu\nu}$ and $R=20k$, and AdS$_5$ of radius $\ell$ has $k=-\ell^{-2}$. The Weyl tensor is
\begin{equation}
C_{\mu\nu\rho\sigma}=R_{\mu\nu\rho\sigma}-\tfrac13\big(g_{\mu\rho}R_{\nu\sigma}-g_{\nu\rho}R_{\mu\sigma}-g_{\mu\sigma}R_{\nu\rho}+g_{\nu\sigma}R_{\mu\rho}\big)+\tfrac1{12}\big(g_{\mu\rho}g_{\nu\sigma}-g_{\mu\sigma}g_{\nu\rho}\big)R\,,
\end{equation}
so that $C_{\mu\nu\rho\sigma}C^{\mu\nu\rho\sigma}=R_{\mu\nu\rho\sigma}R^{\mu\nu\rho\sigma}-\tfrac43R_{\mu\nu}R^{\mu\nu}+\tfrac16R^2$. The superconformal covariant derivatives $\mathcal D_\mu$ and the fields $T_{ab}$, $\chi^i$ and $D$ of the standard Weyl multiplet, which appear in Appendix~\ref{app:composite}, are those of \cite{Bergshoeff:2001hc}, and the quantities $\widehat E_a$, $\widehat F_{\mu\nu}$ and $\widehat R_{\mu\nu}{}^{ij}(V)$ there are supercovariant.

In Section~\ref{sec:GBprod}, we use Poincar\'e coordinates on AdS$_3$, $ds^2({\rm AdS}_3)=\ell^2z^{-2}\big(-dt^2+dx^2+dz^2\big)$, with the frame $e^0=\ell z^{-1}dt$, $e^1=\ell z^{-1}dx$ and $e^2=\ell z^{-1}dz$, and $e^3$, $e^4$ is an orthonormal frame on $\Sigma_2$, so that $\epsilon^{01234}=1$ fixes the orientation of AdS$_3\times\Sigma_2$. Finally, a bar has the following meanings. On a spinor, it denotes the Majorana conjugate. On $Z_\mu$, $z_\mu$ and their field strengths, it denotes complex conjugation. In Section~\ref{sec:GBprod}, $\bar h$ and $\bar T$ denote the right-moving weight and stress tensor. Otherwise, it denotes a quantity evaluated on the background, such as $\bar g_{\mu\nu}$, $\bar\nabla_\mu$ or $\bar m$. A superscript $(1)$ denotes the linearization of a quantity, and the superscript ${\rm TT}$ the transverse traceless part of the metric fluctuation.

\section{The Composite Vector Multiplet}
\label{app:composite}

In this appendix, we present the composite expressions for the components of an off-shell vector multiplet in terms of the components of a linear multiplet $(L^{ij},\varphi^i,\widehat E_a,N)$ in the standard Weyl multiplet background. These expressions are the main ingredients of the linear multiplet action, and therefore of the two-derivative Lagrangian (\ref{eq:LLR}), as well as of the Ricci scalar squared invariant (\ref{eq:LR2}) \cite{Coomans:2012cf,Ozkan:2013uk,Ozkan:2013nwa}. Their bosonic parts are sufficient for the bosonic Lagrangians of Section~\ref{sec:theory}, while their fermionic parts determine the fermionic terms of these Lagrangians given in \cite{repo}.

Here $\widehat E_a$ is the supercovariant vector of the linear multiplet, which is subject to the constraint $\mathcal D_a\widehat E^a=0$. This constraint is solved in terms of the three-form $E_{\mu\nu\rho}$, and the solution relates $\widehat E^a$ to the bosonic dual field strength $E^a=e_\mu{}^aE^\mu$ defined in (\ref{eq:strengths}) as \cite{Ozkan:2016csy}
\begin{equation}
E^a=\widehat E^a+\tfrac12\rmi\bar\psi_b^i\gamma^{ba}\varphi_i+\tfrac14\bar\psi_b^i\gamma^{abc}\psi_c^jL_{ij}\,.
\label{eq:EhatE}
\end{equation}
The two vectors coincide when the fermions vanish. For this reason, the bosonic Lagrangians of Section~\ref{sec:theory} are written in terms of $E^a$, whereas the composite expressions below, which are covariant under supersymmetry, are written in terms of $\widehat E_a$. Notice that in the action for a vector multiplet coupled to a linear multiplet \cite{Coomans:2012cf,Ozkan:2016csy}, the vector field couples to the bosonic dual through the term $eA_\mu E^\mu$, which is gauge invariant up to a total derivative since $\partial_\mu(eE^\mu)=0$ identically. Therefore, the relation (\ref{eq:EhatE}) is needed when the fermionic terms of (\ref{eq:LLR}) and (\ref{eq:LR2}) are expressed in terms of the independent fields of Section~\ref{sec:theory}.

The supersymmetry transformation rules of the scalar and the fermion of a vector multiplet $(\rho,A_\mu,\lambda^i,Y^{ij})$ are given by \cite{Bergshoeff:2002qk,Ozkan:2016csy}
\begin{align}
\delta\rho={}&\tfrac12\rmi\bar\epsilon\lambda\,,\nonumber\\*
\delta\lambda^i={}&-\tfrac14\gamma\cdot\widehat F\epsilon^i-\tfrac12\rmi\slashed{\mathcal D}\rho\,\epsilon^i+\rho\gamma\cdot T\epsilon^i-Y^{ij}\epsilon_j+\rho\eta^i\,.
\label{eq:vecsusy}
\end{align}
The lowest component of the vector multiplet can be expressed in terms of the components of the linear multiplet as $\rho=2L^{-1}N+\rmi L^{-3}L_{ij}\bar\varphi^i\varphi^j$, where $L^2=L_{ij}L^{ij}$. Applying a sequence of supersymmetry transformations, we obtain the composite expressions for the remaining components
\begin{align}
\lambda_i={}&-2\rmi L^{-1}\slashed{\mathcal D}\varphi_i+16L^{-1}L_{ij}\chi^j+4L^{-1}\gamma\cdot T\varphi_i-2L^{-3}NL_{ij}\varphi^j\nonumber\\
&+2\rmi L^{-3}\big(L^{jk}\slashed{\mathcal D}L_{ij}\varphi_k-L_{ij}\widehat{\slashed{E}}\varphi^j\big)+2\rmi L^{-3}\varphi^j\bar\varphi_i\varphi_j-6\rmi L^{-5}L_{ij}L^{kl}\varphi^j\bar\varphi_k\varphi_l\,,\label{eq:compl}\\[4pt]
Y_{ij}={}&L^{-1}\Box^cL_{ij}-L^{-3}L^{km}\mathcal D_aL_{k(i}\mathcal D^aL_{j)m}-L^{-3}L_{ij}\big(N^2+\widehat E_a\widehat E^a\big)\nonumber\\
&+2L^{-3}\widehat E_aL_{k(i}\mathcal D^aL_{j)}{}^k+\tfrac83L^{-1}L_{ij}T_{ab}T^{ab}+4L^{-1}L_{ij}D+\rmi L^{-3}N\bar\varphi_{(i}\varphi_{j)}\nonumber\\
&-3\rmi L^{-5}NL_{ij}L^{kl}\bar\varphi_k\varphi_l-L^{-3}\bar\varphi_{(i}\widehat{\slashed{E}}\varphi_{j)}+3L^{-5}L_{ij}L^{kl}\bar\varphi_k\widehat{\slashed{E}}\varphi_l\nonumber\\
&-16\rmi L^{-1}\bar\chi_{(i}\varphi_{j)}+16\rmi L^{-3}L_{ij}L_{kl}\bar\chi^k\varphi^l+2L^{-3}L_{k(i}\bar\varphi^k\slashed{\mathcal D}\varphi_{j)}\nonumber\\
&+2\rmi L^{-3}L_{ij}\bar\varphi\gamma\cdot T\varphi-3L^{-5}L_{k(i}L^{lm}\bar\varphi^k\slashed{\mathcal D}L_{j)l}\varphi_m+\tfrac12L^{-3}\bar\varphi^k\slashed{\mathcal D}L_{k(i}\varphi_{j)}\nonumber\\
&+\tfrac{15}{2}L^{-7}L_{ij}\big(L^{kl}\bar\varphi_k\varphi_l\big)^2-3L^{-5}L^{kl}\bar\varphi_k\varphi_l\,\bar\varphi_{(i}\varphi_{j)}-\tfrac32L^{-5}L_{ij}\,\bar\varphi_k\varphi_l\,\bar\varphi^k\varphi^l\,,\label{eq:compY}\\[4pt]
\widehat F_{\mu\nu}={}&4\mathcal D_{[\mu}\big(L^{-1}\widehat E_{\nu]}\big)+2L^{-1}L_{ij}\widehat R_{\mu\nu}{}^{ij}(V)-2L^{-3}L_k{}^l\mathcal D_{[\mu}L^{kp}\mathcal D_{\nu]}L_{lp}\nonumber\\
&-2\mathcal D_{[\mu}\big(L^{-3}L_{ij}\bar\varphi^i\gamma_{\nu]}\varphi^j\big)-\rmi L^{-1}\bar\varphi^i\widehat R_{\mu\nu i}(Q)\,.\label{eq:compF}
\end{align}
Here $T_{ab}$, $\chi^i$ and $D$ are the matter fields of the standard Weyl multiplet \cite{Bergshoeff:2001hc}, which become composite in the dilaton Weyl multiplet \cite{Bergshoeff:2001hc,Coomans:2012cf}, $\widehat R_{\mu\nu}{}^{ij}(V)$ and $\widehat R_{\mu\nu}{}^i(Q)$ are the supercovariant curvatures of $V_\mu{}^{ij}$ and $\psi_\mu^i$, and we use the notation $\gamma\cdot T=\gamma^{ab}T_{ab}$, $\widehat{\slashed{E}}=\gamma^a\widehat E_a$ and $\bar\varphi\gamma\cdot T\varphi=\bar\varphi^i\gamma\cdot T\varphi_i$. The supercovariant derivatives of the components of the linear multiplet are given by \cite{Coomans:2012cf,Ozkan:2016csy}
\begin{align}
\mathcal D_\mu L^{ij}={}&\partial_\mu L^{ij}-3b_\mu L^{ij}+2V_\mu{}^{(i}{}_kL^{j)k}-\rmi\bar\psi_\mu^{(i}\varphi^{j)}\,,\nonumber\\
\mathcal D_\mu\varphi^i={}&\big(\partial_\mu-\tfrac72b_\mu+\tfrac14\omega_\mu{}^{ab}\gamma_{ab}\big)\varphi^i-V_\mu{}^{ij}\varphi_j+\tfrac12\rmi\big(\slashed{\mathcal D}L^{ij}\big)\psi_{\mu j}\nonumber\\
&+\tfrac12\rmi\widehat{\slashed{E}}\psi_\mu^i-\tfrac12N\psi_\mu^i+\gamma\cdot T L^{ij}\psi_{\mu j}-3L^{ij}\phi_{\mu j}\,,\nonumber\\
\mathcal D_\mu\widehat E_a={}&\partial_\mu\widehat E_a-4b_\mu\widehat E_a+\omega_{\mu ab}\widehat E^b+\tfrac12\rmi\bar\psi_\mu\gamma_{ab}\mathcal D^b\varphi-2\bar\psi_\mu\gamma^b\varphi\,T_{ab}+2\bar\phi_\mu\gamma_a\varphi\,,\label{eq:Dlin}
\end{align}
where $b_\mu$ and $\phi_\mu^i$ are the gauge fields of dilatations and $S$-supersymmetry. The coefficient of the term $\big(\slashed{\mathcal D}L^{ij}\big)\psi_{\mu j}$ in $\mathcal D_\mu\varphi^i$ is fixed by supercovariance. The supersymmetry transformation rule of $\varphi^i$ contains $\delta\varphi^i\supset-\tfrac12\rmi\big(\slashed{\mathcal D}L^{ij}\big)\epsilon_j$ \cite{Ozkan:2016csy}, and since $\delta\psi_\mu^i\supset\partial_\mu\epsilon^i$, the terms proportional to $\big(\slashed{\mathcal D}L^{ij}\big)\partial_\mu\epsilon_j$ cancel in $\delta\big(\mathcal D_\mu\varphi^i\big)$ only if this coefficient is $\tfrac12\rmi$. This corrects the coefficient $\tfrac12$ given in eq.~(3.7) of \cite{Ozkan:2016csy}. The superconformal d'Alembertian of $L^{ij}$ is given by \cite{Ozkan:2016csy}
\begin{align}
\Box^cL^{ij}={}&\big(\partial^a-4b^a+\omega_b{}^{ba}\big)\mathcal D_aL^{ij}+2V_a{}^{(i}{}_k\mathcal D^aL^{j)k}+6f_a{}^aL^{ij}\nonumber\\*
&-\rmi\bar\psi^{a(i}\mathcal D_a\varphi^{j)}+4\bar\psi^{a(i}\gamma_a\chi^{k)}L^j{}_k+4\bar\psi^{a(j}\gamma_a\chi^{k)}L^i{}_k\nonumber\\*
&-6L^{ij}\bar\psi^a\gamma_a\chi-\bar\varphi^{(i}\gamma\cdot T\gamma^a\psi_a^{j)}+\bar\varphi^{(i}\gamma^a\phi_a^{j)}\,,\label{eq:boxc}
\end{align}
where $f_\mu{}^a$ is the gauge field of special conformal transformations. The dependent gauge fields $\omega_\mu{}^{ab}$, $\phi_\mu^i$ and $f_\mu{}^a$ are given in \cite{Bergshoeff:2001hc}. The composite expressions (\ref{eq:compl})--(\ref{eq:compF}) and the d'Alembertian (\ref{eq:boxc}) are based on the corrected derivative $\mathcal D_\mu\varphi^i$ in (\ref{eq:Dlin}). It can be checked that the supersymmetry transformation of (\ref{eq:compl}) reproduces the transformation rule (\ref{eq:vecsusy}) with $Y_{ij}$ and $\widehat F_{\mu\nu}$ given in (\ref{eq:compY}) and (\ref{eq:compF}).

On the AdS$_5$ backgrounds considered in this paper, the fermions as well as $\mathcal D_aL^{ij}$, $\widehat E_a$ and $\widehat F_{\mu\nu}$ vanish. Consequently, the term $\big(\slashed{\mathcal D}L^{ij}\big)\psi_{\mu j}$ in (\ref{eq:Dlin}), which is fermionic and quadratic in the fluctuations, and the fermion bilinears in (\ref{eq:EhatE}), which enter the action multiplied by fields that vanish on these backgrounds, contribute only to terms that are at least cubic in the fluctuations. Therefore, neither the correction of $\mathcal D_\mu\varphi^i$ nor the distinction between $E^a$ and $\widehat E^a$ affects the quadratic actions around these backgrounds.

These expressions agree with the single linear multiplet case, $\mathcal F_{AB}=\delta_{AB}L^{-1}$, of the general composite expressions given in \cite{Ozkan:2016csy}. The expressions for $\rho$, $\lambda_i$ and $\widehat F_{\mu\nu}$ coincide with those given in eq.~(3.4) of \cite{Ozkan:2013uk}, as does the bosonic part of $Y_{ij}$. However, the terms in $Y_{ij}$ that are quadratic in $\varphi^i$ and proportional to $N$, $\widehat E_a$ or $\mathcal D_aL_{ij}$, as well as the quartic fermion terms, differ from those given there.\footnote{In the basis used in \cite{Ozkan:2013uk}, the coefficient of $L^{-5}NL_{k(i}L_{j)m}\bar\varphi^k\varphi^m$ should read $-3\rmi$, the coefficients of $L^{-3}\bar\varphi_{(i}\widehat{\slashed{E}}\varphi_{j)}$ and $L^{-5}L_{k(i}L_{j)m}\bar\varphi^k\widehat{\slashed{E}}\varphi^m$ should read $\tfrac12$ and $3$, and the coefficients of $L^{-3}\bar\varphi_{(i}\slashed{\mathcal D}L_{j)k}\varphi^k$ and $L^{-5}L_{mn}L^k{}_{(i}\bar\varphi^m\slashed{\mathcal D}L_{j)k}\varphi^n$ should read $-1$ and $3$. The quartic fermion terms cannot be written in the basis used there and are given by the last line of (\ref{eq:compY}).} The first of these terms contributes to the mass terms of the spin-1/2 fields around the AdS$_5$ vacua studied in this paper, where $N\neq0$. The fermionic Lagrangians given in \cite{repo} are based on (\ref{eq:compl})--(\ref{eq:compF}).

\section{Polynomials of Section \ref{sec:GBprod}}
\label{app:quintic}

In this appendix, we collect the longer polynomials of Section~\ref{sec:GBprod}. The polynomial $Y$ that enters $\mu\ell_s$ in (\ref{eq:munuf}) and $\kappa_{\rm eff}$ in (\ref{eq:kappaeff}) is
\begin{equation}
Y=128-224k+16k^2-80kL+120k^2L+24k^2L^2-20k^3L^2-6k^3L^3+k^4L^4\,,
\label{eq:Ypoly}
\end{equation}
and the polynomial $X$ that enters the central charge $c_L$ in (\ref{eq:cLcR}) is
\begin{equation}
X=64-128k+16k^2-48kL+72k^2L+20k^2L^2-14k^3L^2-6k^3L^3+k^4L^4\,.
\label{eq:Xpoly}
\end{equation}
On the torus, where $k$ is given by (\ref{eq:T2curve}), the cubic $R(D)$ that determines the $(1,0)$ multiplets at the constant level reduces to
\begin{multline}
27(L-4)R(D)=27(L-4)D^3-9\big(3L^3-30L^2+98L-96\big)D^2\\
-3L(L-2)\big(6L^2-53L+114\big)D-4L^2(L-2)(L-3)(L-6)\,.
\label{eq:T2cubic}
\end{multline}

Finally, we give the polynomial (\ref{eq:quintic}) explicitly. We write it as
\begin{align}
F(D,M^2)={}&D^5+f_4D^4+f_3D^3+f_2D^2+f_1D\nonumber\\*
&+M^2\big(b_3D^3+b_2D^2+b_1D+b_0\big)+M^4\,\frac{P_4P_5}{P_1^2P_2^2}\Big(D-\nu_f+1\Big)\,,
\end{align}
with $P_i$ defined in (\ref{eq:Pdef}), $\nu_f$ in (\ref{eq:munuf}) and $k=\lambda g^2$. The coefficients of the level-independent part are
\begin{align}
f_4&=\frac{\Pi_4}{kP_1P_3P_4}\,,&
f_3&=\frac{\Pi_3}{kP_1P_3^2P_4}\,,\nonumber\\*
f_2&=\frac{2P_2\,\Pi_2}{kP_1P_3^3P_4}\,,&
f_1&=\frac{16L^2(3k-2)P_2^3P_5}{kP_1P_3^3P_4}\,,
\end{align}
where
\begin{align}
\Pi_4={}&-256-256k+448kL+64k^2+352k^2L-368k^2L^2\nonumber\\
&-64k^3L-120k^3L^2+136k^3L^3+24k^4L^3-24k^4L^4+k^5L^5\,,\nonumber\\
\Pi_3={}&2048L+2048k-2560kL-3328kL^2-3840k^2L+4736k^2L^2\nonumber\\
&+2048k^2L^3+256k^3L+1280k^3L^2-2208k^3L^3-816k^3L^4\nonumber\\
&-208k^4L^3+440k^4L^4+232k^4L^5-24k^5L^5-42k^5L^6+3k^6L^7\,,\nonumber\\
\Pi_2={}&2048+512L^2-1024k-1024kL-1280kL^2-1024kL^3+2048k^2L\nonumber\\
&-1664k^2L^2+2048k^2L^3+1056k^2L^4-320k^3L^2+384k^3L^3-784k^3L^4\nonumber\\
&-688k^3L^5-32k^4L^4+144k^4L^5+244k^4L^6-8k^5L^6-44k^5L^7+3k^6L^8\,.
\end{align}
These coefficients satisfy $F(2,0)=0$, so that $F(D,0)=D(D-2)R(D)$ with
\begin{equation}
R(D)=D^3+\big(f_4+2\big)D^2+\big(f_3+2f_4+4\big)D+f_2+2f_3+4f_4+8\,.
\end{equation}
The coefficients of the terms linear in $M^2$ are
\begin{align}
b_3&=-1-\frac{P_4P_5}{P_1^2P_2^2}\,,&
b_2&=\frac{4\,\Pi_{21}}{kP_1^2P_3}\,,\nonumber\\*
b_1&=-\frac{8L\big(8-8k-4kL+3k^2L\big)P_5}{kP_1^2P_3^2}\,,&
b_0&=\frac{16LP_2P_5\big(2L-8+3kL-kL^2\big)}{kP_1P_3^3}\,,
\end{align}
with
\begin{equation}
\Pi_{21}=32+24k+12k^2-56kL-28k^2L+38k^2L^2+5k^3L^2-11k^3L^3+k^4L^4\,.
\end{equation}
Notice that $2L-8+3kL-kL^2$ is the factor that determines the sign of the curvature $v$ in (\ref{eq:prodsol}). In the two-derivative limit, $f_4$, $f_3$, $f_2$, $f_1$ and $b_2,b_1,b_0$ diverge as $1/k$, and the polynomial $kF$ has degree four in $D$, which reflects the decoupling of one $(1,0)$ multiplet.

\clearpage
\providecommand{\href}[2]{#2}\begingroup\raggedright\endgroup

\end{document}